\documentclass[final,number,sort&compress,5p,times,twocolumn]{elsarticle}

\usepackage[]{graphicx}
\usepackage{booktabs}
\usepackage[draft]{todonotes}   

\usepackage{amssymb}
\usepackage{eurosym}

\usepackage{multicol}

\usepackage{subfigure}

\journal{Nuclear Instruments and Methods in Physics Research Section A}

\usepackage{etoolbox}
\makeatletter
    \patchcmd{\tnotemark}{\ding{73}}{*}{}{\@latex@error{Failed to path \string\tnotemark\space for \string\ding{73}}}
    \patchcmd{\tnotemark}{\ding{73}\ding{73}}{\dag}{}{\@latex@error{Failed to path \string\tnotemark\space for \string\ding{73}\string\ding{73}}}
    \patchcmd{\tnotetext}{\ding{73}}{*}{}{\@latex@error{Failed to path \string\tnotetext\space for \string\ding{73}}}
    \patchcmd{\tnotetext}{\ding{73}\ding{73}}{\dag}{}{\@latex@error{Failed to path \string\tnotetext\space for \string\ding{73}\string\ding{73}}}
\makeatother

\usepackage{etoolbox}
\makeatletter
\patchcmd{\ps@pprintTitle}
  {Preprint submitted to}
  {Accepted manuscript by}
  {}{}
\makeatother

\begin{document}


\begin{frontmatter}

\title{Characterisation and Testing of CHEC-M -- a camera prototype for the Small-Sized Telescopes of the Cherenkov Telescope Array}

\author[mpik]{J.~Zorn\tnoteref{cor}}
\ead{justus.zorn@mpi-hd.mpg.de}
\author[mpik]{R.~White\tnoteref{cor}}
\ead{richard.white@mpi-hd.mpg.de}
\author[uoo]{J.J.~Watson\tnoteref{cor}}
\ead{jason.watson@physics.ox.ac.uk}

\author[uoo,Durham]{T.P.~Armstrong}
\author[uva]{A.~Balzer}
\author[mpik]{M.~Barcelo}
\author[uva]{D.~Berge\fnref{fn1}}
\author[washu]{R.~Bose}
\author[Durham]{A.M.~Brown}
\author[uva]{M.~Bryan}
\author[Durham]{P.M.~Chadwick}
\author[Durham]{P.~Clark}
\author[cppm]{H.~Costantini}
\author[uoo]{G.~Cotter}
\author[paris]{L.~Dangeon}
\author[uoli]{M.~Daniel\fnref{fn2}}
\author[uoo]{A.~De Franco}
\author[ecap]{P.~Deiml}
\author[paris]{G.~Fasola}
\author[ecap,slac]{S.~Funk}
\author[uva]{M.~Gebyehu}
\author[dt-insu]{J.~Gironnet}
\author[Durham]{J.A.~Graham}
\author[uoli]{T.~Greenshaw}
\author[mpik]{J.A.~Hinton}
\author[ecap]{M.~Kraus}
\author[uol]{J.S.~Lapington}
\author[paris]{P.~Laporte}
\author[uol]{S.A.~Leach}
\author[paris]{O.~Le Blanc}
\author[Adelaide]{A.~Malouf}
\author[uol]{P.~Molyneux}
\author[washu]{P.~Moore\tnoteref{tn}}
\author[uva]{H.~Prokoph\fnref{fn1}}
\author[nau]{A.~Okumura}
\author[uol]{D.~Ross}
\author[Adelaide]{G.~Rowell}
\author[slac]{L.~Sapozhnikov}
\author[mpik]{H.~Schoorlemmer}
\author[paris]{H.~Sol}
\author[uva]{M.~Stephan}
\author[nau]{H.~Tajima}
\author[mpik]{L.~Tibaldo\fnref{fn3}}
\author[hawai]{G.~Varner}
\author[ecap]{A.~Zink}

\address[mpik]{Max-Planck-Institut f\"{u}̈r Kernphysik, P.O. Box 103980, 69029 Heidelberg, Germany}
\address[uoo]{Department of Physics, University of Oxford, Keble Road, Oxford OX1 3RH, UK}
\address[Durham]{Department of Physics and Centre for Advanced Instrumentation, Durham University, South Road, Durham DH1 3LE, UK}
\address[uva]{GRAPPA, University of Amsterdam, Science Park 904, 1098 XH Amsterdam, The Netherlands}
\address[washu]{Department of Physics, Washington University, St. Louis, MO 63130, USA}
\address[cppm]{Aix Marseille Universit\'{e}, CNRS/IN2P3, CPPM, 163 avenue de Luminy, case 902, 13288 Marseille, France}
\address[paris]{Observatoire de Paris, CNRS, PSL University, LUTH \& GEPI, Place J. Janssen, 92195, Meudon cedex, France}
\address[uoli]{University of Liverpool, Oliver Lodge Laboratory, P.O. Box 147, Oxford Street, Liverpool L69 3BX, UK}
\address[ecap]{Erlangen Centre for Astroparticle Physics (ECAP), Erwin-Rommel-Str. 1, D 91058 Erlangen, Germany}
\address[slac]{Kavli Institute for Particle Astrophysics and Cosmology, Department of Physics and SLAC National
Accelerator Laboratory, Stanford University, 2575 Sand Hill Road, Menlo Park, CA 94025, USA}
\address[dt-insu]{CNRS, Division technique DT-INSU, 1 Place Aristide Briand, 92190 Meudon, France}
\address[uol]{Department of Physics and Astronomy, University of Leicester, University Road, Leicester, LE1 7RH, UK}
\address[Adelaide]{School of Physical Sciences, University of Adelaide, Adelaide5005, Australia}
\address[nau]{Institute for Space--Earth Environmental Research, Nagoya University, Furo-cho, Chikusa-ku, Nagoya, Aichi 464-8601, Japan}
\address[hawai]{University of Hawai'i at Manoa, 2500 Campus Rd, Honolulu, HI, 96822, USA}

\tnotetext[cor]{Corresponding author}
\tnotetext[tn]{deceased on 20/05/2017}

\fntext[fn1]{now at: Deutsches Elektronen-Synchrotron, Platanenallee 6, 15738 Zeuthen, Germany}
\fntext[fn2]{now at: Harvard-Smithsonian Center for Astrophysics, 60 Garden Street, Cambridge, MA 02138, USA}
\fntext[fn3]{now at: Institut de Recherche en Astrophysique et Plan\'{e}tologie, CNRS-INSU, Universit\'{e} Paul Sabatier, 9
avenue Colonel Roche, BP 44346, 31028 Toulouse Cedex 4, France}

\begin{abstract}
%
The Compact High Energy Camera (CHEC) is a camera design for the Small-Sized Telescopes (SSTs; 4 m diameter mirror) of the Cherenkov Telescope Array (CTA). The SSTs are focused on very-high-energy $\gamma$-ray detection via atmospheric Cherenkov light detection over a very large area. This implies many individual units and hence cost-effective implementation, as well as shower detection at large impact distance, and hence large field of view (FoV), and efficient image capture in the presence of large time gradients in the shower image detected by the camera. 
CHEC relies on dual-mirror optics to reduce the plate-scale and make use of 6~$\times$~6~mm$^2$ pixels, leading to a low-cost ($\sim$150~k\euro), compact (0.5~m~$\times$~0.5~m), and light ($\sim$45~kg) camera with 2048 pixels 
providing a camera FoV of $\sim$9 degrees. The CHEC electronics are based on custom TARGET (TeV array readout with GSa/s sampling and event trigger) application-specific integrated circuits (ASICs) and field programmable gate arrays (FPGAs) 
sampling incoming signals at a gigasample per second, with flexible camera-level triggering within a single backplane FPGA. CHEC is designed to observe in the $\gamma$-ray energy range of 1--300~TeV, and at impact distances up to $\sim$500~m. 
To accommodate this and provide full flexibility for later data analysis, full waveforms with 96 samples for all 2048 pixels can be read out at rates up to $\sim$900~Hz. 
The first prototype, CHEC-M, based on multi-anode photomultipliers (MAPMs) as photosensors, was commissioned and characterised in the laboratory and during two measurement campaigns on a telescope structure at the Paris Observatory in Meudon. In this paper, the results and conclusions from the laboratory and on-site testing of CHEC-M are presented. They have provided essential input on the system design and on operational and data analysis procedures for a camera of this type.
A second full-camera prototype based on Silicon photomultipliers (SiPMs), addressing 
the drawbacks of CHEC-M identified during the first prototype phase, has already been built and is currently being commissioned and tested in the laboratory.

\end{abstract}


\end{frontmatter}

\section{Introduction}
\label{introduction}
%
The current energy frontier for high-energy astronomy lies at around 50~TeV. Above this energy existing instruments detect only a handful of photons. So far, no photons with energies of $>$100~TeV have been observed from any source and only a few sources with $\gamma$-ray energies above 30~TeV have been detected, e.g.~the supernova remnant RX J1713.7-3946, the pulsar wind nebulae Crab and Vela-X, and the extended sources MGROJ2031+41, MGROJ2019+37 and MGROJ1908+06. Increased collection area is therefore a key requirement for future high-energy $\gamma$-ray instruments. Beside ground-based extensive air shower (EAS) arrays like HAWC \cite{Abeysekara:2017mjj} and LHAASO \cite{DiSciascio:2016rgi}, arrays of imaging atmospheric Cherenkov telescopes (IACTs) with an area of $\sim$10~km$^{2}$ are the most promising candidates for instruments being able to push the upper energy frontier to higher energies while still maintaining the necessary level of background suppression and providing excellent angular resolution \cite{2011ExA....32..193A,2008NIMPA.588...48R}. Despite the significant technical challenges and potential cost implications of such a large area instrument, it remains a highly attractive scientific prospect, for example in the search for the sources of Galactic cosmic rays up to the so-called {\it knee} in the cosmic-ray spectrum at PeV energies, and the search for new physics including Lorentz invariance violation and axion-like particles. 
When observing with a wide-energy coverage instrument, such as the Cherenkov Telescope Array (CTA) \citep{2013APh....43....1H,2011ExA....32..193A}, the wide spectral range makes it possible to remove ambiguities on the nature of the radiating particles, with inverse Compton emission strongly suppressed by the Klein-Nishina effect at $>$100~TeV.

IACTs record the Cherenkov light emitted by the secondary particles of the EAS initiated by very-high-energy ($>$100~GeV) $\gamma$-rays or by charged cosmic rays. Since at these energies cosmic-ray events are several hundred times (the exact number being dependent on the energy of the primary particle) more abundant than $\gamma$-ray events, they are the main background and need to be excluded in the analysis. Showers imaged by IACTs reach maximum development at heights of $\sim$10~km above the observer. Due to the height dependence of the Cherenkov emission angle in the atmosphere ranging from about 0.8$^\circ$ at a height of 10 km a.s.l.~up to a maximum of 1.4$^\circ$ at sea level, the radius of the Cherenkov light pool on the ground at 1500 m a.s.l.~is about 120~m (for vertical showers)\footnote{This can also be roughly calculated using geometry and the emission angle at 10 km a.s.l.: $(10000$~m$-1500$~m$)\,\tan(0.8^\circ)\approx$ 120~m.}. Assuming the shower axis being parallel to the telescope axis, the maximum of the observed Cherenkov emission is displaced by angles of $\approx0.8^{\circ}\times(\frac{d}{120~\mathrm{m}})$ from the telescope axis (using small-angle approximations, $\tan \alpha \approx \alpha$ for angles $\alpha \lesssim 1^\circ$), where $d$ is the distance of the shower core position on ground to the telescope. Two approaches therefore exist for the design of a multi-km$^2$ array: either (1) a large number of closely spaced telescopes with a modest field of view (FoV) (e.g.~as in existing IACTs like H.E.S.S.~\citep{2003APh....20..129C,2004NewAR..48..331H} with an inter-telescope spacing of $\sim$100~m and a FoV of $\sim$5$^\circ$) or (2) wider inter-telescope spacings (e.g.~$>$800~m) with larger mirrors (to achieve the same lower energy threshold as in option 1) providing a wide FoV (e.g.~$>$10$^\circ$, see \citep{2006APh....26...69D}).

CTA is a multi-km$^2$ array following the former approach with telescope spacings of 100--200~m. To cover a wide energy range (from 20~GeV to 300~TeV) it uses three telescope classes (referred to as small-, medium-, and large-sized) with mirror diameters of 4 to 23~m and different quantities of telescopes of each class. Furthermore, CTA will consist of a northern and a southern hemisphere site to cover the whole sky. $\gamma$-rays at TeV energies are few in number but initiate EASs which produce a high number of Cherenkov photons (compared to $\gamma$-rays at energies of a few tens of GeV being high in number but initiating a relatively low number of Cherenkov photons, see e.g.~\citep{1997APh.....8....1D,Akhperjanian1997}). Since they are expected to mostly origin from Galactic sources, which can be better observed (at higher elevation) from the southern hemisphere, the southern hemisphere site is envisaged to host 70 small-sized telescopes (SSTs). For a conventional IACT design with the size of an SST (mirror diameter of $\sim$4~m), following single-mirror-dish-design (parabolic reflector or Davies--Cotton~\citep{1957SoEn....1...16D} design) such as H.E.S.S., VERITAS, and MAGIC ~\citep{2002APh....17..221W,2011ICRC...12..137H,Bigongiari:2005sw}, the cost of the camera dominates that of the telescope. The use of a secondary reflector and aspherical Schwarzschild--Couder (SC) \citep{1905MiGoe...9....1S, Couder1926} optics enables a reduction in the plate scale of a $9^{\circ}$-FoV telescope by a factor of $\sim$3 \cite{2007APh....28...10V}. The reduced plate scale introduces rather novel options for the camera photosensor technology, including multi-anode photomultipliers (MAPMs) and Silicon photomultipliers (SiPMs). These offer considerably reduced cost with respect to photomultiplier tubes (PMTs) used in cameras of conventional IACTs. Thus, the overall camera cost is reduced allowing a larger array of SSTs, and therefore increased area coverage at fixed cost and telescope spacing. Two such dual-mirror optical designs are currently being prototyped for the SSTs of CTA: GCT \cite{2012SPIE.8444E..3AL,Dournaux:2017vtk} and ASTRI \cite{2013HEAD...1312331V}.

The Compact High Energy Camera (CHEC) is a proposed camera suitable for use in both of these CTA dual-mirror SST designs and under development by groups from Australia, Germany, Japan, the Netherlands, the UK, and the US. The required compact nature of CHEC hinges on the use of commercially available multi-pixel photon-counting photosensors and the custom TARGET (TeV array readout with GSa/s sampling and event trigger) application-specific integrated circuits (ASICs) \cite{2012APh....36..156B} to provide a high-performance low-cost solution.  Research and development for CHEC is progressing via the development of two prototype cameras: CHEC-M, based on MAPMs, and CHEC-S, based on SiPMs. In this paper, we present results from CHEC-M, the first CHEC prototype. The prototype design and laboratory test results are presented. On-sky Cherenkov data is shown from CHEC-M installed on a GCT prototype telescope in Meudon near Paris, and upcoming prospects towards a camera design for the production phase of CTA are discussed.

\section{Concept}
\label{concept}
%
%
Above a few TeV, the Cherenkov light intensity is such that showers can also be detected outside the light pool of fairly uniform illumination of about 200--250~m diameter. Thus, an energy threshold of around 1~TeV can be achieved with a telescope spacing of $\sim$250~m and a telescope diameter $D$ of only $\sim$4~m. An angular pixel size of $\sim$0.2$^\circ$ is required to be less than the full width half maximum (FWHM) of a typical 1~TeV $\gamma$-ray image \citep{2011ExA....32..193A,2013APh....43..171B}. It may then be matched to low-cost photosensors of $\sim$6~mm diameter, setting the telescope focal length $f$ to $\sim$2~m. The resulting $f/D$ ratio of $\sim$0.5 can be achieved using a dual-mirror telescope design based on the SC optics. A 9$^{\circ}$ FoV can then be realised with a camera with $\sim$2000 pixels and with a diameter of only $\sim$35~cm \footnote{The result of rough calculations is rather $\sim$30~cm but more detailed calculations and simulations, taking into account the final optical specifications of the telescope (focal length of 2.283~m) and a correction due to distortion, results in a camera diameter of 36.2~cm.}. The use of an SC design requires a focal plane with a curved surface in contrast to conventional one-mirror designs which use cameras with a plane focal plane.

Cherenkov light from EASs peaks at a wavelength of $\sim$350~nm at ground level with a flash duration of typically only a few to a few tens of nanoseconds. The use of fast and blue-sensitive photosensors and high-speed digitising electronics is therefore required. To make maximal use of the information contained in the time evolution of the Cherenkov signal, full waveform digitisation and readout is desirable per camera pixel. 
Given the expected range in impact distance from the telescope, EASs from primary particles with energies of 1--300~TeV produce Cherenkov images ranging in amplitude from around 250 to many thousands of photons between 300 and 500~nm. With an effective quantum efficiency of the MAPMs in the given wavelength range ($\sim$20\%), this results in a dynamic range from around 50 photoelectrons (p.e.) up to a few thousands of p.e.~requiring each camera pixel to cover a dynamic range of about three orders of magnitude.
 
The $\gamma$-ray flux from typical astrophysical sources and the cosmic-ray background both exhibit power-law spectra, falling off with increasing energy and resulting in an event rate above 1~TeV that implies a maximum mean trigger and readout rate per SST of 600~Hz. However, the night sky background (NSB) contributes with a Poisson-distributed incoming stream of single photons independently to each camera pixel at an expected rate of tens to hundreds of MHz. A fast topological trigger is therefore needed to efficiently record Cherenkov images whilst rejecting signals due to NSB photons.

An intelligent safety system should ensure the camera being operated within defined limits (e.g.~by measuring temperature, humidity, current etc.) and taking action if needed. Additionally, hardware allowing in-situ gain calibration, flat-fielding measurements, and regular monitoring of the camera and telescope response is required.

\section{Technical Design}
\label{design}
%

\begin{figure*}[tb]
\begin{center}
  \makebox[\textwidth][c]{\includegraphics[trim=0.0cm 3.5cm 0.0cm 0.5cm,
    clip=true, width=1\textwidth]{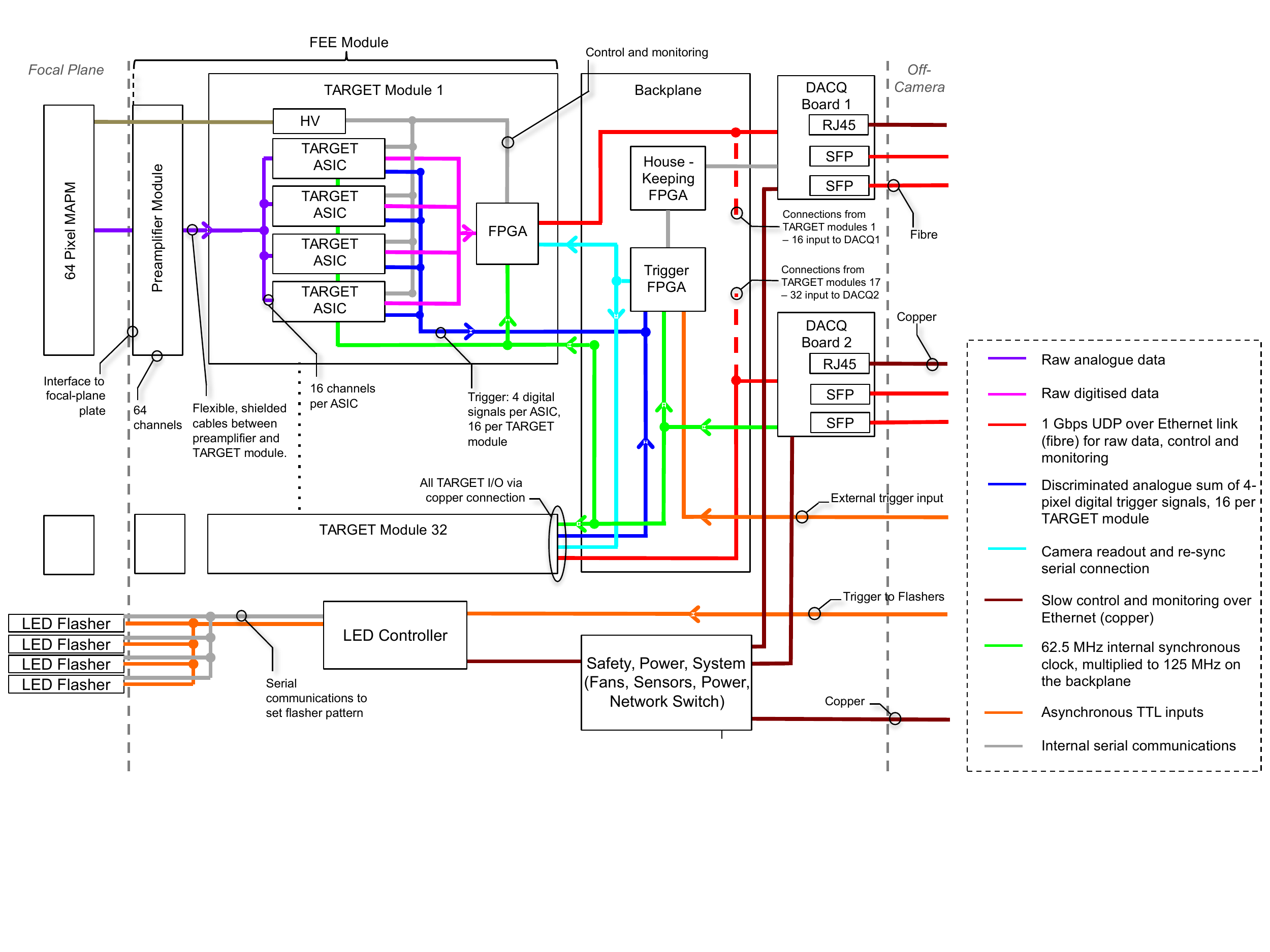}}
 \caption{Schematic showing the logical elements of CHEC-M,
    the communication between those elements, the raw data flow
    through the camera, the trigger architecture, and the clock
    distribution scheme (reproduced from \citep{DeFranco:2015gza}).}
  \label{fig:architecture}
\end{center}
\vspace{-0.4cm}
\end{figure*}

CHEC-M contains 2048 pixels instrumented as 32 Hamamatsu H10966B MAPMs each comprising 64 pixels of $\sim$6$\times$6~mm$^2$ and arranged in the curved focal plane to approximate the required radius of curvature of 1~m. The camera architecture is shown in Fig.~\ref{fig:architecture}.
Front-end electronics (FEE) modules connect to each photosensor providing full-waveform digitisation for every channel and the first level of camera trigger. A backplane forms a (second-level) camera trigger decision based on the trigger signals from all FEE modules. Data is read out from all FEE modules to data-acquisition (DACQ) boards routed off-camera via four 1~Gbps fibre-optic links. A safety board intelligently controls power to camera components based on monitored environmental conditions whilst LED flashers located in each corner of the focal plane provide a calibration source. An internal network switch provides control connections to the safety board, LED flashers, and DACQ boards. A camera server PC off-camera runs software to collect data, control, and monitor the camera. 

\subsection{Camera Mechanics and Thermal Control}

\subsubsection{Camera Mechanics}
\label{mec}

\begin{figure}[tb]
	\centering
	\includegraphics[angle=0,trim=4cm 2cm 3.4cm 1.5cm,
	clip=true,width=0.5\textwidth]{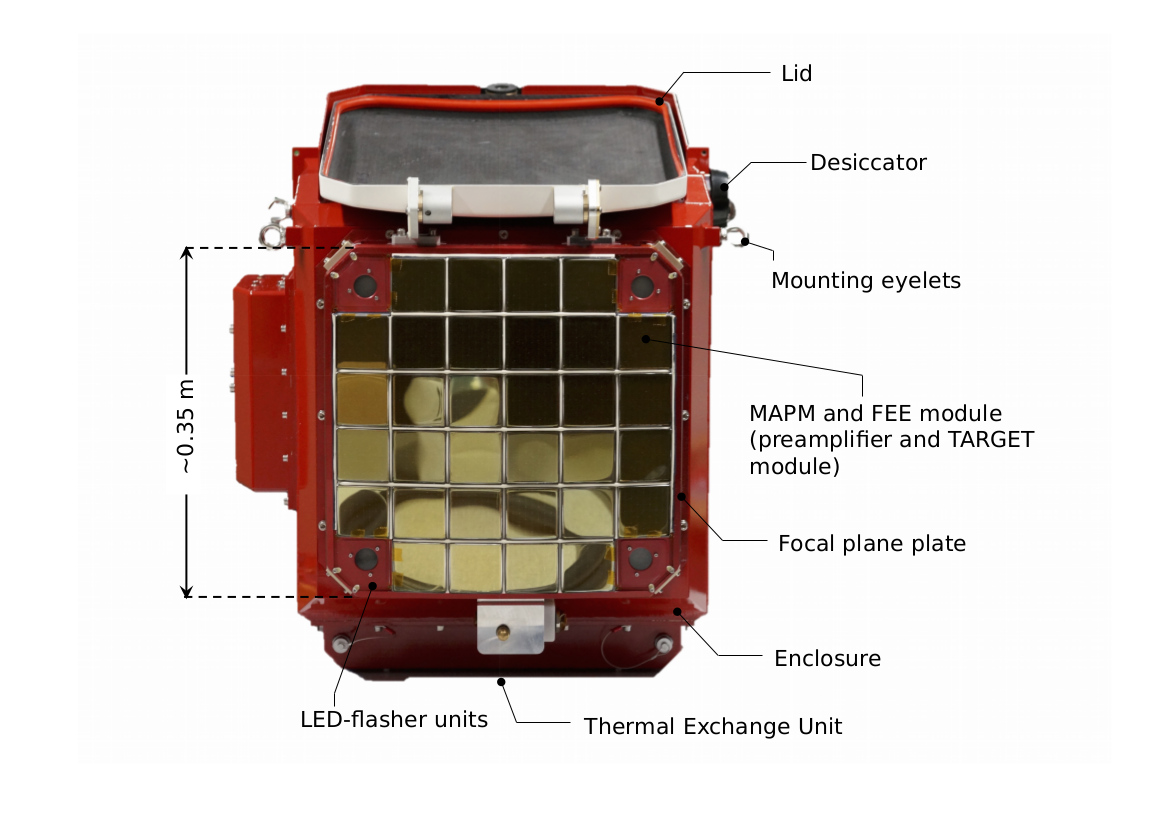}
	\caption[]{The CHEC-M prototype camera, with major elements indicated.}
\label{fig:checm}
\end{figure}

The mechanical structure of CHEC-M is manufactured entirely from aluminium and consists of an external enclosure with focal-plane plate, an internal rack, a thermal exchange unit, and a manual lid assembly. Fig.~\ref{fig:checm} shows an annotated view of CHEC-M with the major mechanical elements highlighted.

\begin{figure}[tb]
	\centering
	\includegraphics[angle=0,trim=5.5cm 3cm 2cm 1cm, clip=true,width=0.5\textwidth]{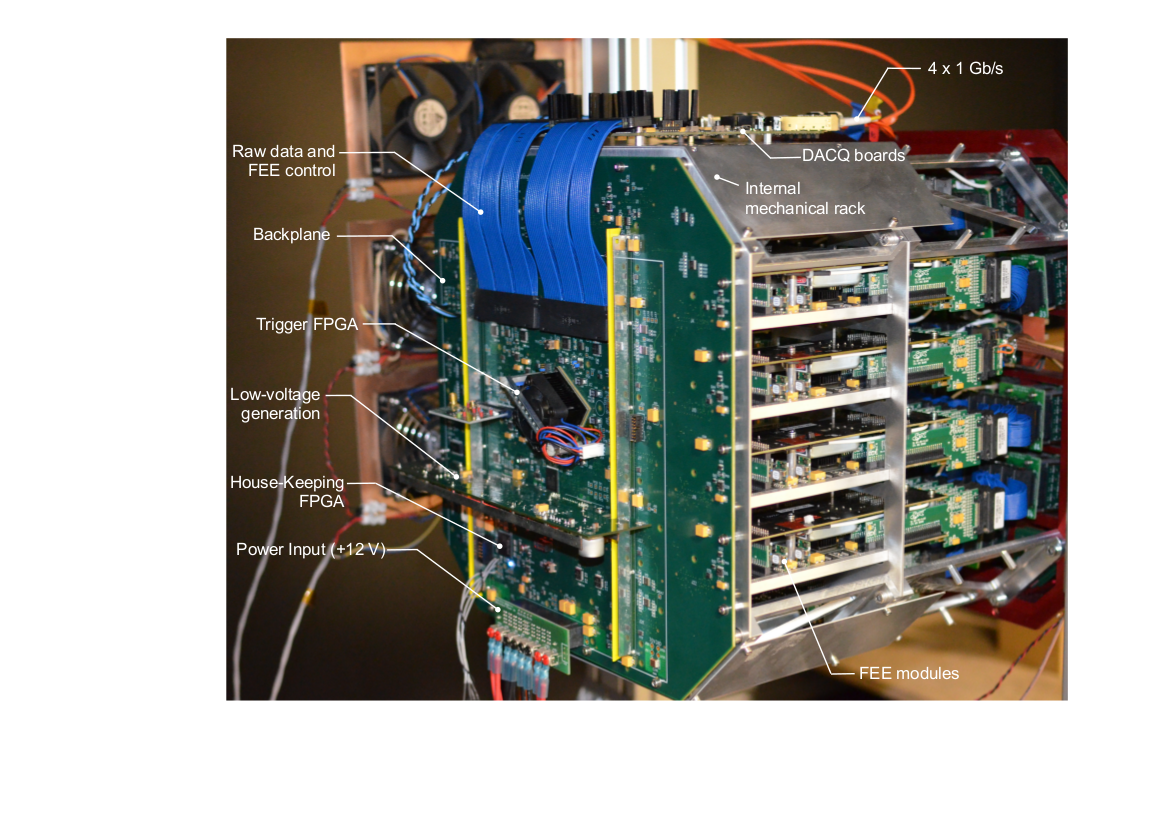}
	\caption[]{A photograph of the CHEC-M camera without the external enclosure in place, taken from the back to highlight the backplane. The TARGET modules can be seen inserted into the internal rack mechanics, whilst the DACQ boards can be seen at the top of the rack, attached to the backplane via two large Samtec ribbon cables.}
	\label{fig:checm_without_enclosure}
\end{figure}

The focal-plane plate located at the front of the camera is responsible for the accurate positioning of the photodetectors. As shown in Fig.~\ref{fig:checm_without_enclosure}, the FEE modules are slotted through this plate and into the internal rack.
On the rear of the rack is an aluminium plate with through-holes and screws for securing the FEE modules and for electrical connectors. Once all FEE modules are secured, the photodetector units are attached. The interface backplate at the rear of the camera provides a stable mounting point for attachment to the telescope structure.

There are two access panels in the enclosure sides, into which machined removable aluminium panels are fastened. One access panel contains the feed-through for power cables and optical fibres via bulk-head connectors, whilst the other houses the thermal-control assembly. CHEC-M includes a manually operated prototype lid to protect against dust and liquid ingress. A wind shield on one side of the enclosure provides protection for the lid whilst open. The camera is painted with corrosion-resistant automotive paint.

Overall rigidity holds the focal plane position stable to $\pm$0.2~mm. Measurements on the camera mechanics indicate that the centre of all MAPMs may be placed to within 0.35~mm of the ideal position in the direction of the optical axis (Z). This misalignment may be compensated for by a translation and rotation at the rear of the camera when installed on-telescope using an adjustable camera mounting mechanism and actuators on the telescope secondary reflector. This would reduce the spread between the centres of the MAPMs in the Z direction down to 0.035~mm for an ideal set of MAPMs. In reality the measured range in the depth of the 32 MAPMs purchased for CHEC-M is $\pm$0.26~mm. The tiling of the curved focal plane with flat MAPMs creates an additional maximum shift of 0.45~mm in Z from the ideal position for the edge pixels. Combining these tolerances implies a less than 10\% degradation in telescope point spread function during operation.

\subsubsection{Thermal Control}
\label{chiller}

The total power dissipation of CHEC-M during normal operation is $\sim$450~W. The thermal control system is designed to keep the camera temperature stable over a wide range of ambient temperatures up
to the maximum required 25$^{\circ}$C during normal operation and to protect the camera electronics up to an ambient temperature of 45$^{\circ}$C. A breather desiccator removes humidity from the camera interior.

The thermal control system consists of four fans coupled to a liquid-cooled heat sink. The fans, together with a system of baffles, provide a recirculating airflow within the sealed camera enclosure. A commercially available chilling unit (Rittal SK 3336.209) provides a flowing liquid (R134a: water glycol mixture) of controllable temperature, delivers a cooling power of $\sim$1.5~kW, and can operate over an ambient temperature range of $-20$ to 45$^{\circ}$C. 
The unit weighs 97~kg (without liquid) and measures 485~mm $\times$ 965~mm $\times$ 650~mm.
It is installed at the azimuth axis of the telescope and is connected by 3/4'' pipes to the thermal exchange unit inside the camera. Quick-release non-leak couplings allow the chiller and the camera to be disconnected quickly from the telescope structure whilst preventing fluid loss / spillage.

\subsection{Photosensors}
\label{mapm}

The Hamamatsu H10966B MAPM has a super-bialkali photocathode with a spectral response between 300 and 650~nm, peaking at $\sim$340 nm with a quantum efficiency of $\sim$30\%. An MAPM consists of 64 pixels of size 6~mm $\times$ 6~mm corresponding to an average angular size of $0.15^{\circ}$ when installed on the GCT telescope structure. The value of $D_{80}$ (the diameter which contains 80\% of the light resulting from a point source) for the telescope design is smaller than 6~mm over the full camera FoV once the telescope mirrors have been aligned. Due to the arrangement in a curved focal plane, a gap of $\sim$2~mm between the front of the MAPMs is required to accommodate their depth of 25.8$\pm$0.26~mm. When combined with the dead space at the edges of each MAPM, a total maximum dead space of $\sim$5~mm (corresponding to the gap between the corners of two MAPMs) is achieved. Each MAPM accepts a single high voltage (HV) source of 800--1100~V (generated on the TARGET modules and routed to the corresponding MAPM by an insulated cable)\footnote{The MAPM operates with negative HV, i.e.~the anode is held at negative potential while the cathode is held at ground. However, for simplicity, positive HV values are used throughout the paper.}. The gain ranges from 4$\times$10$^4$ to 6$\times$10$^5$, adjustable by setting the HV. Due to the background rate associated with exposure to the night sky (expected to lie between 15 and 500~MHz at the CTA site depending on the observation conditions\footnote{Throughout this paper, the NSB rate is defined as number of p.e.~induced by NSB photons per pixel and per second.}) the gain for nominal operation is 8$\times$10$^4$. The 10--90\% risetime of the MAPM is $\sim$0.4~ns while the transit time is $\sim$4~ns with a spread of 5--10\%. Thus, the MAPM produces pulses with an FWHM of about 1~ns. The response of the MAPM extends from a single p.e./pixel~to thousands of p.e./pixel (cf.~Sec.~\ref{dynamic_range}).

\subsection{Front-end electronics}
\label{fee}

\begin{figure*}
	\begin{center}
		\makebox[\textwidth][c]{\includegraphics[trim=1cm 3.5cm 1cm 2.5cm,
			clip=true, width=1\textwidth]{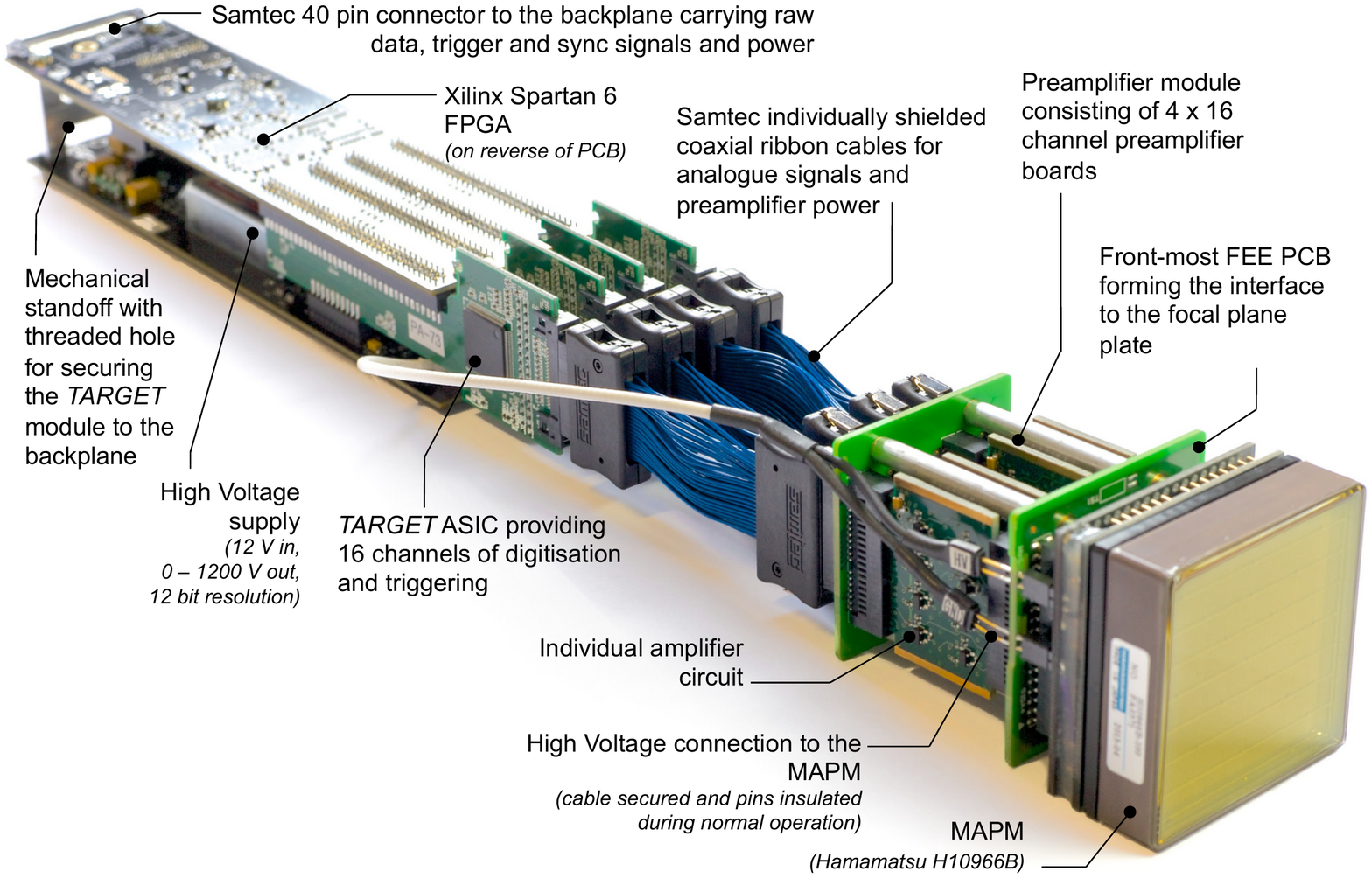}}
		\caption{An MAPM attached to a CHEC-M FEE module consisting of preamplifier module, ribbon cables and TARGET module based around four TARGET 5 ASICs (reproduced from \citep{DeFranco:2015gza}).}
		\label{fig:fee}
	\end{center}
	\vspace{-0.4cm}
\end{figure*}

The FEE module developed for CHEC-M, and shown in Fig.~\ref{fig:fee}, consists of a preamplifier module connected to a TARGET module based around four TARGET 5 ASICs, 16-channel devices combining digitisation and trigger functionalities~\cite{2017APh....92...49A}.

The MAPMs produce narrow pulses that must be shaped to optimise the camera trigger performance. Simulations show that the optimal pulse FWHM for triggering is around 5 to 10~ns with a 10-90\% risetime of 2 to 6 ns. If the pulses are faster, the time gradient of Cherenkov images across neighbouring pixels forming the analogue sum prevents pile-up to reach the trigger threshold. If they are slower, NSB photons limit the performance of the camera trigger.

The preamplifier module connects directly to the photosensor to amplify and shape the signals and to provide noise immunity for signal transport to the TARGET module. The preamplifier circuit contains an AD8014 operational amplifier operated in trans-impedance mode and consumes $\sim$1~mA quiescent current. The overall power consumption of the preamplifier lies between 9 and 20~mW per channel depending on the incoming photon rate. Individually shielded ribbon cables minimise the influence of noise and provide the connection to the curved focal plane, allowing the use of a planar internal rack to house the modules. 

Each TARGET module provides 64 channels of digitisation and first-level triggering. For this purpose, both the signal and a reference signal (an input pedestal voltage, Vped, supplied by an external digital-to-analogue converter (DAC), and used for common-mode noise rejection and as a reference to fix the trigger threshold) are input to each ASIC and simultaneously processed for sampling (data path) and triggering (trigger path). The TARGET 5 ASIC is an analogue sampling chip capable of digitising signals with 12-bit resolution. When used within CHEC-M, it provides an effective dynamic range of 1 to $\sim$500~p.e./pixel (with the recovery of larger signals offline possible due to the full-waveform digitisation). The sampling rate is tunable, but is set to 1~GSa/s for CHEC-M. TARGET 5 contains two capacitor arrays -- a 64~ns deep analogue sampling array followed by a storage array with a maximum depth of 16384~ns -- to simultaneously achieve a large analogue bandwidth and a deep buffer. Acquisition occurs in one group of 32 cells in the sampling array while the charge of the cells of the other group is transferred to the storage array cells. Such a ping-pong approach provides continuous sampling (cf.~\cite{2012APh....36..156B} for further details). The position of the readout window digitised from storage array is selectable with 8~ns resolution\footnote{This means that the start of the readout window occurs on an 8~ns edge with respect to the nanosecond accurate event trigger (known as TACK, see Sec.~\ref{bee}). This start position is identical for all pixels in a given event. The TACK is used offline to correct the waveforms (shift them by up to 8~ns) such that a recovered pulse can always be found at the ``correct'' position.} with a size settable in 32~ns blocks, nominally set to 96~ns for CHEC-M (chosen to capture high-energy, off-axis events and/or events with a high impact parameter as they transit through the FoV). 

A Xilinx Spartan-6 field programmable gate array (FPGA) on board each TARGET module is used to configure the ASICs and other module components, to read out raw data from the ASICs, and to package and buffer raw data for output from the module. Module control and raw data output are managed via user datagram protocol (UDP) over a 1~Gbps Ethernet link at the rear of the modules. The TARGET 5 ASIC is continuously sampling and dead time free, i.e.~sampling continues while data is being digitised. 

The TARGET 5 ASICs also provide the first level of triggering for the camera. The trigger consists of the analogue sum of a square of four neighbouring pixels (referred to herein as a superpixel), which is then discriminated. Each ASIC outputs four digital trigger signals, which are routed through the module to the backplane, resulting in 16 differential LVDS trigger signals per module and 512 (32~$\times$~16) in total for the whole camera.

Each FEE module accepts a 12~V input for all electronics use and consumes roughly 7--8~W of power during full operation. 

\subsection{Back-end electronics}
\label{bee}

The back-end electronics (BEE) for CHEC-M consist of a backplane and two DACQ boards.

\subsubsection{Backplane}
\label{bee-bp}

The backplane provides the power, clock, trigger, and data interface to the FEE modules. Data links to the FEE modules are routed via the backplane to DACQ boards. 

The backplane triggering scheme is implemented in a single Xilinx Virtex-6 FPGA, referred to herein as the trigger FPGA (TFPGA). The TFPGA accepts all 512 first-level trigger lines from the FEE modules (their signal width currently being set to 30~ns) and implements a camera-level trigger algorithm (currently requiring a coincidence between two neighbouring superpixels). Following a successful camera trigger, a readout request consisting of a serial message with a 64-bit nanosecond counter (known as a TACK message) is sent to the FEE modules to initiate a full camera readout. On the FEE modules the TACK is compared to a local counter to determine a look-back time in the ASIC buffers. The TACK is added to the raw data event on each FEE module and -- as already mentioned in Sec.~\ref{fee} -- is later used to shift the waveform and thus correct for the 8~ns resolution of the starting position of the readout window. The TFPGA provides functionality to individually disable any of the 512 FEE module trigger inputs from the camera trigger decision to prevent noisy trigger patches from dominating the event rate. The 512-bit pattern causing the last camera trigger is accessible, and 512 counters provide a method to monitor trigger rates across the camera.

Rate control is implemented via a settable minimal time between consecutive camera triggers (trigger hold-off time). During commissioning it was observed that the TARGET 5 ASIC produces noise whilst digitising previously sampled and stored analogue data. This noise corrupts any new data sampled and stored in the ASIC whilst this digitisation is ongoing. This is why in CHEC-M digitising is only enabled when a readout is requested and all data sampled within the time of the digitising process ($\sim$20~$\mu$s) has to be discarded from the analysis. Furthermore, it was observed that triggering with the TARGET 5 ASICs based on signal discrimination leads to additional, false triggers due to pick-up by the trigger circuitry of the serial data signals from the FPGA used to read out the ASICs. Thus, a trigger hold-off time between triggers of 80~$\mu$s (corresponding to slightly longer than that required to readout the TARGET 5 ASICs) is used in CHEC-M to allow stable operation.  Whilst both of these problems (noise due to digitising and readout) are solved by design in future FEE iterations (see Sec.~\ref{outlook}), they lead to an overall dead time of CHEC-M of 80~$\mu$s in nominal operation.

A second, smaller house-keeping FPGA (HKFPGA), Actel A3P400, provides access to status and monitoring registers on the TFPGA and monitors the current and voltage supplied to the FEE modules. Control and monitoring of both FPGAs is provided via a serial peripheral interface (SPI) link routed to one of the DACQ boards. 

Clocks between the backplane and the FEE modules are kept in-sync through a low-skew fan-out network and signals between the backplane and the FEE modules are used to synchronise local time counters. A reference clock with a frequency of 62.5~MHz is provided to the backplane from a DACQ board. The backplane utilises a programmable quad clock generator to generate a 125~MHz clock from the reference clock for distribution to the TFPGA, HKFPGA, and the FEE modules. Each of the four outputs is phase programmable in 20~ps increments and is routed to a 1:16 fanout buffer, specified to introduce no more than 25~ps delay between the outputs. While absolute time synchronisation will be present in the final CHEC design (see Sec.~\ref{outlook}), it is not present in CHEC-M.

The camera may be externally triggered via an external pulses input to an SMA connector on the backplane, and routed to a bulk-head connector on the camera enclosure. For power, the backplane accepts a single 12~V input and generates all required voltages on a daughter board mounted perpendicular to the main printed circuit board (PCB).

\subsubsection{DACQ Boards}

The DACQ boards form a link for raw data and communications between the FEE modules and the camera server PC. Each board connects 16 FEE modules via wired 1~Gbps Ethernet links to two 1~Gbps fibre-optic links to the PC. Network interface cards (2 $\times$ Intel PRO/1000 Dual Port PCIe) are used on the PC to connect the fibres. Data is sent to and from the FEE modules via a custom format over UDP. Jumbo frames are used to minimise the number of packets sent per raw data event. For each event, an FEE module serialises data from 64 pixels into two UDP packets. Buffering on the FEE and controlled delays between packet sending prevents the 1~Gbps (up-)links to the PC from being saturated by the traffic from the 32 1~Gbps links to the FEE modules. The DACQ boards each act as a layer-2 switch, with the MAC addresses of eight FEE modules hard-mapped to a single 1~Gbps link to the PC. Each board is based around a Xilinx Virtex-6 
FPGA providing 18 GTX serial transceivers and an ARM Atmel microprocessor running an entire light-weight Linux system for managing purposes. An Ethernet connection to each DACQ board enables controlling and monitoring using a UDP server which runs on the microprocessor. The SPI control of the backplane is provided through another UDP server running only on one of the two DACQ boards. The boards are custom-made revisions of network switches based on White Rabbit technology \citep{WR2009} which are also commercially available from the company Seven Solutions.

\subsection{LED Calibration Flashers}
\label{led}

CHEC-M is equipped with four flasher units, each containing ten LEDs of different brightnesses placed in the corners of the focal plane to illuminate the photodetectors via reflection from the telescope secondary mirror (cf.~\citep{Brown:2015nca}). A Thorlabs ED1-C20 one-inch circle-pattern engineered diffuser is mounted in front of the flashing LEDs. The LED flasher units are based around fast gated TTL drive pulses and 3~mm, low self-capacitance, Bivar UV3TZ-400-15 LEDs, with a peak wavelength of 400~nm. The Bivar LEDs are enabled/disabled by an on-board microcontroller and triggered via an external TTL pulse. An LED controller based on an Arduino Leonardo ETH board connects all flasher units and provides an interface to set the LED pattern and to fan out a trigger signal. The TTL trigger signal is in-turn input to the LED controller from an SMA connector mounted on the camera chassis. While in CHEC-M, the flasher units can only be triggered from an external source, this will be different in the final CHEC design where triggers can also be provided by an internal device (cf.~Sec.~\ref{outlook}). Communication with the LED controller is via a network switch installed on the internal camera rack and based on the same UDP scheme as used for the communication with the TARGET modules and for several other devices in CHEC-M (cf.~Sec.~\ref{software}).

The LED flasher units are designed to flat-field the camera across a wide dynamic range, providing optical pulses of width $\sim$4.5~ns (FWHM) at 400~nm from 0.1~p.e./pixel, for absolute single-p.e.~calibration measurements, to over 1000~p.e./pixel, to characterise the camera up to and at saturation (cf.~Sec.~\ref{flashers} for results of characterisation measurement). Since the time distribution of the flasher signals in the camera pixels can be calibrated, absolute single p.e.~calibration using the flasher units is expected to be possible even under the presence of NSB with an expected nominal rate of about 15--25~MHz on the CTA site. In addition, one of the SST prototypes features a shelter to protect the telescope and camera from the environment. In such a case, flasher calibration measurements would be performed with closed shelter, i.e.~without the presence of NSB.

\subsection{Safety, Power, and Control}
\label{safety_power_control}

\subsubsection{Safety system}
\label{safety_system}

The camera safety system provides the capability to remotely control power to camera components, control and monitor fan speeds, monitor component supply voltage and current draw as well as internal camera temperature and humidity. Furthermore, it monitors the status of the camera subsystems to prevent or reject actions taken by the user that would endanger the camera, issues alerts when certain conditions are met (e.g.~temperature limits exceeded or communication lost), and automatically takes actions to minimise the risk of damage if the situation persists (e.g.~switch off camera), i.e.~if the user or software has not taken any action first to change the situation within a defined time window (alert-action-feedback).

The safety system consists of a power board and safety board mounted internally in the camera on the side of the FEE rack. 
The power board distributes 12~V to camera components via relays controlled from the safety board and provides analogue monitoring of camera component voltage supply and current draw to the safety board. The safety board contains a microprocessor controlling the power board relays, the digitisation of current and voltage readings from the power board, the reading of sensors, and the alert-action-feedback. An external high-current relay mounted on the internal camera chassis controls power to the backplane and FEE. Communication with the safety board is based on the same UDP interface as used for the LED controller. 

\subsubsection{Power supply}
\label{power_supply}

A single power supply from the company ARTESYN (iMP series) provides CHEC-M with 12~V at up to 60~A for all electronics. The power supply contains two individually controllable 12~V units. One unit provides power to the camera safety system for fans, safety and power board. The power board then distributes 12~V to the camera internal network switch, DACQ boards, and LED controller via relays controlled from the safety board. The second 12~V unit of the power supply provides power to the backplane and FEE via a high-current relay, also controlled from the safety board. The division of power distribution in this way allows safety-critical systems to be controlled independently from high-current components.  

The power supply weighs $\sim$1~kg and measures only 60~mm~$\times$~120~mm~$\times$~250~mm. As such it can easily be housed at the rear of the secondary mirror of the telescope. A ``sense'' feed-back input from the camera ensures the desired voltage at the camera. A single Chainflex CF10 12-way cable with outer diameter of 19~mm connects the power supply to the camera and is flexible down to $-35^{\circ}$C. The power supply can be externally controlled and monitored using basic SMBus protocols built on top of I$^2$C. An Ethernet-SMBus interface board has been implemented, based on an Arduino Leonardo ETH board, to allow easy Ethernet control using the same user-defined protocol based on UDP as for the LED controller and safety board.

\section{Operation Procedures}
\label{procedures}
%
%
%
In this section, the camera operation procedures are outlined. Operation procedures cover the standard control of the camera, the data acquisition procedure, and the waveform processing methodology. In general, there are two different types of measurements which are
\begin{itemize}
\item EAS data taking and tests being executed every time the camera is booted (like specific software and hardware tests as well as pedestal and transfer function measurements, cf.~Sec.~\ref{waveform_processing}) and
\item specific measurements performed while commissioning like flat fielding measurements, trigger threshold and hardware temperature dependence determination. Such specific commissioning measurements and their results are described in detail in Sec.~\ref{testing}.
\end{itemize}

As mentioned previously in Sec.~\ref{mapm}, an MAPM gain of 8$\times$10$^4$, corresponding to an HV of $\sim$800~V, is expected to be used for nominal data runs at the CTA site. However, due to the intrinsic performance of MAPMs, single photoelectron (SPE) can only be resolved when operating at a gain higher than nominal, corresponding to 1100V. Hence, most of  the characterisation and performance tests (described in Sec.~\ref{testing}) were done at this HV\footnote{Such an approach of increasing the HV, especially for SPE measurements, is a common procedure also used by operating IACTs like VERITAS, cf.~\cite{2008ICRC....3.1417H}.}. The effect of lowering the gain is discussed in Sec.~\ref{testing} for each performance parameter separately if relevant.

\subsection{Control software and data acquisition}
\label{software}
\begin{figure*}[tb]
\centering
\includegraphics[trim=0cm 2cm 0cm 2.8cm,
    clip=true, angle=0,width=1. \textwidth]{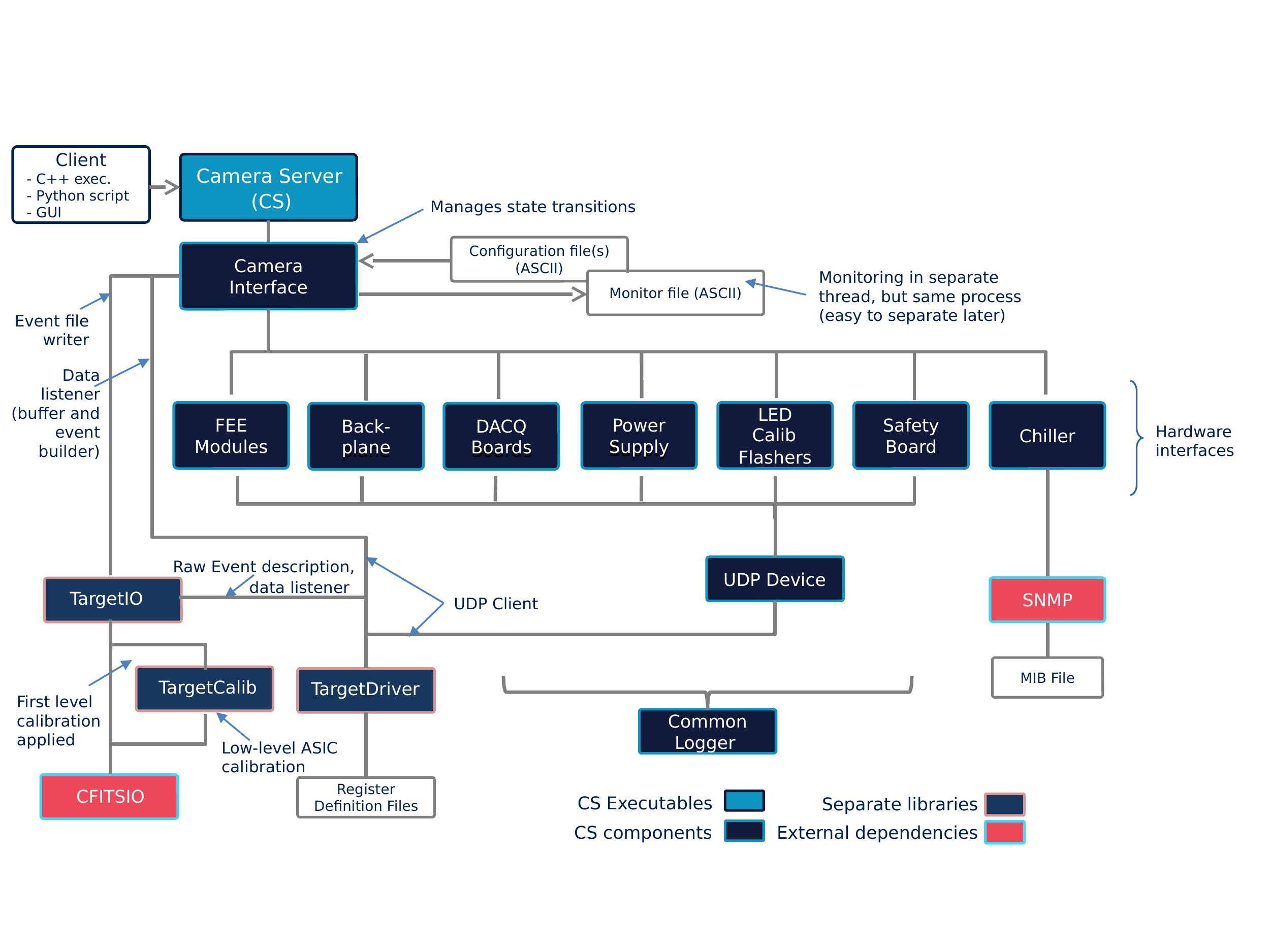}
\caption[]{Diagram showing different classes, their dependencies, and network protocols used in the camera control software CHECInterface.}
\label{fig:software_overview}
\end{figure*} 
The camera control and readout are managed by a C++ software package referred to as CHECInterface. It is designed to be maintainable, simple, and robust, and to fulfil the following requirements:
\begin{enumerate}
\item The number of dependencies on external software is at a minimum.
\item Only minimal code changes are necessary when a camera hardware component is upgraded.
\item The code is written in C++.
\item Scripts and programs written in other programming languages are only allowed for user interface, test programs, and other executables, but not for core-functionality.
\item Continuously buffering and writing of events without loss is ensured up to a rate of 600~Hz (corresponding to 2.5~Gbps).
\item The software is easy to adapt as soon as a final data format or pipeline for CTA is in place.
\end{enumerate}

Two external libraries (CFITSIO~\cite{refId0} and simple network management protocol (SNMP)~\citep{Cas90}) and three self-developed libraries are used within the software: TargetDriver (for control and readout of TARGET modules), TargetCalib (for applying TARGET module calibration), and TargetIO (for reading and writing data from TARGET modules).
The software is structured so that each hardware component inside (e.g.~TARGET modules) and outside (e.g.~chiller) the camera is represented by its own lower level class. The main class CameraInterface serves as interface to all lower level classes. Fig.~\ref{fig:software_overview} illustrates the software architecture with dependencies and libraries used. The user / client, which can be represented by a C++ executable, a Python script, or a GUI, can connect to the Camera Server (running on the camera server PC physically connected to the camera) via Ethernet. The server is implemented as a state machine, linking the client commands to functions and state transitions defined in the main interface class. The states are  
\begin{multicols}{3}
\begin{enumerate}
\item Off
\item Safe
\item StandBy
\item Ready
\item Observing
\item Engineering
\end{enumerate}
\end{multicols}
\setlength\parindent{0pt}
and serve as preliminary placeholders for final camera states to be defined by CTA. Several safety features are implemented in the software, e.g.~timeouts, guaranteeing in addition to the safety system that the camera is operated in safe conditions.

The hardware components are configured through ASCII files following a custom, but simple, format. Communication with all hardware components except the chiller (which uses SNMP) is via a simple custom protocol based on UDP (originally designed for communication with the TARGET modules). 

The camera readout and event building based on the 2048 full waveforms provided by the TARGET ASICs of the 32 TARGET modules is implemented in and managed by the TargetDriver and TargetIO libraries. Events arrive at the camera server PC in asynchronous sets of 64 UDP packets (2 packets per event per module). They are first buffered and then assembled into associated events based on the TACK in each packet header which serves as a unique event identifier.
A timeout is used to prevent the PC buffer filling up if (in an unexpected case) events with missing packets arrive or if the event building takes longer than expected. Missing packets are not requested again and incomplete events are discarded. In a subsequent step, the events are written to disk as FITS files \cite{refId0}. Once CTA is operating and a final CTA data framework exists, the data will instead be further processed in a pipeline. However,  even then it is planned to continue developing and using TargetDriver and TargetIO for single test purposes.

Due to the modest event rate per SST in CTA (600~Hz, cf.~Sec.~\ref{concept}), no inter-telescope hardware array trigger is required. When a telescope triggers, all data is read out and transferred from the camera to be processed by the software array trigger system. Decisions on whether to proceed with the ``array event building'' in software including other camera events from neighbouring telescopes will then be based on the different camera event timestamps.

\subsection{Data calibration and waveform processing}
\label{waveform_processing}

\begin{figure}[tb]
\centering
\includegraphics[trim=0.5cm 1cm 4cm 1cm,
    clip=true,angle=0,width=0.5 \textwidth]{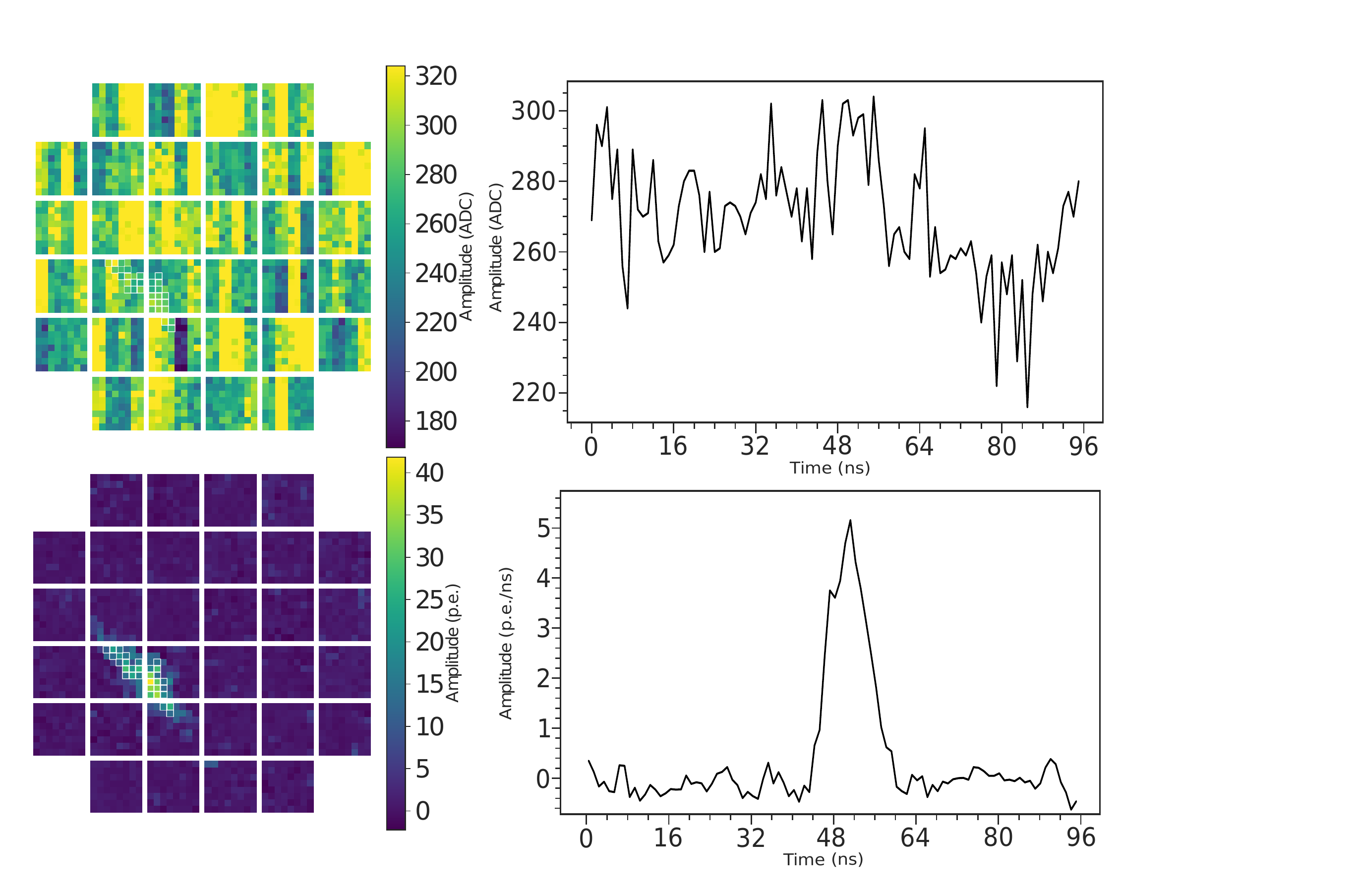}
\caption[]{Camera image and waveform for different steps in the data calibration and processing for the same Cherenkov event. Uncalibrated camera image at $t=48$\,ns (top left) and uncalibrated (raw data) waveform (top right) of pixel 1162. Calibrated camera image after charge extraction via the ``neighbour peak finding'' method (bottom left) and calibrated waveform (bottom right) with pedestal subtraction, transfer function correction, and signal-to-p.e.~conversion applied for pixel 1162. The procedure of calibration and charge extraction is explained in the text. To get the y-unit of p.e./ns, the samples (in units of V) are divided by the SPE value (in units of V\,ns/p.e.), determined in single-p.e.~measurements (cf.~Sec.~\ref{spe_gain}). In both camera images, white squares indicate pixels which survive image cleaning (see text for reference).}
\label{fig:waveform_processing}
\end{figure}

The 2048 raw data waveforms are calibrated and processed in different steps. These consist of (1) applying ASIC specific calibration, (2) signal charge extraction and conversion to p.e., and (3) image cleaning and data reduction. Fig.~\ref{fig:waveform_processing} shows camera images and waveforms for the different steps in the data calibration and processing and impressively illustrates the need for calibration.

\begin{figure}[tb]
\centering
\includegraphics[angle=0,width=0.5 \textwidth]{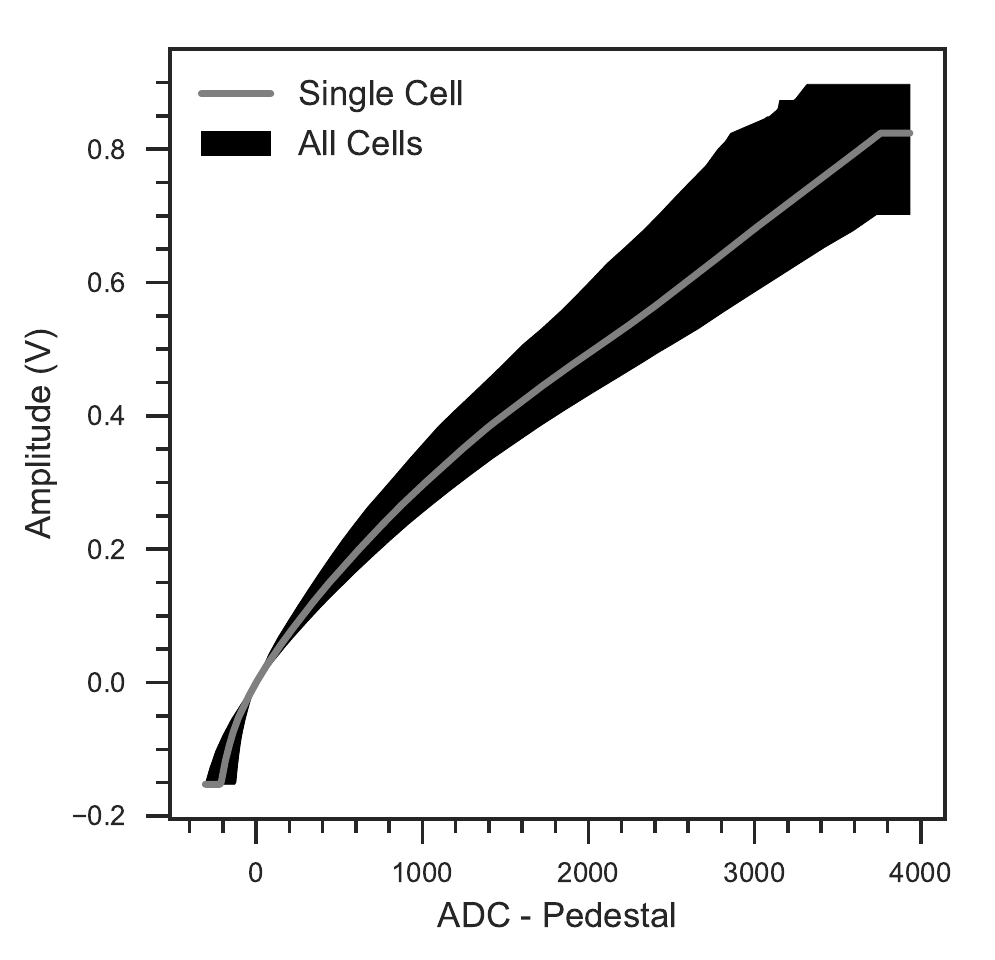}
\caption[]{Transfer function (conversion between ADC value and amplitude in volts). The range of all ASIC sampling cells of all pixels and an example of a single cell are shown.}
\label{fig:tf_camera}
\end{figure}

(1) Since the response of each of the 16384 ASIC \emph{storage} array cells (one storage array per pixel, cf.~Sec.~\ref{fee}) to the reference voltage Vped (cf.~Sec.~\ref{fee}) is different, the pedestal of each cell has to be measured and then subtracted from the raw data. A run with 20000 externally triggered events provides enough hits per cell to calculate the mean pedestal of each storage array cell. In addition to that, the response of each of the 64 ASIC \emph{sampling} array cells (one sampling array per pixel, cf.~Sec.~\ref{fee}) on the supplied voltage is different. Thus, depending on which sampling array cell is hit, another conversion between measured ADC and signal voltage has to be used. This conversion is measured by recording the sampling array cell pedestal (in ADC) as function of the supplied cell voltage (given by Vped). A run with 1000 externally triggered events per Vped provides enough hits per cell to calculate its mean. Such a measurement results in the so-called transfer function (see Fig.~\ref{fig:tf_camera}). Its slope depends on the ASIC parameter Isel \cite{2017APh....92...49A} which is adjusted to maximise the range in which the transfer function is linear. Once determined, the values of Vped and Isel are set to default values and do not need to be adjusted for different measurement purposes.

(2) After subtracting the pedestal and correcting for the transfer functions, the samples of the waveform are converted into units of p.e./ns. In order to perform this conversion the SPE value (in units of V\,ns/p.e.) for each pixel is used. This value is determined from the analysis of single-p.e.~measurements, which were processed using the same procedures as described in this section, but with this conversion omitted. The SPE measurements are described in Sec.~\ref{spe_gain}. For on-site calibration, LED calibration flasher runs described in Sec.~\ref{flashers} will instead be used to obtain the SPE value per pixel. Following that, the signal charge per pixel is extracted by integrating the pulse signal. Several algorithms to find the pulse in the waveform are currently under investigation. One of them is the ``neighbour peak finding'' algorithm which averages the waveforms in the pixels neighbouring the pixel of interest and then takes the time of the maximum value in that averaged waveform as peak time for the pixel of interest. Then the signal is integrated in a defined window around the peak. The default size of the window is 7~ns, with a shift to the left of the peak time of 3~ns. A correction is then applied that uses a single reference pulse shape for the camera to determine the percentage of the pulse outside of the integration window. In doing so, the method is less dependent on the integration window size used for the determination of the number of photoelectrons inside the pulse, however all the analysis performed for this paper did keep the same window size of 7~ns.

(3) Similarly to the charge extraction methods, several algorithms for image cleaning and data reduction are under investigation. A possible procedure is the tail-cut approach, where all pixels containing a signal greater than a threshold are retained, and all other pixels above a second lower threshold are retained if they are a neighbour to a pixel that satisfies the first criterion.

The results of calibration and waveform processing are waveforms and camera images as shown in the lower panels of Fig.~\ref{fig:waveform_processing}. The software being used in (2) and (3) is the low-level data processing pipeline software ``ctapipe'' \cite{ctapipe}, currently under development for CTA.

\section{Characterisation and Testing}
\label{testing}
%
In this section, the camera characterisation based on results of camera lab tests during commissioning and their impacts on the camera performance and characterisation are presented. For illumination measurements, a laser of wavelength 398~nm with a pulse FWHM of 80~ps is used\footnote{Note that for all laser illumination measurements presented in this paper, the deliberate shaping of electrical pulses by the camera preamplifier circuit dominates the intrinsic width of the laser pulse so that the FWHM of the pulses measured by the camera is quite stable as long as no saturation effects occur (cf.~Sec.~\ref{pulse_shape}).}. Its intensity can be regulated using a filter wheel settable to attenuations in the range of 1 to 5$\times 10^4$ and diffusers are used to obtain a uniform illumination of the camera focal surface. The experimental set-up of the light source is similar to the one presented in \cite{Werner:2016lkr}. The maximum spread of the illumination across the camera focal surface is measured (using an SiPM with known gain and temperature-gain dependence and a robot arm to scan the laser beam) to be $\sim$1\%. 
\subsection{Data rate}
\label{data_rate}
The ability of the camera to read out waveforms from all 2048 pixels and send them in UDP packets to the camera server as function of the trigger rate is assessed in this section.

\begin{figure}[tb]
\centering
\includegraphics[angle=0,width=0.49 \textwidth]{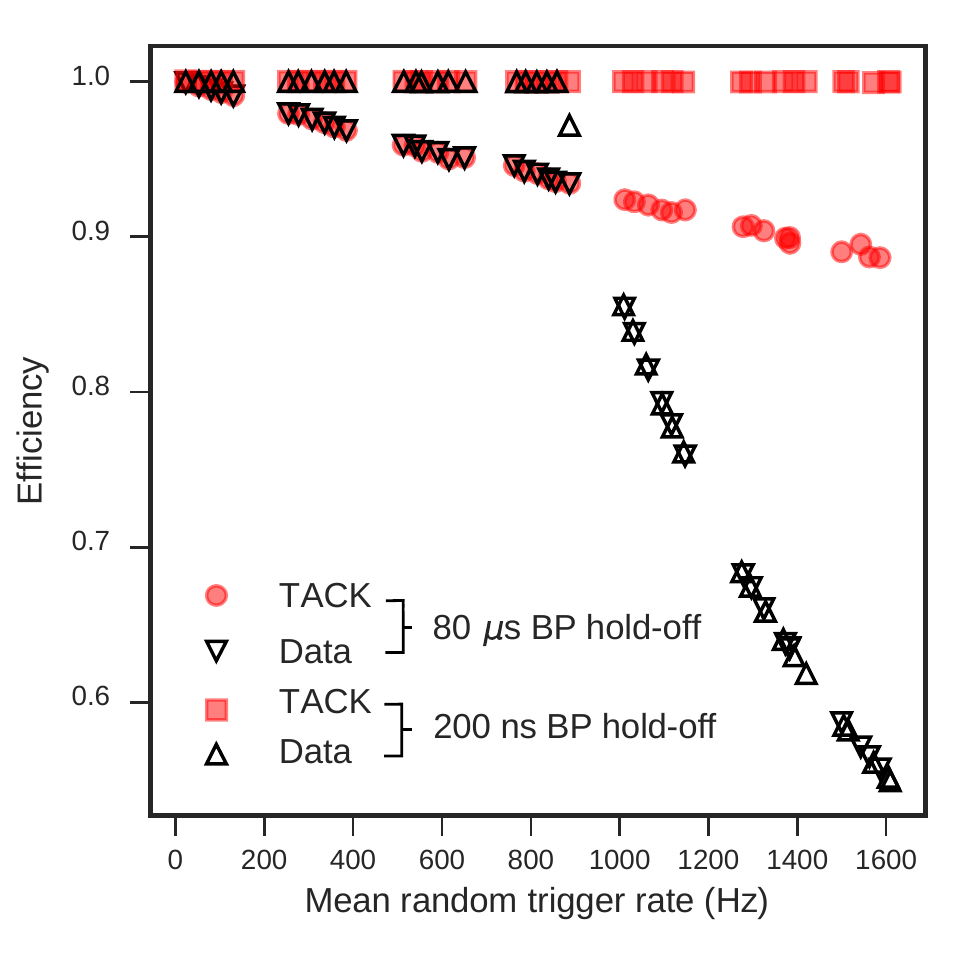}
\caption[]{TACK Efficiency (number of readout requests divided by number of external triggers) and data packet efficiency (number of data packets divided by product of data packets per event (64) and number of external triggers) as a function of mean random trigger rate for two different backplane hold-off times (written in brackets in the legend).}
\label{fig:data_rate_random_200ns}
\end{figure}

To measure the data packet efficiency (number of arrived data packets divided by number of expected data packets\footnote{The number of expected data packets is the product of the number of external triggers and the number of packets per event being 64 (2 packets for each module).}) as a function of the trigger rate, the camera was externally triggered by pulses randomly distributed in time. Cases with two different backplane hold-off times (cf.~Sec.~\ref{bee-bp}) of 80\,$\mu$s and 200\,ns were investigated. Since the camera is triggered externally, no additional triggers due to sampling are expected in the latter one. Fig.~\ref{fig:data_rate_random_200ns} shows the result of such rate measurements.
It can be observed that
\begin{itemize}
\item in case of no (or a very low) artificial backplane hold-off time (of 200\,ns), an efficiency of 1 ($\pm$0.1\%) is observed for camera trigger rates smaller than $\sim$900\,Hz. The uncertainty of 0.1\% is due to the (in)accuracy of counting the pulse generator triggers in a given time window,
\item in case of an artificial backplane hold-off time of 80\,$\mu$s, there is a chance of two or more (random) triggers arriving within that time. This chance increases with increasing mean trigger rate. This is why the packet efficiency is less than 1 at rates below 900\,Hz and decreases consistently with the fraction of readout requests, and
\item the packet efficiency data points of both cases lie above each other at rates higher than $\sim$900\,Hz. At this rate, the efficiency decreases (while the readout request fraction continues with the same slope) because the maximum transfer rate of the DACQ boards is reached causing packet loss being independent of any backplane hold-off time.
\end{itemize}

To conclude, due to the artificial hold-off time of 80~$\mu$s implemented on the backplane to avoid additional triggering on the trigger circuitry, the camera suffers a 5\% data packet loss at the requested mean random rate of 600~Hz in normal operation mode. However, it is shown here that once this problem is solved (which is the case in the next TARGET ASIC generation, cf.~Sec.~\ref{outlook}), no losses will occur at the requested rate of 600~Hz using UDP to send the data to the camera server.
\subsection{Transfer function}
\label{vped_isel}

As mentioned in Sec.~\ref{waveform_processing}, the transfer function provides a look-up table to relate ADC counts to signal amplitude in volts. This look-up table, different for each sampling cell of each pixel, is used in the electronic calibration of the camera raw data. Fig.~\ref{fig:tf_camera} illustrates the range of transfer functions of all ASIC sampling cells of all pixels in the camera for the ASIC parameters Vped\,$=$\,1050 (corresponding to 650~mV) and Isel\,$=$\,2816. The  chosen Vped value minimises the spread at low signal amplitudes while the selected Isel value ensures a high dynamic range of $\sim$3800\,ADC counts with a reasonable linear shape of the transfer function.
\subsection{Single photoelectron measurement and charge extraction}
\label{spe_gain}

\begin{figure*}[tb]
\centering
\subfigure[]{\includegraphics[width=0.49 \textwidth]{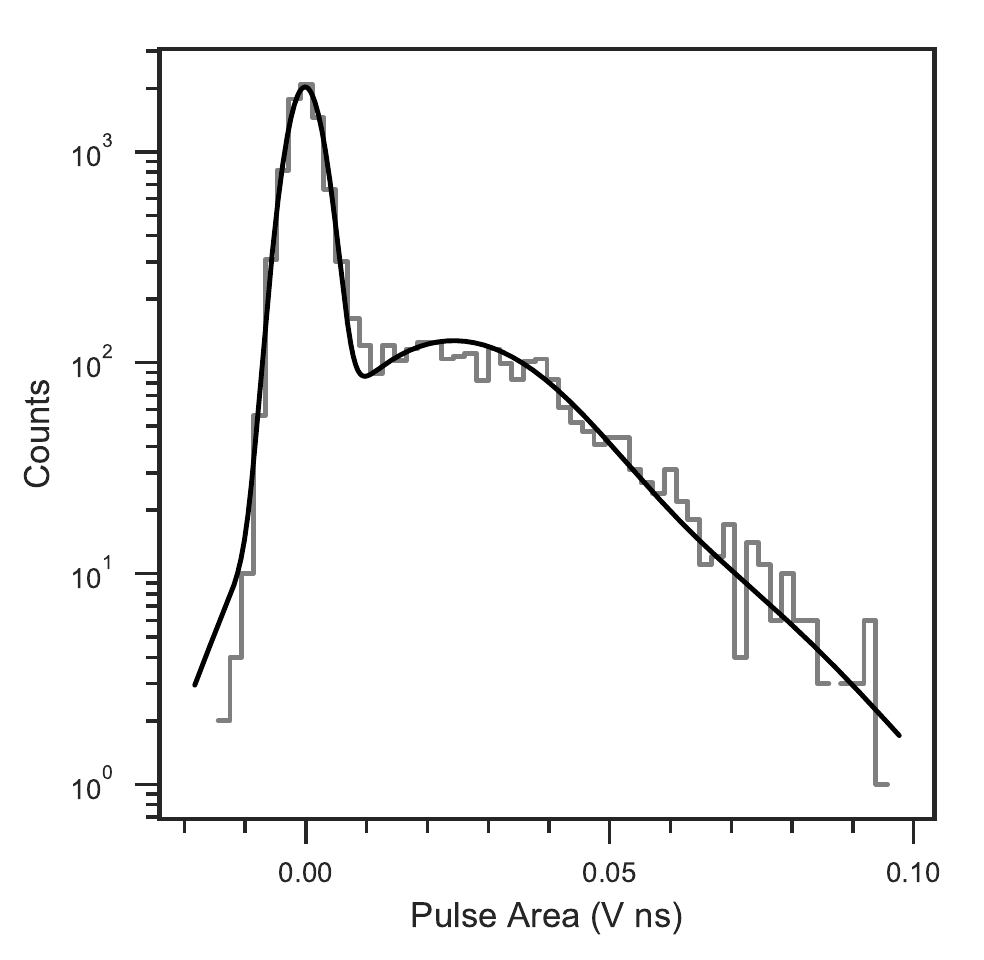}\label{fig:spe_1pixel}}
\subfigure[]{\includegraphics[width=0.49 \textwidth]{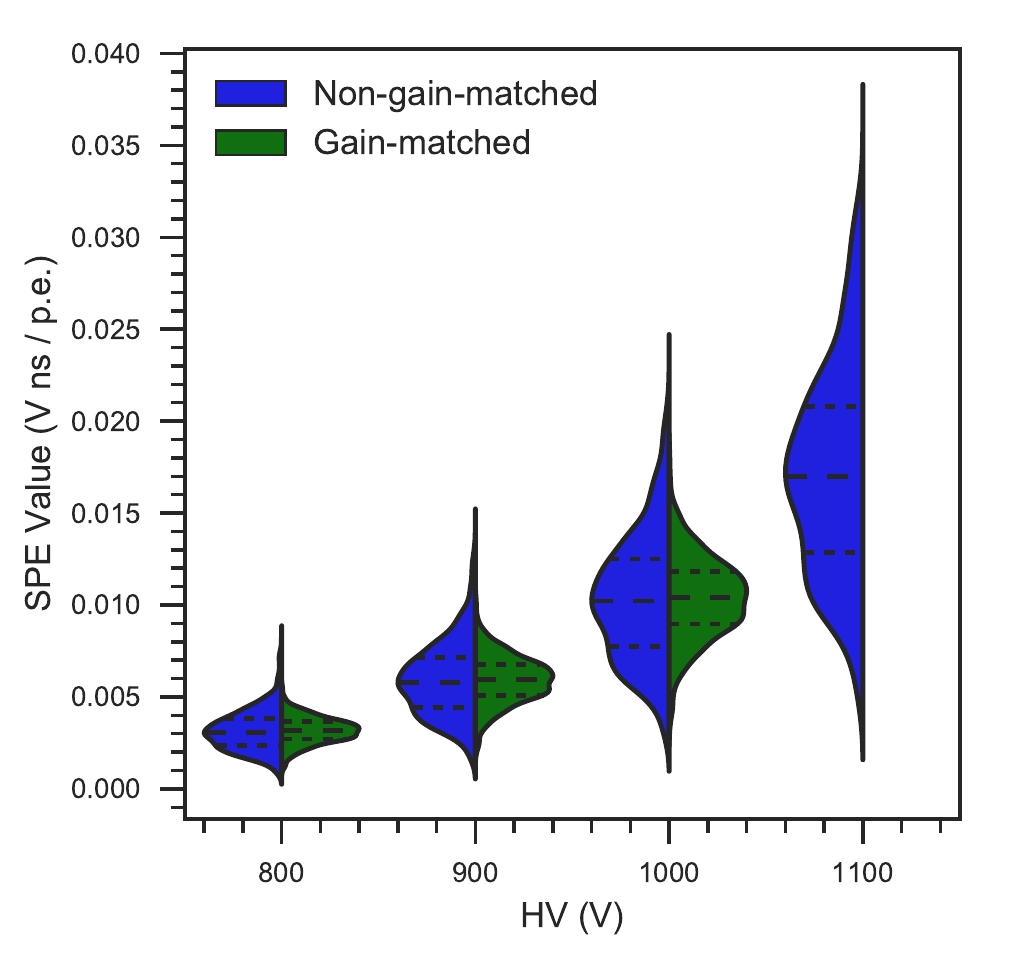}\label{fig:gain_matching}}
\caption[]{(a) SPE spectrum and fit for pixel 1559 at 1100\,V. (b) Distribution of SPE value across the camera for a non-gain-matched camera (all MAPMs set to the same HV indicated on x-axis, blue) and for a gain-matched camera (MAPMs set to different HVs, where mean HV is indicated on x-axis, green). Lines show the median and the interquartile ranges.}
\label{fig:spe_gain}
\end{figure*}

\begin{figure*}[tb]
\centering
\subfigure[]{\includegraphics[width=0.32\textwidth]{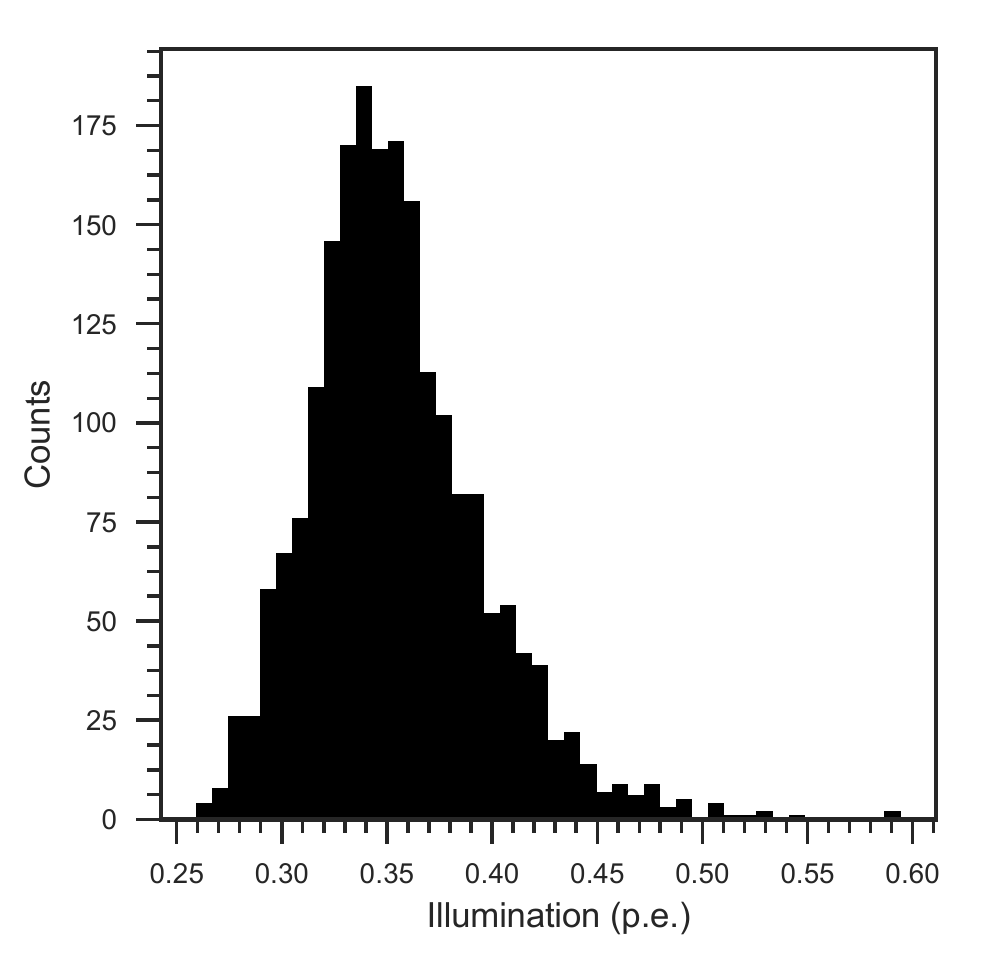}\label{fig:lambda_distribution}}
\subfigure[]{\includegraphics[width=0.32\textwidth]{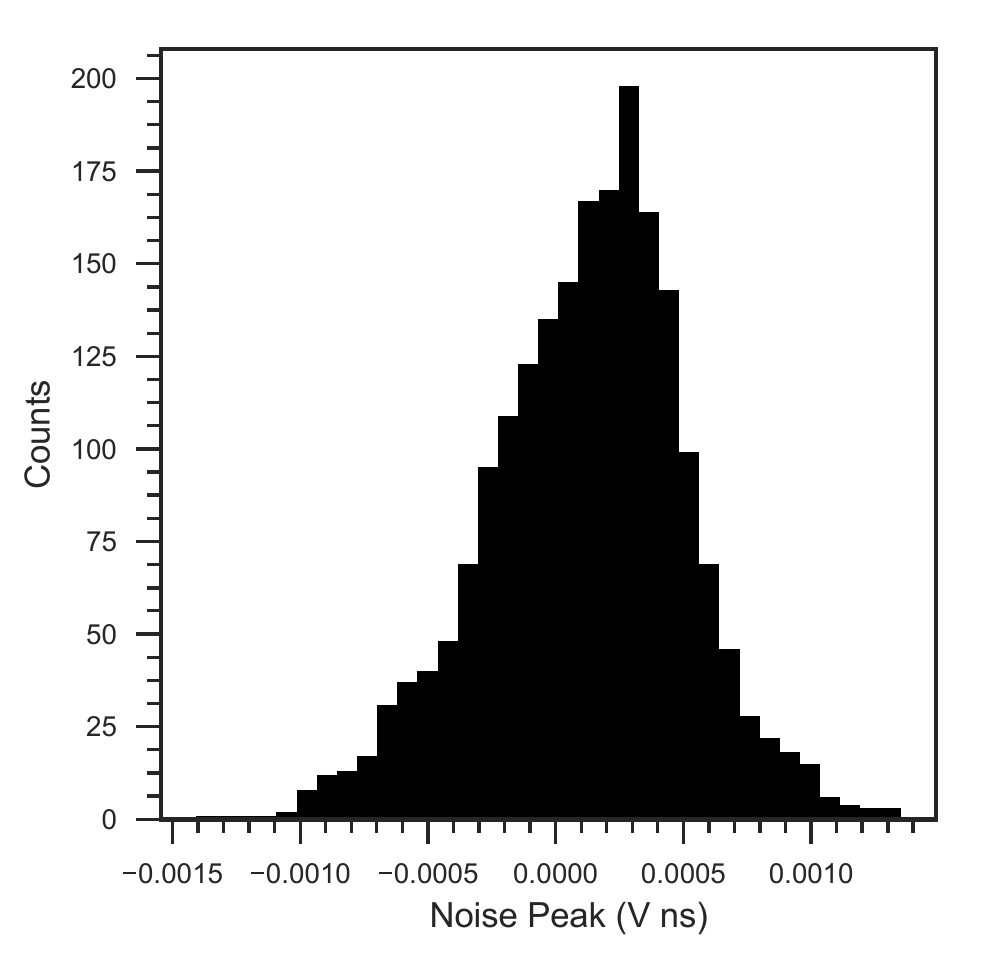}\label{fig:ped_distribution}}
\subfigure[]{\includegraphics[width=0.32\textwidth]{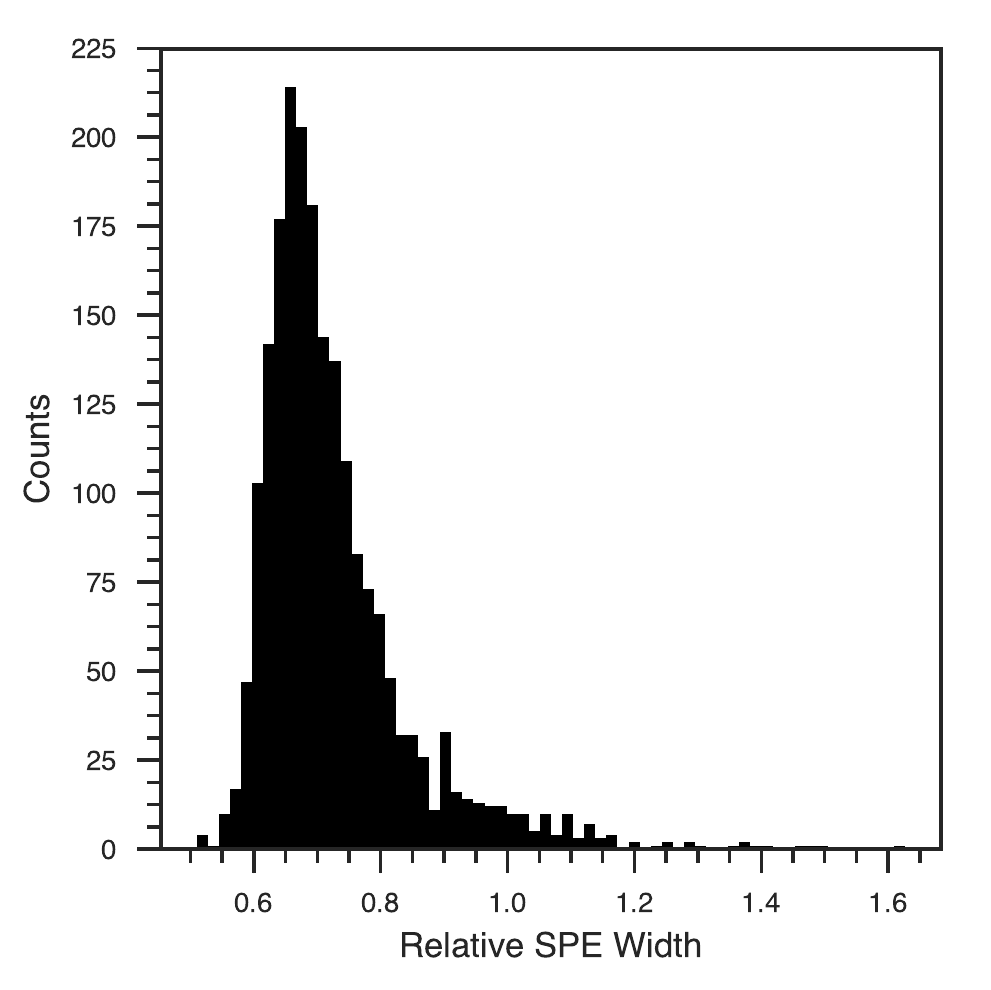}\label{fig:spe-width_distribution}}
\caption[]{Distributions of SPE fit results for all 2048 pixels: (a) mean illumination, (b) noise peak, and (c) relative SPE width.}
\label{fig:spe_fit_results}
\end{figure*}

Measuring the pulse area spectrum on the single p.e.~level is a fundamental step in the calibration to determine the conversion factor between pulse area (in V~ns) and p.e.~(referred to as SPE value). This factor is different for each pixel and depends on the HV the MAPM is supplied with. Due to the intrinsic performance of the MAPM, the most reliable SPE resolution can be obtained at the highest possible MAPM gain, corresponding to an HV of 1100~V. This is why the SPE measurement was done at that HV illuminating the whole camera with a medium light level of $\sim$0.35~p.e./pixel. Fitting the spectrum of each pixel with a Poisson distribution convolved with individual Gaussians for the noise and SPE peak (example for one pixel shown in Fig.~\ref{fig:spe_1pixel}) leads to different fit parameters for all 2048 pixels such as mean illumination level, noise peak\footnote{The noise peak is centred around 0 since the pedestal was subtracted for each ASIC storage cell during calibration of the data (as explained in Sec.~\ref{waveform_processing}).}, SPE value (distance between first (noise) and second (1~p.e.) peak), and relative SPE width. The latter one is proportional to the excess noise factor and defined as $\frac{\sqrt{\sigma_{\rm{Noise}}^2-\sigma_{\rm{SPE}}^2}}{SPE}$, where $\sigma_{\rm{Noise}}$ is the width of the noise peak, $\sigma_{\rm{SPE}}$ the width of the SPE peak, and $SPE$ the SPE value. Distributions of fit parameters including all 2048 pixels are shown in Fig.~\ref{fig:gain_matching} (SPE value, rightmost blue distribution) and Fig.~\ref{fig:spe_fit_results} (illumination, noise peak, and relative SPE width).

To measure SPE values at HVs less than 1100~V, in the first step the camera was illuminated at a higher illumination of $\sim$100~p.e./pixel at 1100~V. The charge in p.e.~was determined by calculating the pulse area and using the previously determined SPE value. In the second step, the HV was reduced keeping the illumination at the same level and thus the number of p.e.~constant. However, since the gain $G$, being the amplification factor of each pixel, decreases with decreasing HV, the pulse area also does so. This results in a lower SPE value, determined by measuring the pulse area at this lower HV and using the known number of p.e.. 
The blue distributions in Fig.~\ref{fig:gain_matching} show the SPE values of all pixels when all MAPMs are set to the same HV (indicated on the x-axis). The spread can be reduced by ``gain matching'' the camera, i.e.~by supplying each MAPM with a different HV. Three mean HV values (800~V, 900~V, and 1000~V), each of them consisting of 32 different HVs providing the same mean gain for all MAPMs, were determined in specific laser measurements where the laser amplitude was kept fixed at a medium illumination level while the HV was altered. The gain spread over the camera after gain matching is reduced, but still around 30\% (see green distributions in Fig.~\ref{fig:gain_matching}). This is due to the fact that the HV can only be set individually for each MAPM, not for each pixel (a fundamental feature of the MAPM design). However, since the camera was illuminated with the uniform light source, the remaining gain differences between the pixels could be used to define pixel dependent flat field coefficients, which are included in the SPE conversion factors described in ~Sec.~\ref{waveform_processing}.

\subsection{Trigger threshold determination}
\label{trigger_threshold}

\begin{figure*}[tb]
\centering
\subfigure[]{\includegraphics[angle=0,width=0.502 \textwidth]{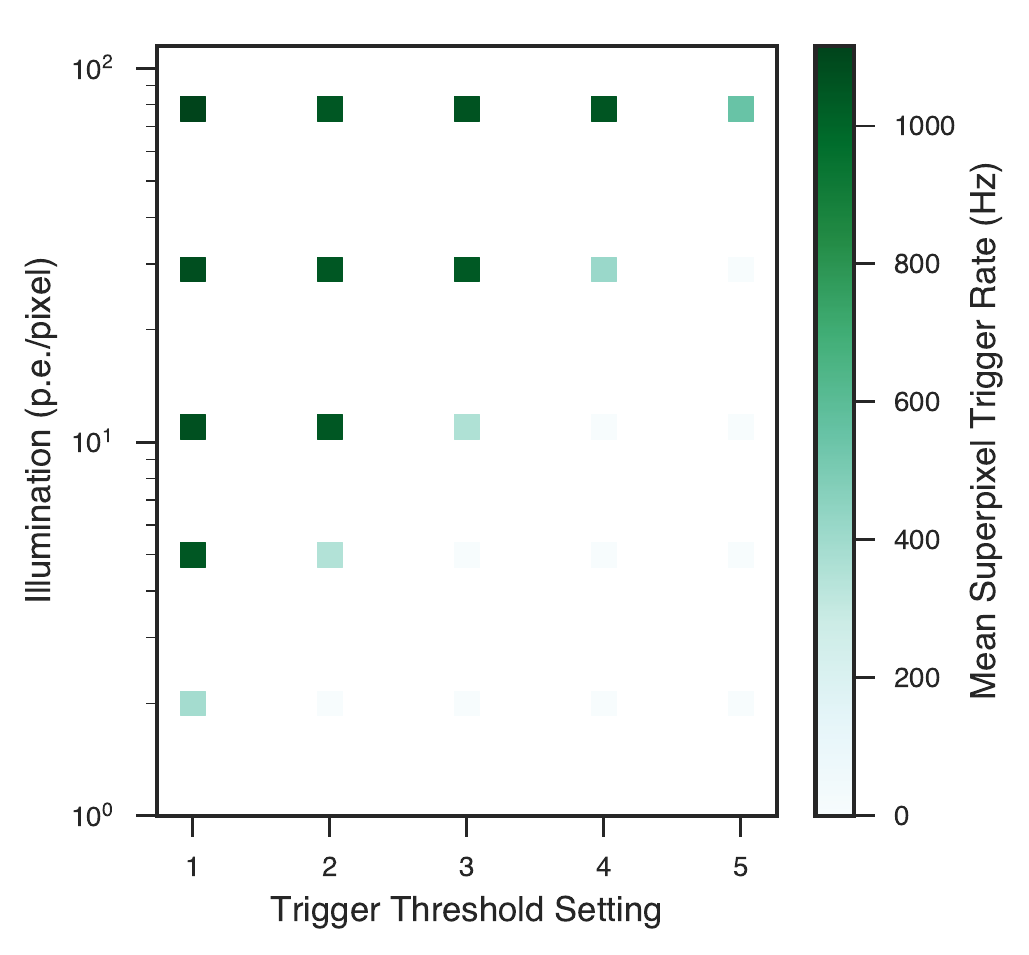}\label{fig:camera_rate_laser_setting}}
\subfigure[]{\includegraphics[width=0.478 \textwidth]{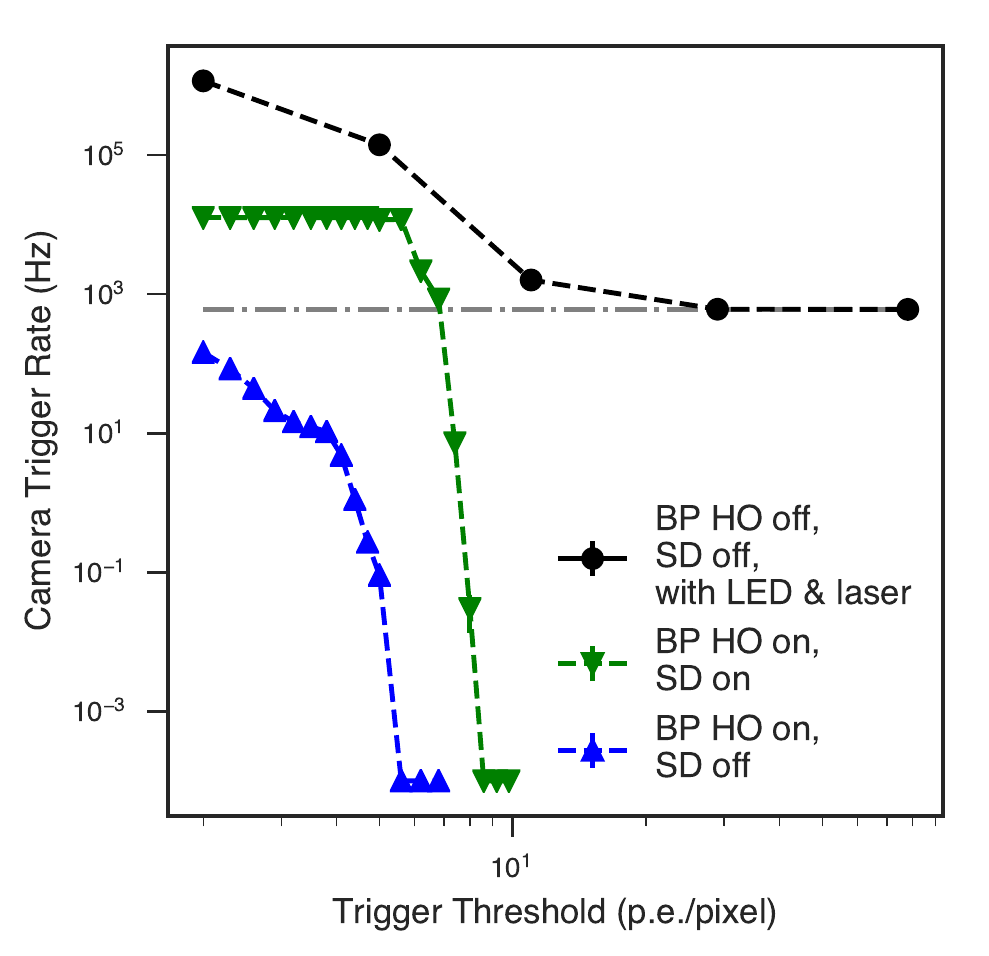}\label{fig:camera_rate_threshold_lab}}
\caption[]{(a) Camera trigger rate as function of the trigger setting and illumination (in p.e./pixel) for a mean HV of 800~V, laser triggering at 1\,kHz. The absolute illumination levels were deduced from the reconstructed mean pixel charges. (b) Camera trigger rate as function of the trigger threshold for three different scenarios all at a mean HV of 800~V: (1) BP HO on and SD off, (2) BP HO on and SD on, and (3) BP HO off, SD off, with white LED of 50~MHz and a laser with a constant rate of 600~Hz and an illumination level of about 200~p.e./pixel -- further explanation given in the text.}
\label{fig:camera_trigger_rate}
\end{figure*}

As described in Sec.~\ref{fee}, the TARGET 5 ASICs provide the first-level trigger of the camera by discriminating the analogue sum of four neighbouring pixels (referred to as superpixel). The threshold for discrimination is set individually for each superpixel by the combination of two ASIC parameters: Pmtref4, setting the reference voltage for the summing amplifier, and Thresh, setting the reference voltage for the comparators. There is a certain range of combinations of those parameters for which the trigger functions properly and different combinations can lead to the same trigger threshold, resulting in slightly different trigger noise levels (cf.~\cite{2017APh....92...49A} for detailed information and test results). 
Thus, each full camera trigger threshold setting consists of 512 (possibly) different pairs of Pmtref4 and Thresh. To identify these values, the camera was uniformly illuminated at different laser amplitudes while counting (with the backplane TFPGA) the number of triggers for each superpixel (first-level trigger) for a given light/HV level and (Pmtref4/Thresh) pair individually. The final (Pmtref4/Thresh) pair was chosen so that the superpixel trigger efficiency is about 50\% for the given laser amplitude in each superpixel. In this way, five different threshold settings (each with 512 pairs of Pmtref4 and Thresh) at mean illumination levels of around 2, 5, 11, 29, and 78 p.e./pixel at each of the three previously defined gain-matched HV settings were defined. Fig.~\ref{fig:camera_rate_laser_setting} shows the resulting mean superpixel trigger rate as a function of the laser illumination for the five threshold settings (at a mean HV of 800~V).

In the next step, the overall camera trigger rate (second-level trigger) was measured as a function of the camera trigger threshold. On telescope, such measurements are typically used to identify the camera trigger threshold operating point for given background light conditions. To understand the influence not only of the NSB but also of the TARGET module sampling and data sending on the camera trigger rate, three different scenarios were investigated in the lab (results shown in Fig.~\ref{fig:camera_rate_threshold_lab}):
\begin{enumerate}
\item Backplane hold-off (BO) time set to 80~$\mu$s as used in CHEC-M for nominal operation (BP HO on) and all TARGET modules configured to not sample and not send data (SD off) -- blue data points in Fig.~\ref{fig:camera_rate_threshold_lab},
\item BP HO on and all TARGET modules configured to sample and send data (SD on) -- green data points in Fig.~\ref{fig:camera_rate_threshold_lab}, and
\item BP HO off and SD off, with a white LED emulating an NSB rate of $\sim$50~MHz (measured with an additional SiPM with known gain and temperature-gain dependence) and a bright laser with a constant rate of 600~Hz (shown as grey line in Fig.~\ref{fig:camera_rate_threshold_lab}) emulating Cherenkov light of about 200~p.e./pixel -- black data points in Fig.~\ref{fig:camera_rate_threshold_lab}.
\end{enumerate}
It was observed that SD produces additional triggers causing the trigger rate to increase by $\sim$4 orders of magnitude at a camera trigger threshold of 5~p.e./pixel compared to the scenario where SD is disabled. This implies that the trigger circuitry picks up noise not only from the FPGA serial data signals used to read out the ASICs (already fixed by using a BP HO time of 80~$\mu$s, cf.~Sec.~\ref{data_rate}) but also from the sampling and data sending process itself. Both issues will be solved by design in future FEE iterations (see Sec.~\ref{outlook}). This is why the third measurement in Fig.~\ref{fig:camera_rate_threshold_lab} (black curve) shows how the rate curve is expected to look like in future FEE iterations under the influence of NSB and Cherenkov showers. In this example, the camera trigger rate first decreases from $\sim$50~MHz at 2 p.e./pixel, where the trigger is completely dominated by the emulated NSB, down to 600~Hz at a camera trigger threshold of 29~p.e./pixel. From this point on, the camera trigger is dominated by the laser, emulating a Cherenkov light signal.

For an SST with a CHEC-M-like camera the NSB level on the CTA site is expected to lie between 15 and 25~MHz (lower than in the lab measurements shown in Fig.~\ref{fig:camera_rate_threshold_lab}). Thus, the rate curve is expected to flatten at lower trigger threshold compared to Fig.~\ref{fig:camera_rate_threshold_lab}, which means that a range of trigger threshold settings between 2 and 100~p.e./pixel should be sufficient. It is useful to determine additional intermediate trigger threshold settings. This can be done either by performing a finer and thus more time-consuming (Pmtref4/Thresh) scan or by interpolating the (Pmtref4/Thresh) values between two settings. The latter method was used in two of the three trigger rate measurements presented above (blue and green data points in Fig.~\ref{fig:camera_rate_threshold_lab}, intermediate steps between the previously determined five trigger threshold settings).

This procedure for trigger threshold determination was used to produce fine-grained steps of approximately 20 per decade in threshold for the use of on-sky measurements. Excessively noisy trigger pixels were disabled at each threshold setting until the second-level trigger rate stabilised. 
For the next CHEC camera prototype (cf.~Sec.~\ref{outlook}), the procedure will be different: The individual ASIC parameters (Pmtref4 and Thresh) will be  determined during the commissioning of the new TARGET modules by injecting electrical signals. Thus, the trigger threshold determination is disentangled from photosensor characteristics like gain, quantum efficiency, etc.~and a camera trigger threshold determination to a precision of $<$1~p.e./pixel will be possible.
\subsection{Pulse shape characteristics}
\label{pulse_shape}

\begin{figure*}[tb]
\centering
\includegraphics[width=1. \textwidth]{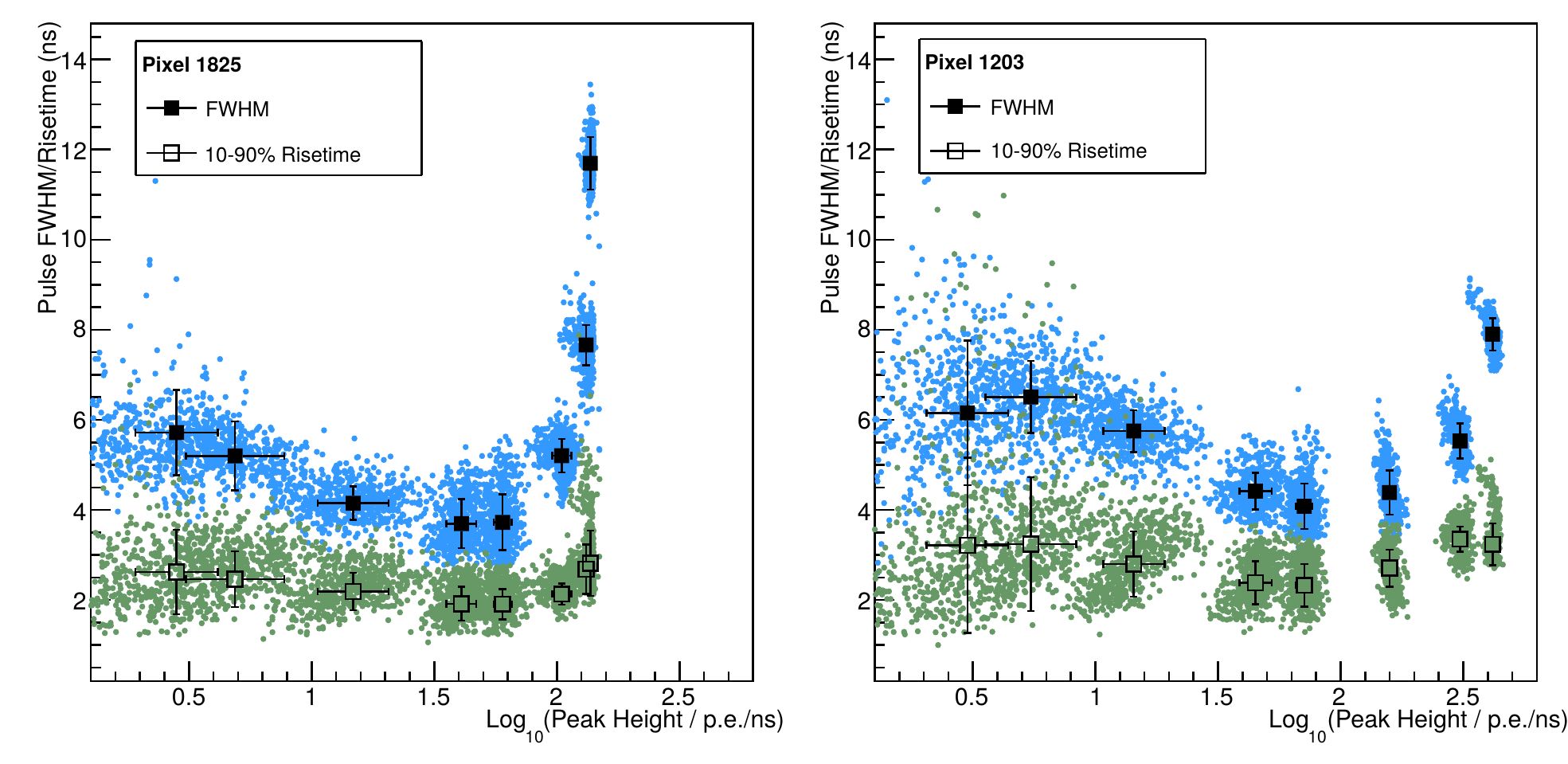}
\caption[]{Pulse FWHM (blue points) and 10--90\% risetime (green points) as function of measured peak height for camera pixel 1825 (left) and 1203 (right). The peak height (in units of p.e./ns) was extracted from the data using the procedure explained in Sec.~\ref{waveform_processing}. The squares indicate different laser illumination levels, their y-position the mean pulse FWHM/risetime, and the bars the standard deviation in measured FWHM/risetime and peak height at each laser brightness.}
\label{fig:fwhm_risetime_vs_illumination}
\end{figure*}

As explained in Sec.~\ref{fee}, to optimise the trigger performance, the signal pulse FWHM and 10--90\% pulse risetime should lie between 5 and 10~ns and between 2 and 6~ns, respectively, over the whole SST energy range. Fig.~\ref{fig:fwhm_risetime_vs_illumination} illustrates the FWHM and risetime for two pixels\footnote{These two pixels were chosen on a semi-random basis as qualitative representatives of all camera pixels. They show neither particularly good nor poor characteristics and are geometrically well separated (one is near the edge of the camera and one at the centre), rather than e.g.~being of the same FEE module.} as function of the illumination level for laser data taken at different laser brightnesses all at 1100~V.
It shows that the pulse shape requirements for optimal triggering are met at low illumination levels and that the pulse shape is stable up to an illumination of 200--400~p.e./pixel depending on the camera pixel. At these illuminations, saturation effects can be observed, i.e.~while the pulse peak height stops to increase the FWHM continues to increase with increasing laser amplitude. The amplitude at which saturation occurs is different for each pixel due to different gains at 1100~V and/or different quantum or collection efficiency. Since no quantum or collection efficiency spread between different MAPMs and pixels are reported by the manufacturer, the influence of the latter aspect is expected to be small and much lower than the gain spread at 1100~V.

\begin{figure}[tb]
\centering
\includegraphics[width=0.5 \textwidth]{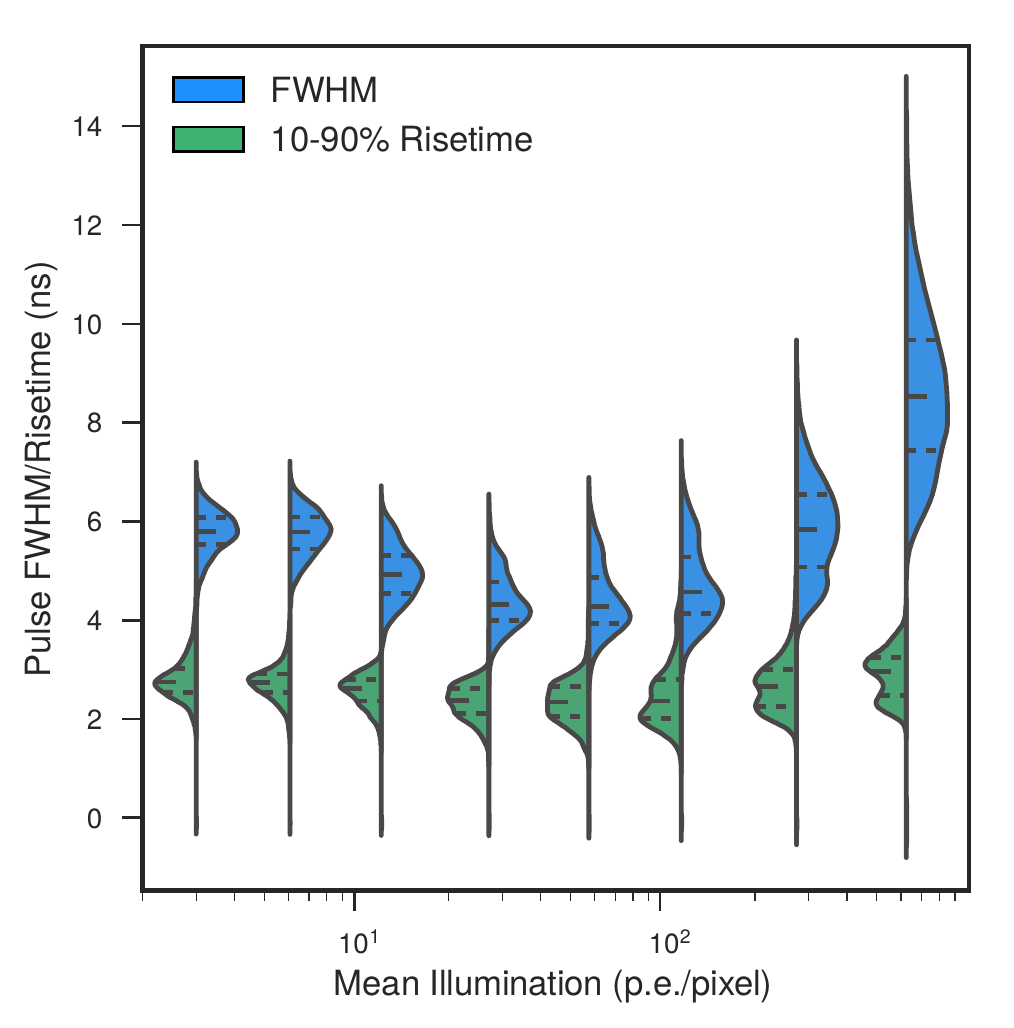}
\caption[]{Pixel distributions (mean for each pixel) for pulse FWHM (blue) and 10--90\% risetime (green) for different illumination levels. Lines show the median and the interquartile ranges.}
\label{fig:fwhm_risetime_vs_illumination_pixel_distribution}
\end{figure}

The influence of saturation can also be observed in Fig.~\ref{fig:fwhm_risetime_vs_illumination_pixel_distribution} showing the FWHM and 10--90\% risetime distribution for all 2048 pixels at different illumination levels (same data set as used for the sample pixels in Fig.~\ref{fig:fwhm_risetime_vs_illumination}).
As explained above, due to the gain spread at 1100~V, saturation occurs at a different illumination level for each pixel, resulting in a wide spread in FWHM, especially at the highest illumination levels when all pixels are affected by saturation effects. However, for all pixels, both the FWHM and 10--90\% risetime fulfil the requirements for optimal triggering.

The intrinsic MAPM pulses are significantly shorter (FWHM of $\sim$1~ns, cf.~Sec.~\ref{mapm}) than the pulses measured with the whole chain (FWHM between 5 and 10~ns) being dominated by the preamplifier pulse shape characteristics. Thus, the effect of the HV (i.e.~also of a lower HV and gain) on the pulse shape characteristics -- measured at a given signal amplitude in V -- is expected to be negligible\footnote{Of course, since a higher HV/gain results in higher SPE values and saturation effects affect the pulse shape characteristics (as shown in this section), a lower/higher HV affects the pulse shape characteristics, if measured as a function of illumination level, shifting the data points in Fig.~\ref{fig:fwhm_risetime_vs_illumination} \& \ref{fig:fwhm_risetime_vs_illumination_pixel_distribution} to higher/lower illumination levels.}.
\subsection{Crosstalk}
\label{xtalk}
\begin{figure}[tb]
\centering
\includegraphics[width=0.5 \textwidth]{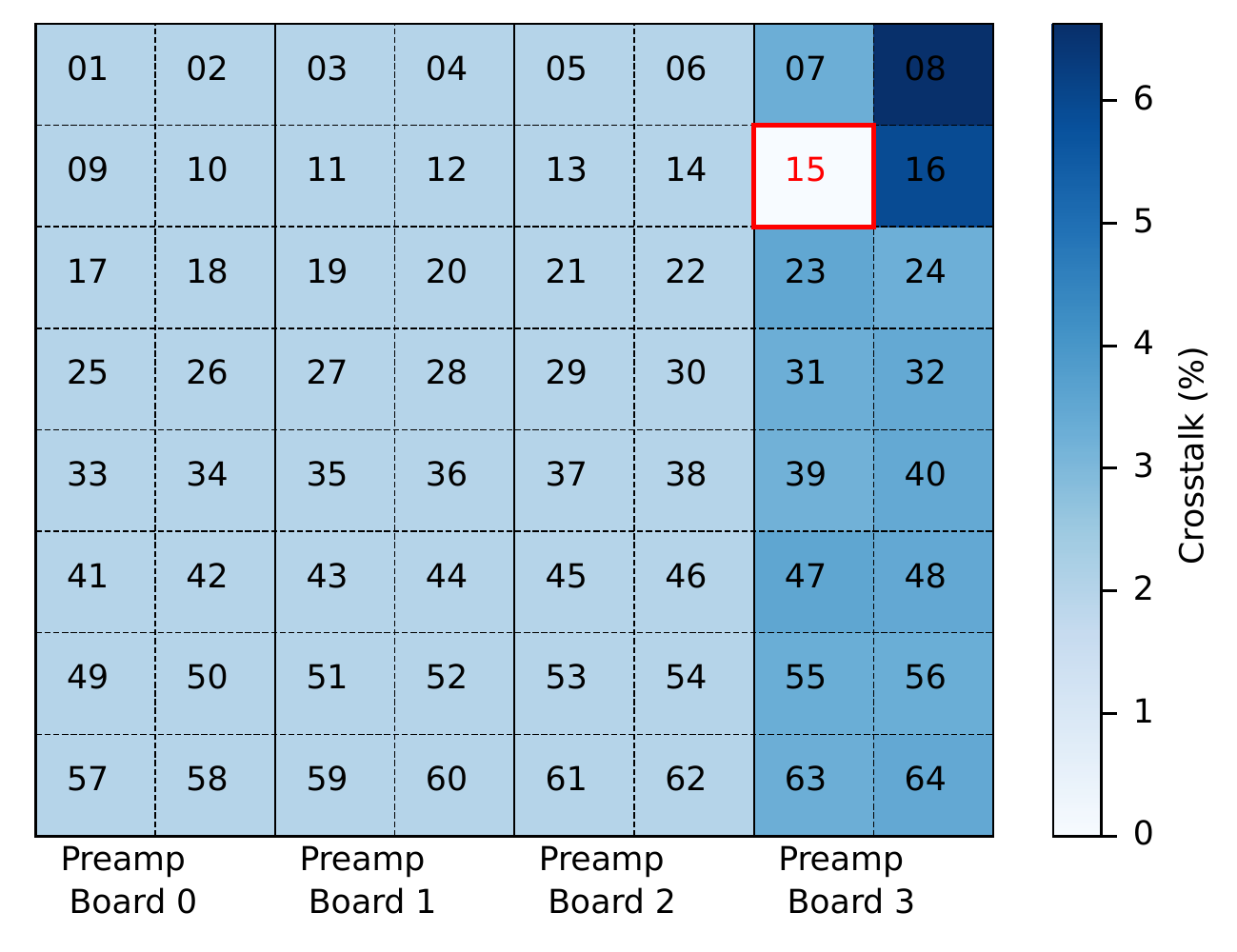}
\caption[]{Crosstalk (in \%, indicated by colorbar) for different pixels of one MAPM at 1100~V routed to different preamplifier (preamp) boards when only pixel 15 (indicated by red square) was illuminated with a laser (all other pixels were physically masked). For more details see text.}
\label{fig:xtalk_1000V}
\end{figure}

The crosstalk measurement was performed at 1100~V with an MAPM connected to a CHEC-M preamplifier module which was in-turn probed with an oscilloscope while only one pixel was illuminated with a laser (all other pixels physically masked). The peak-to-peak voltage of the average pulses from both the signal and candidate pixel were then measured and the ratio taken as an indication of the crosstalk. The 64 pixels of a single MAPM are mapped in groups of 16 to preamplifier boards, and then one-to-one to ASICs. Clear average pulses were seen in all pixels connected to the same board as the signal pixel, resulting in an average crosstalk of 4--5\% and reaching a maximum of 6\% in neighbouring pixels (see Fig.~\ref{fig:xtalk_1000V}).
No discernible pulses were measured in pixels of the other three preamplifier boards and the values shown for boards 0--2 in Fig.~\ref{fig:xtalk_1000V} represent the limit of the measurement technique and should be taken as an indication that no significant crosstalk has been measured. This measurement strongly indicates that the preamplifier PCBs rather than the MAPM are the dominant source of crosstalk in the system. A lower HV/gain is therefore not expected to have a significant impact on the crosstalk. Furthermore, this result is not entirely unexpected since in a compact, high-density system it is inevitable that signals must be routed in close proximity on any given PCB. However, for next CHEC iteration designs, the preamplifier board routing has been optimised to minimise crosstalk.

The results of the crosstalk measurements have to be taken into account in the uncertainty evaluation. A maximum crosstalk of 6\% between neighbouring pixels does not only degrade the charge resolution by the same amount, but also the SPE calibration, both affecting the image reconstruction: Taking into account a possible gain spread of about 30\% between neighbouring pixels, high-gain pixels can bias the SPE calibration of neighbouring low-gain pixels by about 8\% due to crosstalk. In addition, the crosstalk also affects other camera performance aspects like the trigger efficiency.
\subsection{Dynamic range}
\label{dynamic_range}

\begin{figure*}[tb]
\centering
\subfigure[]{\includegraphics[width=0.49 \textwidth]{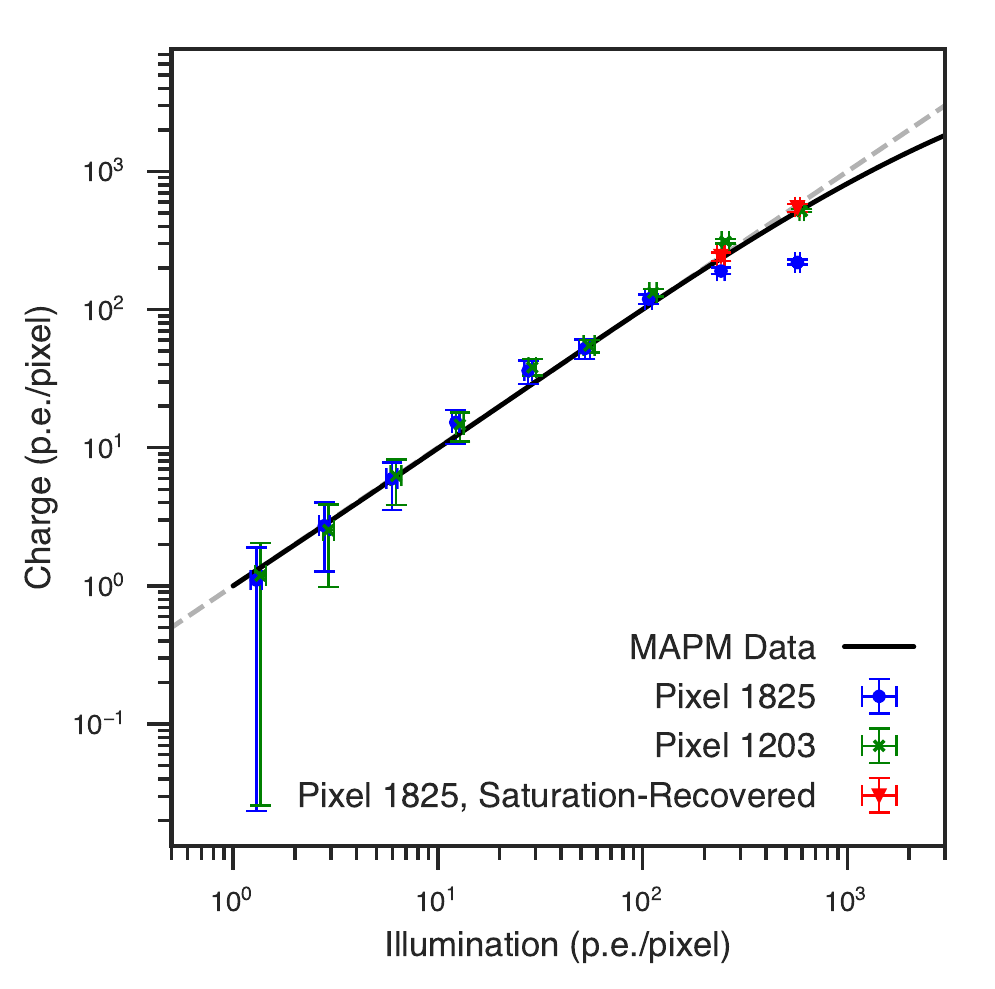}\label{fig:dynamic_range}}
\subfigure[]{\includegraphics[width=0.49 \textwidth]{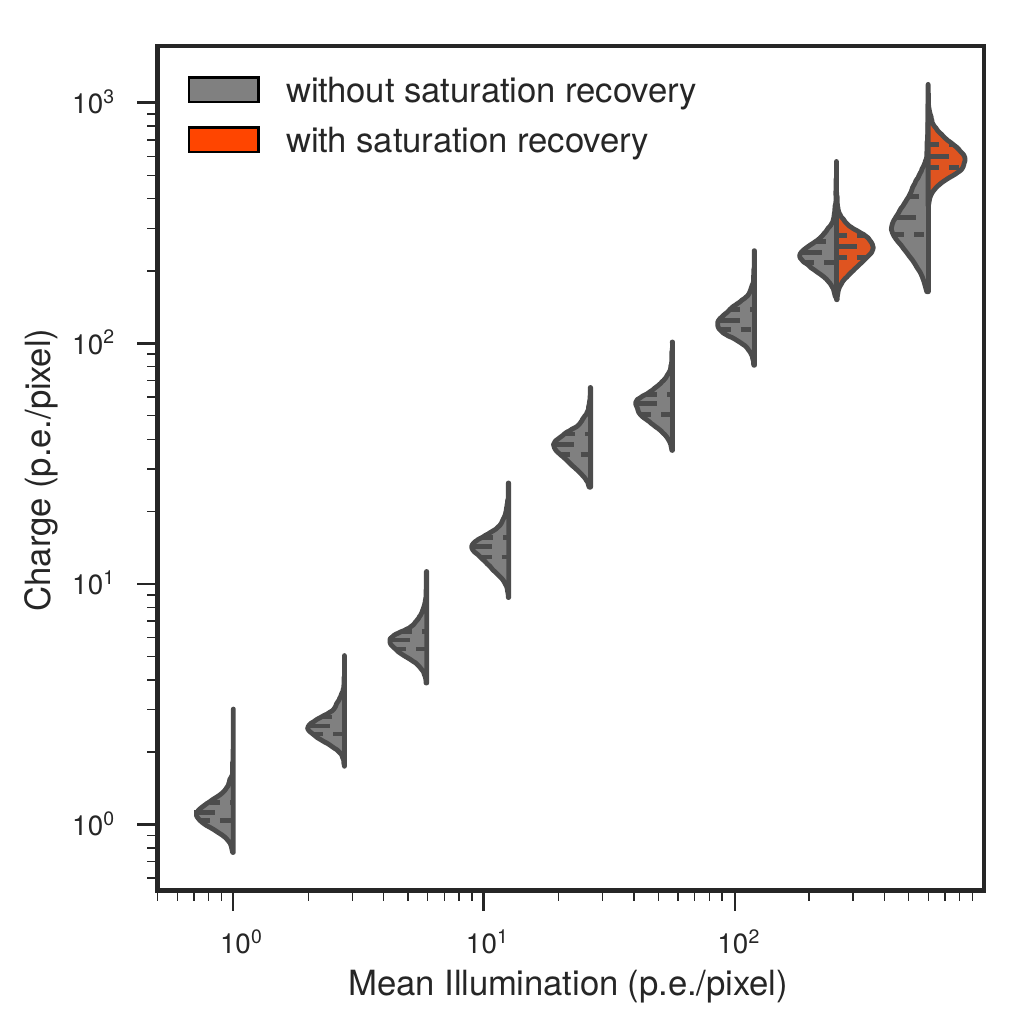}\label{fig:dynamic_range_all}}
\caption[]{(a) Reconstructed charge as function of the camera illumination for two camera pixels with the full chain of MAPM and FEE module, and for the MAPM only. 
Points show the mean, bars the 25$^{\rm{th}}$ and 75$^{\rm{th}}$ percentile of the charge distribution for the given pixel and illumination. A first attempt for saturation recovery using the pulse width at a fixed peak amplitude is shown for pixel 1825 (see text for details). The grey dashed line shows a 1:1 relation between the axes. (b) Pixel distributions (mean for each pixel) for extracted charge at a given mean illumination level with and without attempts to recover from saturation at the highest illumination levels using the pulse width. Lines show the median and the interquartile ranges.}
\end{figure*}

The dynamic range of the signal recording chain (MAPM and FEE module) was assessed by illuminating the entire camera with a uniform light level ranging from below 1 p.e./pixel to several hundreds of p.e./pixel in calibrated steps. The measurements were done supplying all MAPMs with the maximum voltage of 1100~V to be able to resolve SPE at low illumination. Results for two camera pixels (same as used for the FWHM and risetime investigation, cf.~Sec.~\ref{pulse_shape}) and of an MAPM-only measurement for comparison are shown in Fig.~\ref{fig:dynamic_range}.
\begin{figure}[tb]
\centering
\includegraphics[width=0.5 \textwidth]{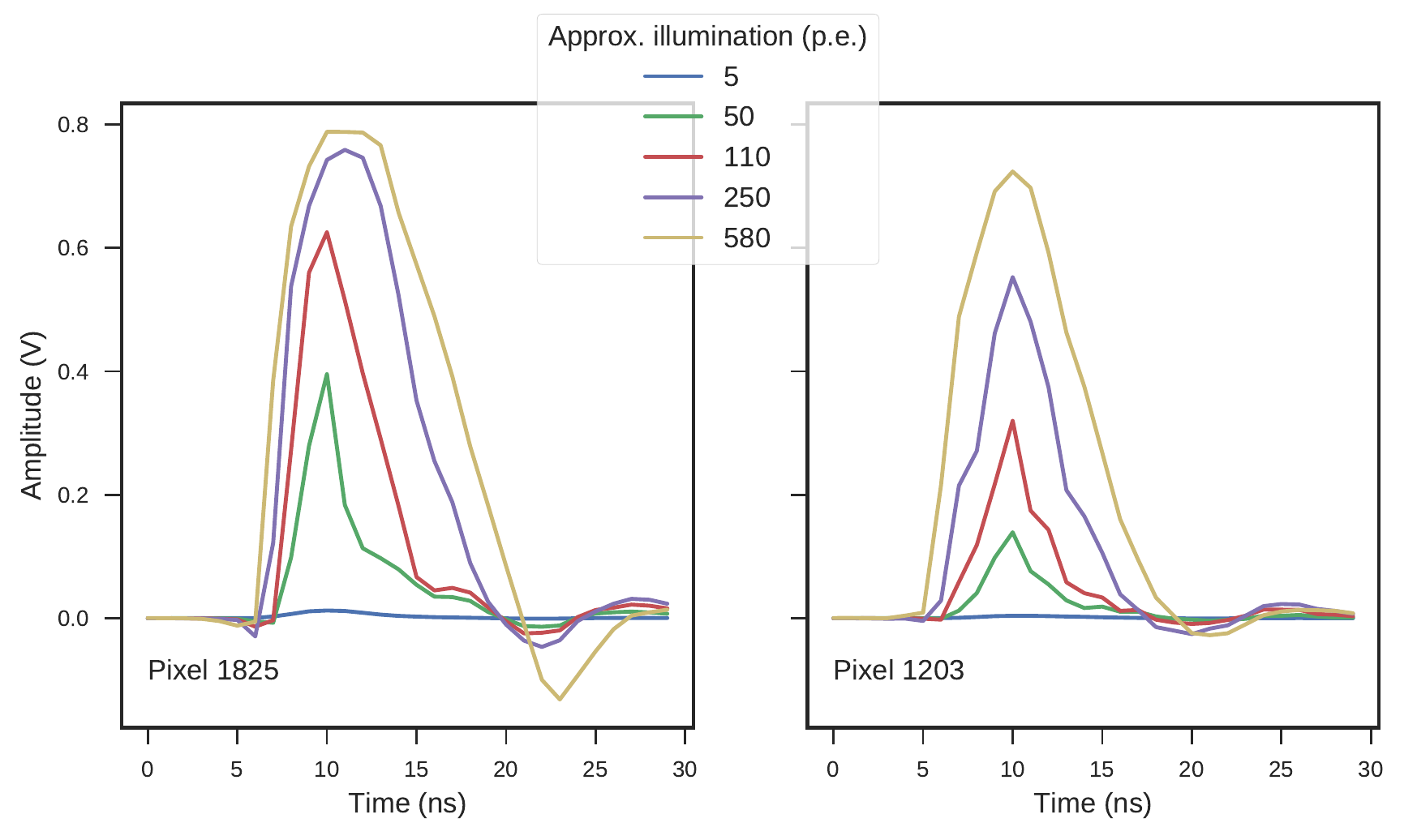}
\caption[]{Calibrated waveforms of camera pixel 1825 (left) and 1203 (right) at different illumination levels. At an illumination of 580~p.e./pixel, the pulse in pixel 1825 shows obvious saturation effects (plateau at the top).}
\label{fig:saturated_waveforms}
\end{figure}
For the camera pixel data points, the charge (y-axis) was reconstructed following the procedure explained in Sec.~\ref{waveform_processing}. Its value at $\sim$50~p.e./pixel was used as anchor for the illumination level on the x-axis to absolute calibrate the laser. Other points on the x-axis were then inferred from the relative calibration of the filter wheel used to adjust the laser intensity, resulting in an illumination in units of p.e./pixel. The MAPM data points were inferred from measurements where the MAPM signal was directly measured with an oscilloscope while the absolute calibration of the x-axis was determined from single-p.e.~spectra fits at the lower end of the range and using the relative calibration of the filter wheel for the rest of the x-axis range.

According to Fig.~\ref{fig:dynamic_range}, clear deviation from a linear correlation between illumination level and reconstructed charge (using the procedure described in Sec.~\ref{waveform_processing}) starts to occur at an illumination of $\sim$250~p.e./pixel for camera pixel 1825 while it is not observed in pixel 1203 over the range of laser brightnesses used in these measurements. This can again (as in Sec.~\ref{pulse_shape}) be explained by saturation effects occurring at different illumination levels due to different gains at 1100~V and due to different quantum and collection efficiencies between the two pixels (less substantial). Furthermore, the full signal recording chain (MAPM and FEE module) has a similar response as the MAPM itself, showing non-linearity effects of about 20\% at 1000~p.e./pixel. Consistent results are obtained when looking at the dynamic range of all pixels in the camera (see Fig.~\ref{fig:dynamic_range_all}).

As can be suggested from the waveforms of the two pixels at different illumination levels, shown in Fig.~\ref{fig:saturated_waveforms}, as well as from the FWHM dependence on the illumination level (cf.~Fig.~\ref{fig:fwhm_risetime_vs_illumination}), the pulse width increases with illumination. Thus, the relationship between pulse width at fixed amplitude (e.g.~at 20~p.e./pixel) and input illumination level can be used as a first attempt for recovery in saturation as shown in Fig.~\ref{fig:dynamic_range} for pixel 1825 and Fig.~\ref{fig:dynamic_range_all} for all pixels.

The overall dynamic range can be shifted to higher illumination levels by reducing the gain. Operating the camera at a mean HV of 800~V instead of 1100~V reduces the gain by a factor of $\sim$6 shifting the upper end of the dynamic range to $\sim$6000\,p.e./pixel, resulting in an overall dynamic range of $\sim$4 orders of magnitude.
\subsection{Timing}
\label{timing}
To investigate time differences between digitised signals of different pixels hit by the same light flash simultaneously, the camera was externally triggered while illuminated by a laser at 1100~V. 
\begin{figure*}[tb]
\centering
\subfigure[]{\includegraphics[width=0.49 \textwidth]{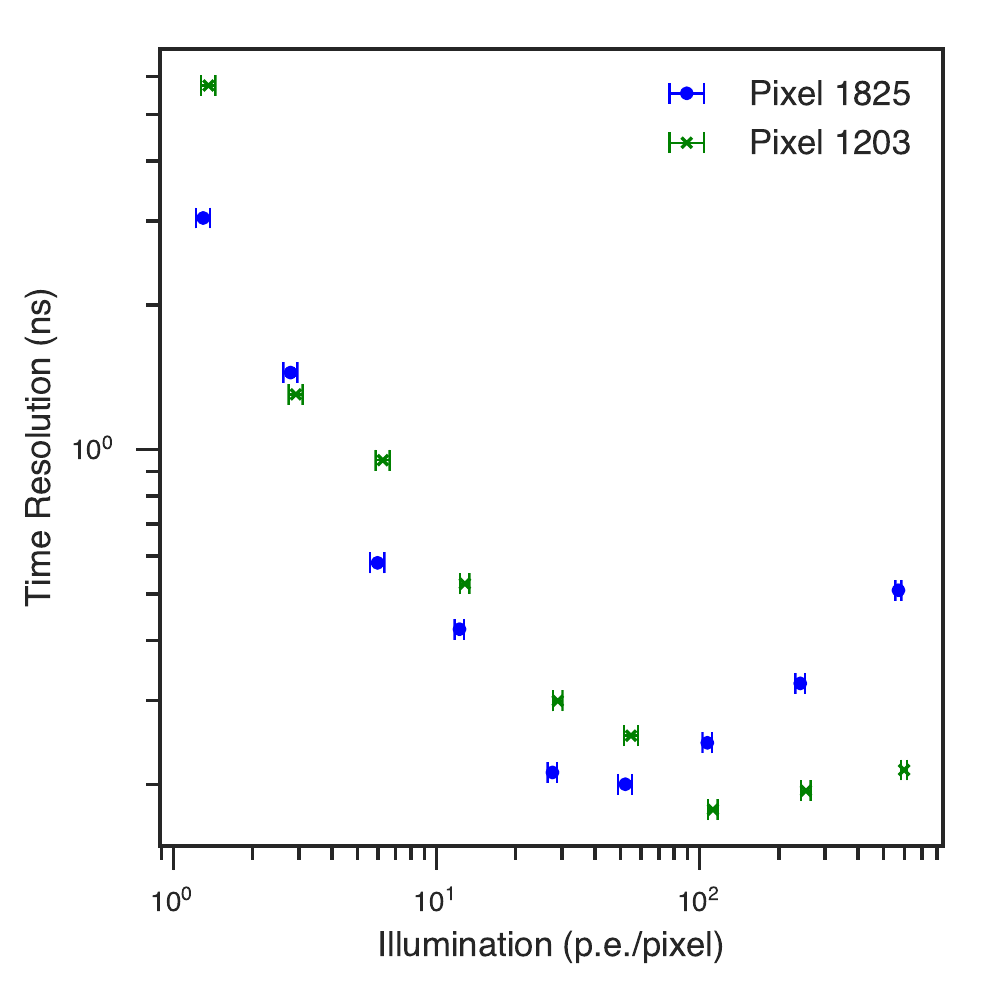}\label{fig:time_res}}
\subfigure[]{\includegraphics[width=0.49 \textwidth]{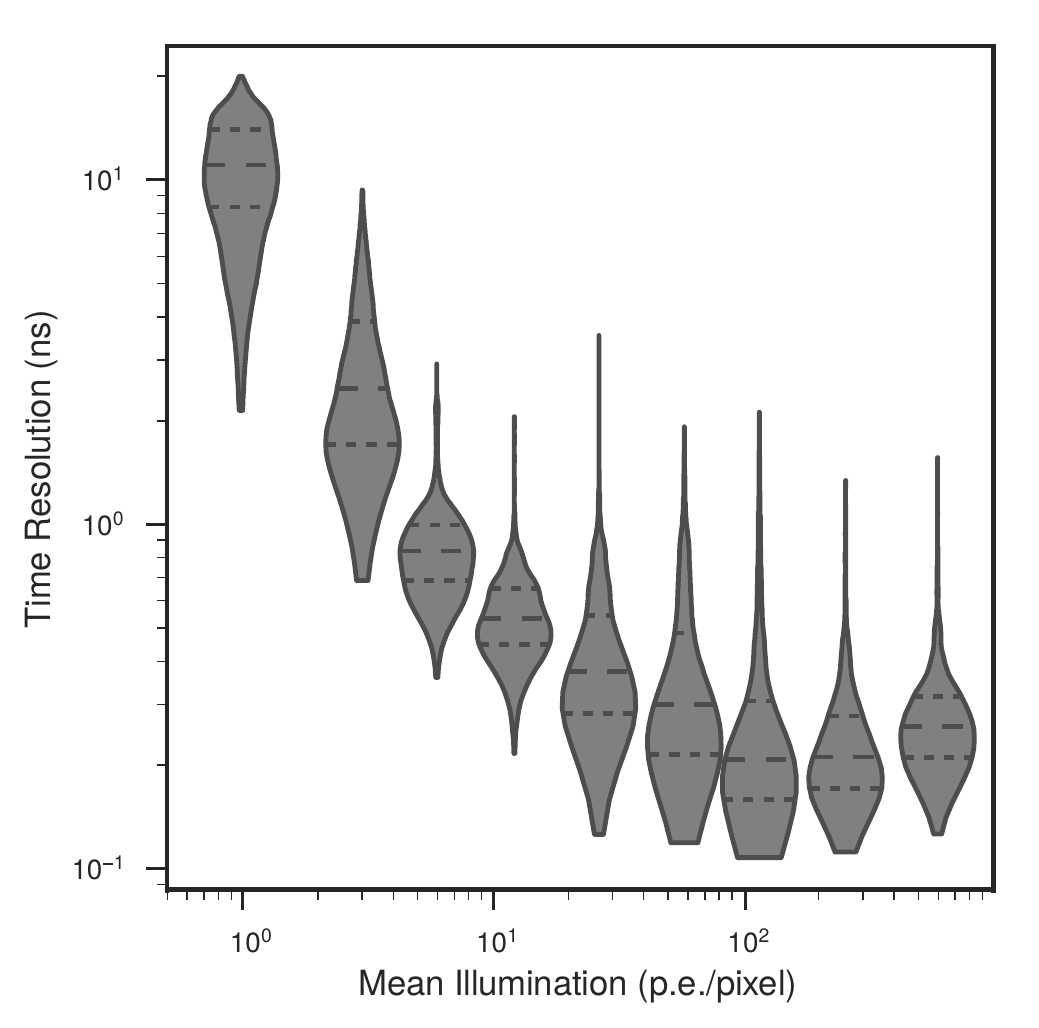}\label{fig:time_res_all}}
\caption[]{Time resolution as function of the camera illumination (a) for camera pixels 1825 and 1203 and (b) for all pixels (distributions) with lines showing the median and the interquartile ranges.}
\label{fig:time_res_overall}
\end{figure*}
For each illumination level and pixel, the pulse peak time distribution was determined out of 500 events, where the individual peak time of each pixel and event was shifted by the camera mean of the given event to overlap different events. The time resolution, defined as standard deviation of the peak time distribution, was measured for different illumination levels (see Fig.~\ref{fig:time_res_overall} for two sample pixels and for the distribution of all pixels, respectively). It improves with increasing illumination due to increasing signal-to-noise ratio and is (for most of the pixels) better than 1~ns for illumination levels $>$6p.e./pixel. Deterioration for illumination levels higher than 110~p.e./pixel is observed in the mean of the all-pixel distribution and for the sample pixel 1825, again due to saturation effects occurring at different illumination levels.

The timing and time resolution could be affected by a changing pulse shape. This could explain the degradation of the time resolution at high illumination levels in the saturation regime. However as explained previously in Sec.~\ref{pulse_shape}, the impact of a lower HV/gain at a given signal amplitude in V is expected to be insignificant on the pulse shape, thus the same holds for the time resolution.
\subsection{LED calibration flashers}
\label{flashers}

The four flasher units were tested for stability and temperature dependence, as well as on their dynamic range. They were operated in a temperature controlled environment while their brightness was measured with an SiPM (with known temperature-gain dependence). The overall dynamic range of one flasher consisting of 10 LEDs is about four orders of magnitude with illumination levels at the camera in the range of sub-p.e./pixel up to a few thousands of p.e./pixel.
\begin{figure*}[tb]
\centering
\subfigure[]{\includegraphics[width=0.49 \textwidth]{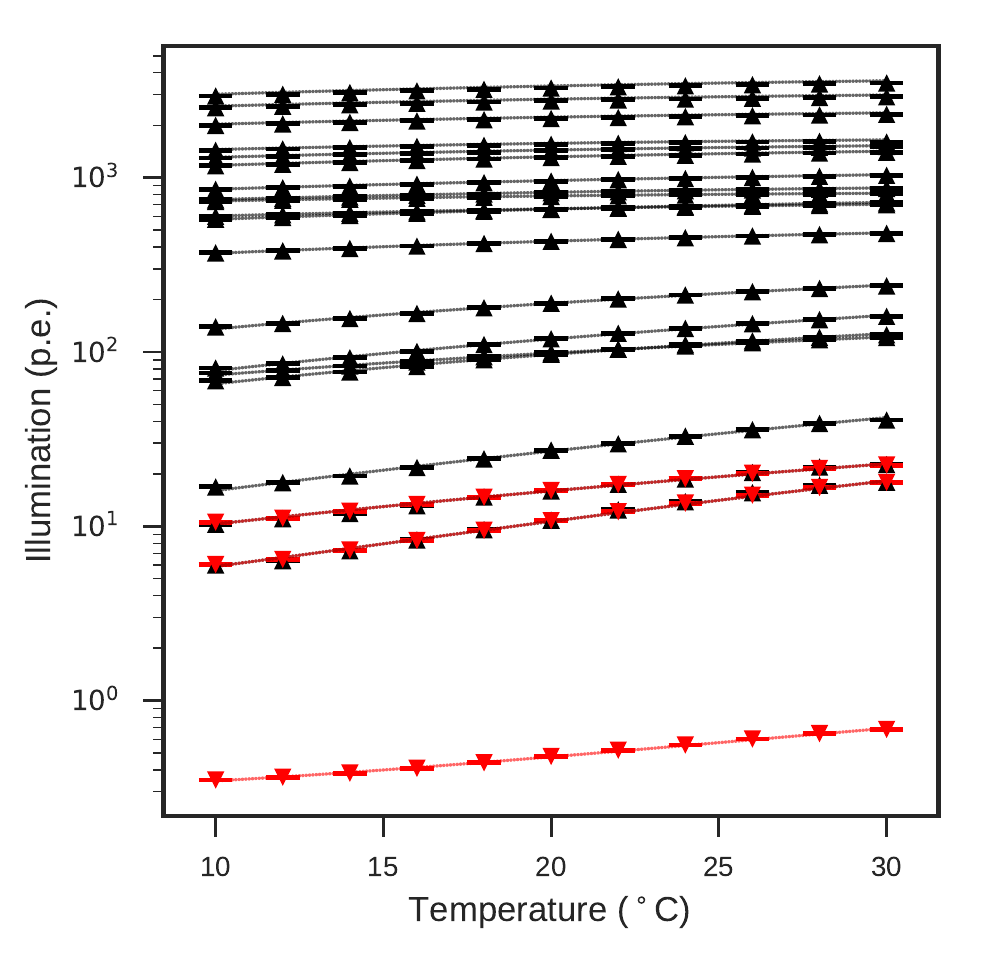}\label{fig:flasher_temp_fit}}
\subfigure[]{\includegraphics[width=0.49 \textwidth]{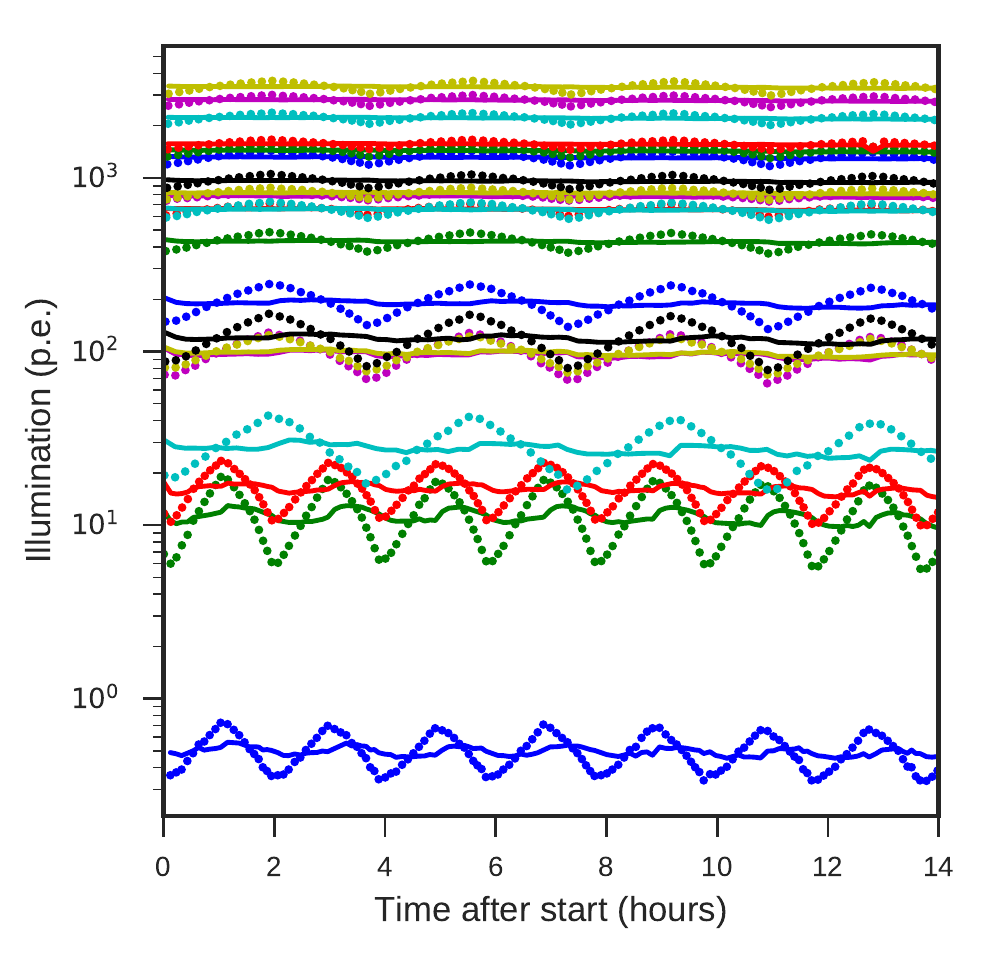}\label{fig:flasher_temp_cycle_correction}}
\caption[]{(a) Flasher brightness, measured with an SiPM with known gain and temperature-gain dependence, and converted into an illumination level per camera pixel as function of ambient temperature for 20 different LED patterns. Each line shows an individual polynomial fit of second order ($f(x) = a\,x^2+b\,x+c$, a, b, c being fit parameters) to the data of each pattern. Black (upper) points: brightness measured with a neutral density (ND) filter of $\sim$10\% transmission in front of the SiPM; red (lower) points: brightness measured without ND filter but additional amplifier to amplify the SiPM signal. Temperature dependence of second and third dimmest pattern measured with both set-ups, third dimmest pattern used to scale. (b) Flasher brightness converted into an illumination level per camera pixel as function of temperature cycling measurements for the same 20 LED combinations as used in (a). Different cycle lengths were used for measurements with and without (three dimmest LED patterns) ND filter. Points showing the data (results of the measurements), lines showing the ``corrected'' data after applying the temperature correction factor deduced from the polynomial fit of the data sets in (a). }
\end{figure*}
The flasher brightness dependence on temperature is different for each combination of the 10 LEDs (pattern). It shows an increase with temperature of $<$1\%/$^\circ$C for the brightest pattern up to 5\%/$^\circ$C for the dimmest (see Fig.~\ref{fig:flasher_temp_fit}). Fitting this dependence with a polynomial of second order gives different fit results for each combination which can be used to correct for temperature effects in other data sets (see Fig.~\ref{fig:flasher_temp_cycle_correction}). The spread after applying the temperature correction is about 3--8\% for the eight dimmest LED combinations decreasing to about 1\% for all other combinations. Whilst this resulting spread results in sufficient stability to utilise the LED flashers over the expected operating range of CHEC-M, some correlation with temperature clearly remains in Fig.~\ref{fig:flasher_temp_cycle_correction}. The resulting residuals from a perfectly stable response are larger whilst the temperature decreases -- implying a hysteresis in the temperature response that is not considered here (either in the derivation or application of temperature coefficients) and will be examined in the future. The long-term stability (measured over the time-scale of several days) shows a decrease in brightness ranging from 0.25\%/hour and $\sim$1\%/hour depending on the LED combination, recovering completely after a power down of one hour.

The measurements showed the flashers being appropriate devices for regular camera calibration with the possibility of absolute gain determination using the dimmest LEDs for SPE measurements and for monitoring changes in dynamic range and linearity of the full signal recording chain (MAPMs and FEE modules). Furthermore, the analytic description of the temperature dependence can be used to correct for changes in illumination in case temperature drifts occur.

\subsection{Camera and temperature stability}
\label{power_stability_cycling}
Several outdoor camera power cycle and temperature stability measurements were done to test the camera reliability and its behaviour with temperature.

\begin{figure*}[tb]
\centering
\subfigure[]{\includegraphics[width=0.49 \textwidth]{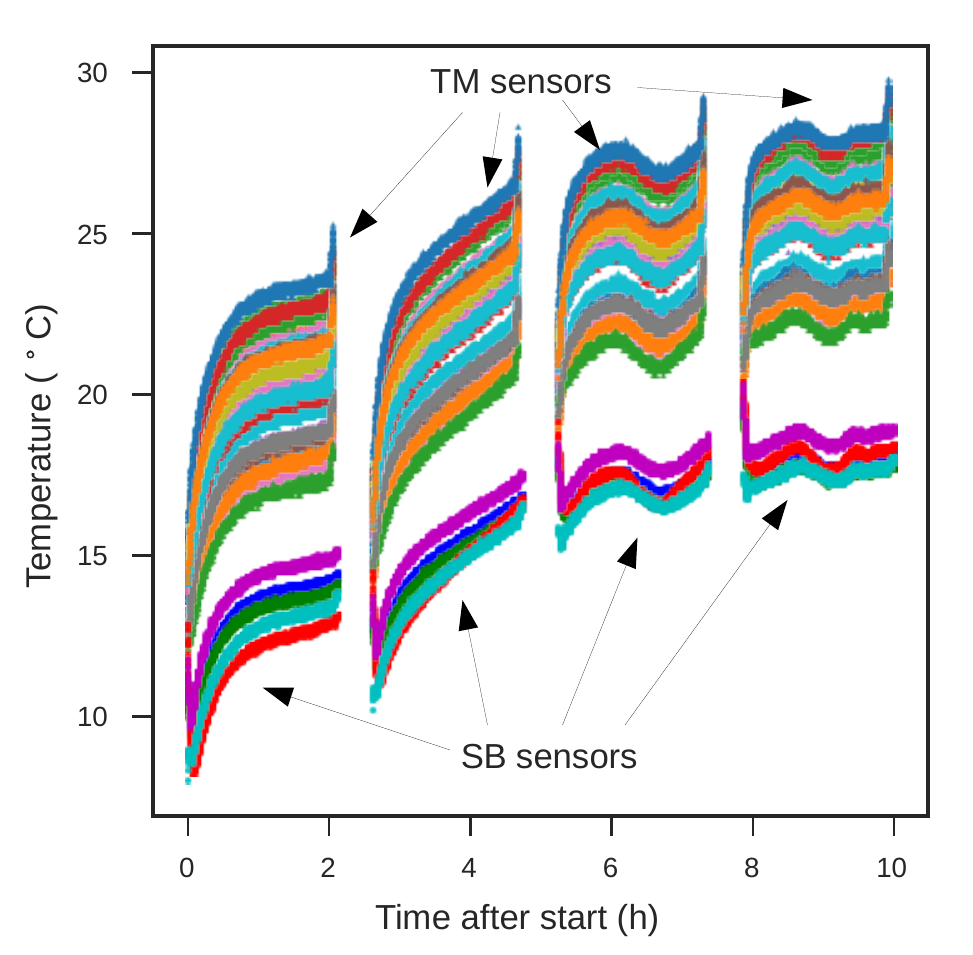}\label{fig:power_cycling_tests}}
\subfigure[]{\includegraphics[width=0.49 \textwidth]{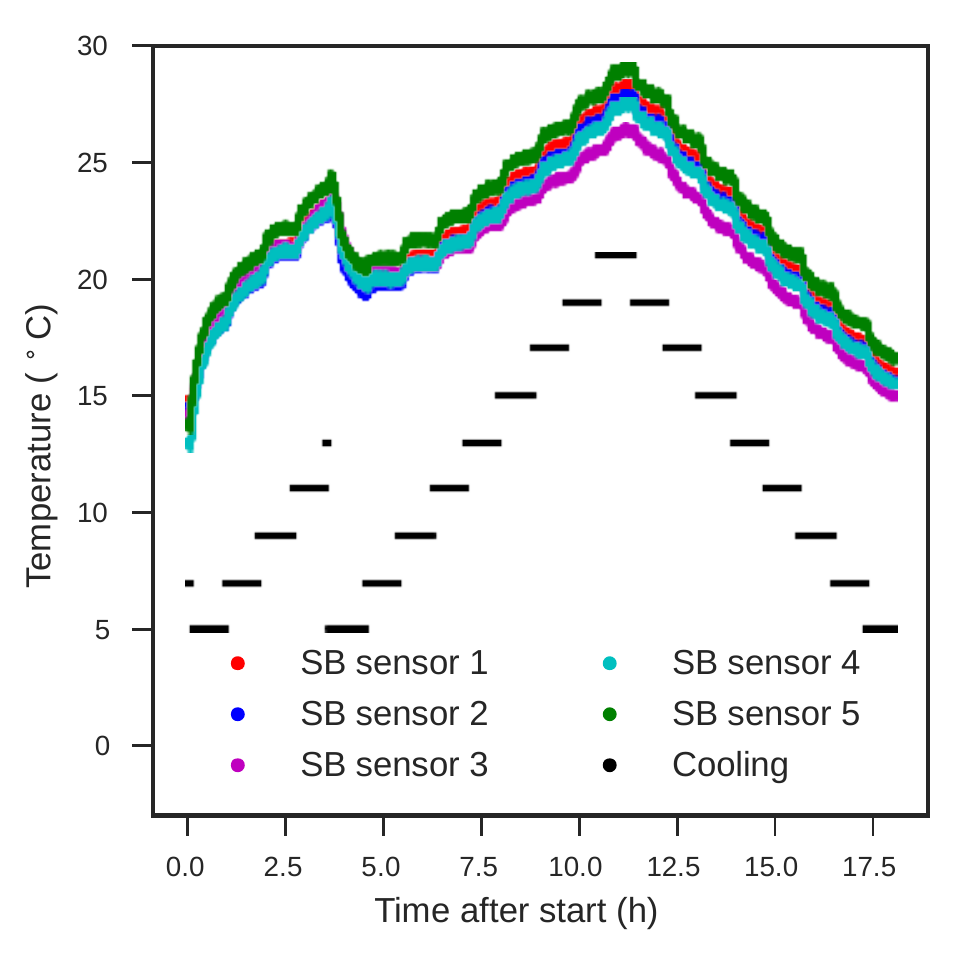}\label{fig:temperature_cycling_tests}}
\caption[]{Camera temperature as a function of time (a) measured with the TARGET module (TM) temperature sensors (upper data points) and the five safety board (SB) sensors (lower data points) during four power cycles keeping the chiller temperature at a fixed value of 5$^\circ$C and (b) measured with the SB sensors when the chiller cooling temperature was changed by one degree every hour while the camera was on and data was taken.}
\end{figure*}

Fig.~\ref{fig:power_cycling_tests} shows four power cycle measurements at a fixed chiller temperature of 5$^\circ$C starting with the first power cycle at sunrise on a spring day at an ambient temperature of $\sim$5$^\circ$C. It can be deduced that the maximum temperature difference between the TARGET modules in the camera at a certain time is about 6$^\circ$C and that the modules with the lowest temperature are those located at the top and bottom of the camera. This is expected since all auxiliary boards like safety, power, and DACQ boards are attached on the sides, while no boards are located at the top and bottom, and the fans are installed at the bottom. Furthermore, it was observed that the mean camera temperature varies over 8$^\circ$C over a complete day with the largest change occurring during sunrise and sunset. This effect could be corrected for by changing the chiller temperature accordingly to the ambient temperature. As shown by temperature cycling tests (Fig.~\ref{fig:temperature_cycling_tests}), the camera temperature can be controlled and maintained on a certain level by adapting the chiller temperature.

\begin{figure}[tb]
\centering
\includegraphics[width=0.5 \textwidth]{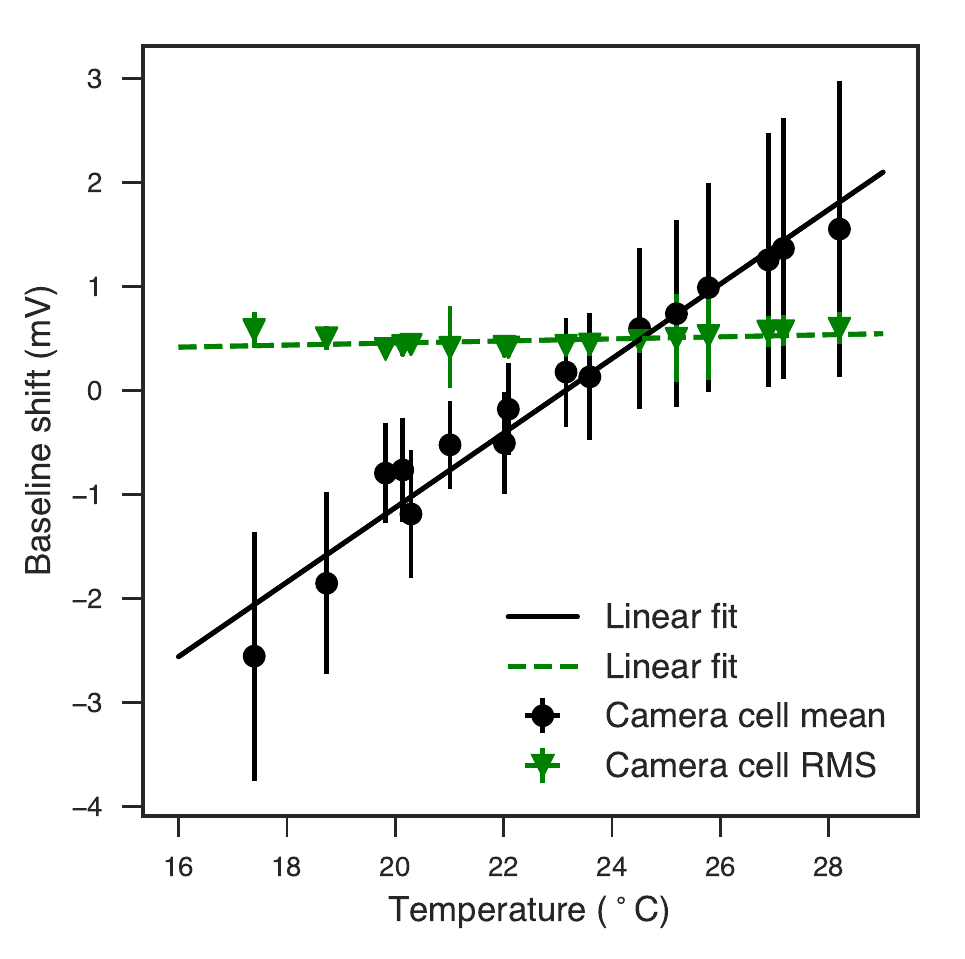}
\caption[]{Camera pedestal mean and standard deviation (std) as function of the temperature for the data set shown in Fig.~\ref{fig:temperature_cycling_tests}. The data is fitted with a linear function $g(T)=a\,T+b$ with temperature $T$ and resulting fit parameters of $a\sim0.36$~mV/$^\circ$C and $b\sim-8.30$~mV for the pedestal mean and $a\sim0.01$~mV/$^\circ$C and $b\sim 0.26$~mV for the standard deviation.}
\label{fig:pedestal_vs_temp}
\end{figure}

\begin{figure*}[tb]
\centering
\subfigure[]{\includegraphics[width=0.49 \textwidth]{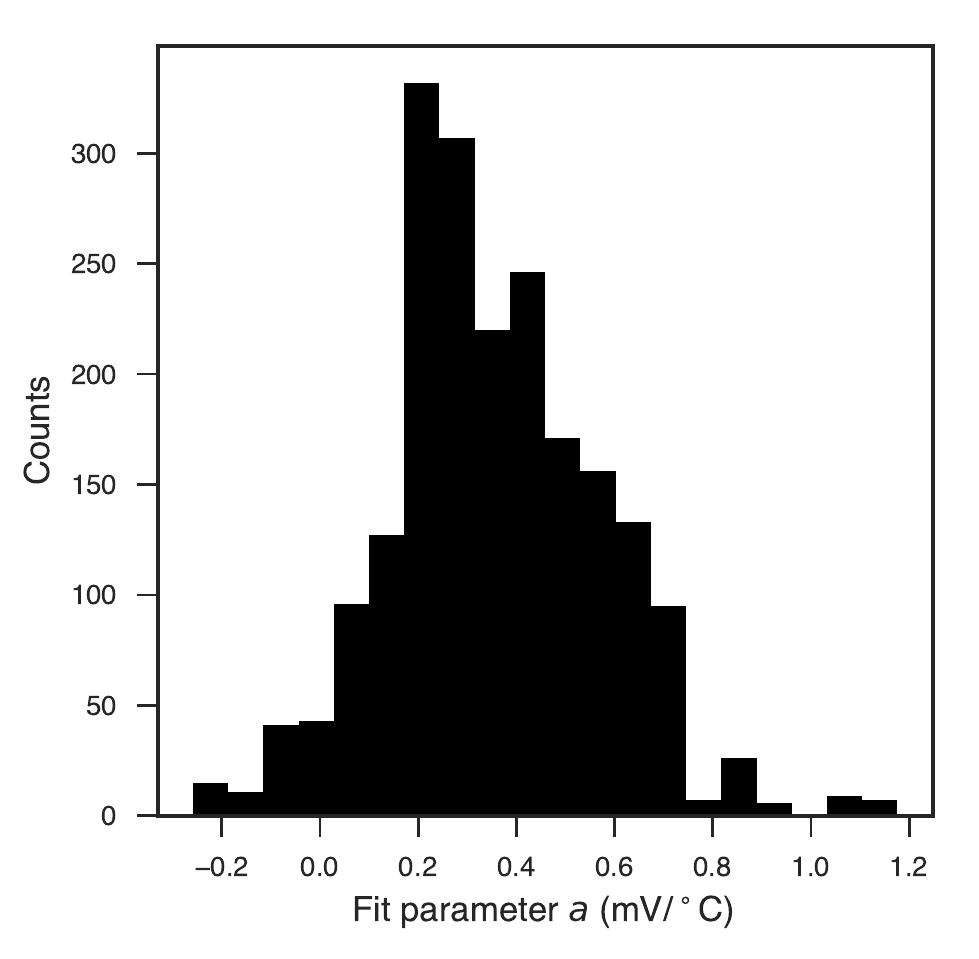}\label{fig:pedestal_vs_temp_parameter_distribution}}
\subfigure[]{\includegraphics[width=0.49 \textwidth]{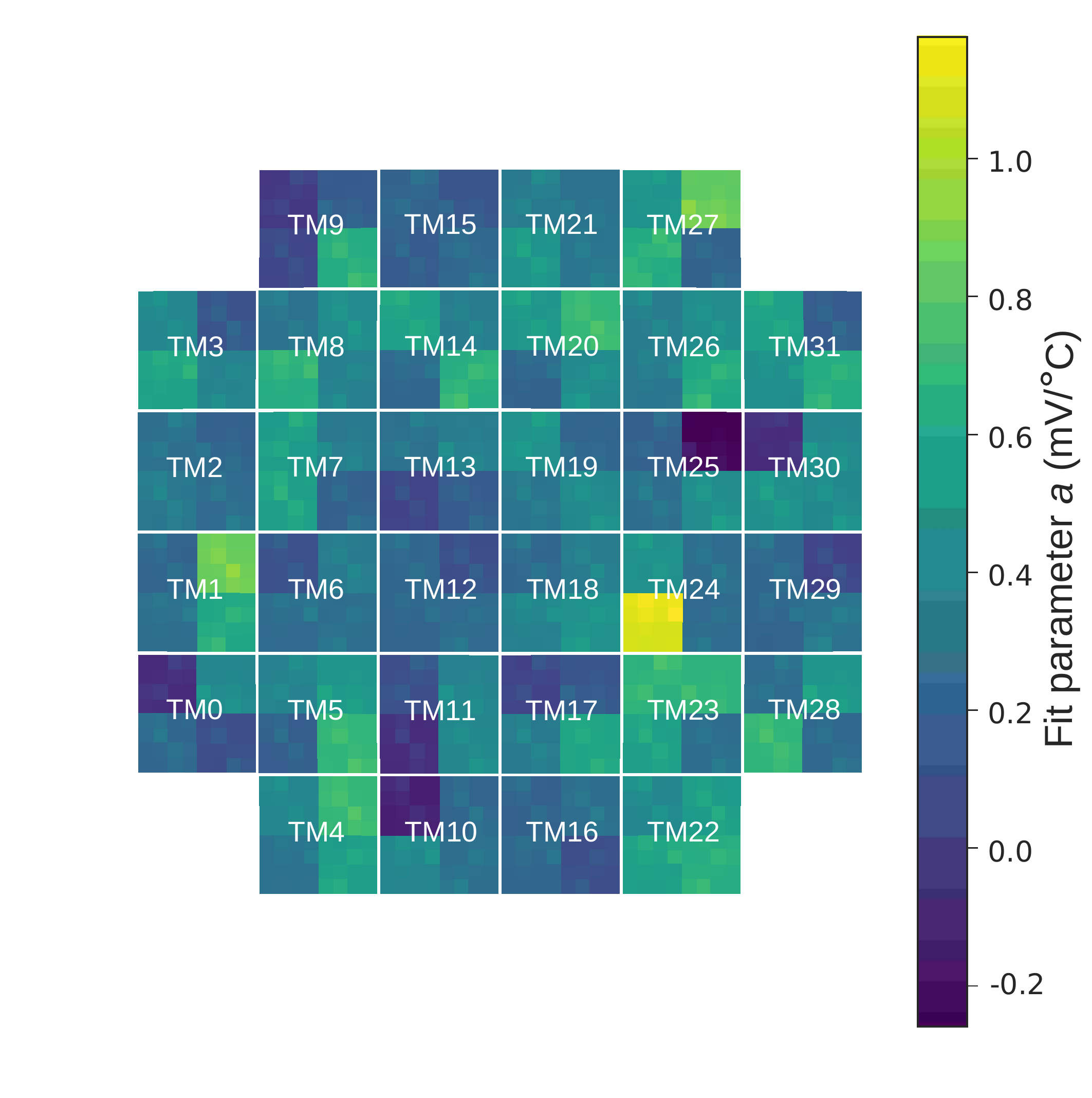}\label{fig:pedestal_vs_temp_parameter_camera}}
\caption[]{(a) Distribution of all 2048 pixel baseline temperature coefficients (fit parameters $a$ in linear fits, $g(T)=a\,T+b$) giving a mean of $\sim$0.35~mV/$^\circ$C and a standard deviation of $\sim$0.22~mV/$^\circ$C. (b) Camera image illustrating the 2048 pixel baseline temperature coefficients. Each block of 16 pixels is connected to the same ASIC and external DAC on the given TARGET module.}
\end{figure*}

The temperature dependence of the baseline of all 2048 pixels was investigated using the same chiller temperature cycle runs shown in Fig.~\ref{fig:temperature_cycling_tests}, in which externally triggered 45-minutes runs at 3~Hz for each chiller temperature were taken. In order to observe a relative change in the baseline with temperature, all 45-minutes run data was subtracted from a fast (600~Hz) reference pedestal run taken at the beginning of the temperature cycle runs at a chiller temperature of 5$^\circ$C. The resulting baseline-shift temperature dependence is fitted with a linear function $g(T)=a\,T+b$ with temperature $T$ and fit parameters $a$ and $b$ for the camera mean and for each pixel individually. The results show that the mean camera baseline shift is about 0.36~mV/$^\circ$C (see Fig.~\ref{fig:pedestal_vs_temp}). A spread in the temperature dependence between individual pixels is observed (some of them even with opposed sign, see Fig.~\ref{fig:pedestal_vs_temp_parameter_distribution}) with minimum and maximum temperature coefficient of $a_{\rm{min}}\sim-0.26$~mV/$^\circ$C and $a_{\rm{max}}\sim1.18$~mV/$^\circ$C, respectively.
This spread can be attributed to different temperature behaviours of either the ASICs or the external DACs providing the input signal Vped to each ASIC\footnote{In the next TARGET module generation, the offset Vped will not be supplied by an external DAC anymore but by the ASIC itself.}. This is illustrated by Fig.~\ref{fig:pedestal_vs_temp_parameter_camera} where it can be observed that pixels connected to the same ASIC and external DAC show similar baseline temperature coefficients. However, even though the baseline of different pixels have different temperature dependencies, only a small temperature dependence of the pixel baseline standard deviation is observed (see Fig.~\ref{fig:pedestal_vs_temp}). This means that all 16384 storage cells of one pixel have similar temperature dependencies. Thus, one linear correction factor per pixel is enough to characterise or correct for the temperature dependence.

An uncertainty on the baseline of $\pm$~0.5~mV is not expected to affect the camera performance in terms of charge resolution. However, in case the camera is not kept at a constant temperature level within $\pm$~1$^\circ$C, either a pixel dependent linear correction factor has to be used or the pedestal must be remeasured every time the camera temperature changes by more than 1$^\circ$C. A pedestal measurement takes about 30~s and the HV needs to be off or the lid closed. Thus, this approach would cause a maximum dead time of $\sim$0.8\% per night (according to a very simplified calculation, assuming a linear temperature drift of 8$^\circ$C within an eight-hours night). To avoid that, a pedestal determination could be done ``online'' during observation instead, either with interleaved events or by using parts of the waveforms/pixels with no signals. The latter option is the long-term plan for next camera iterations.

\begin{figure*}[tb]
\centering
\subfigure[]{\includegraphics[width=0.32 \textwidth]{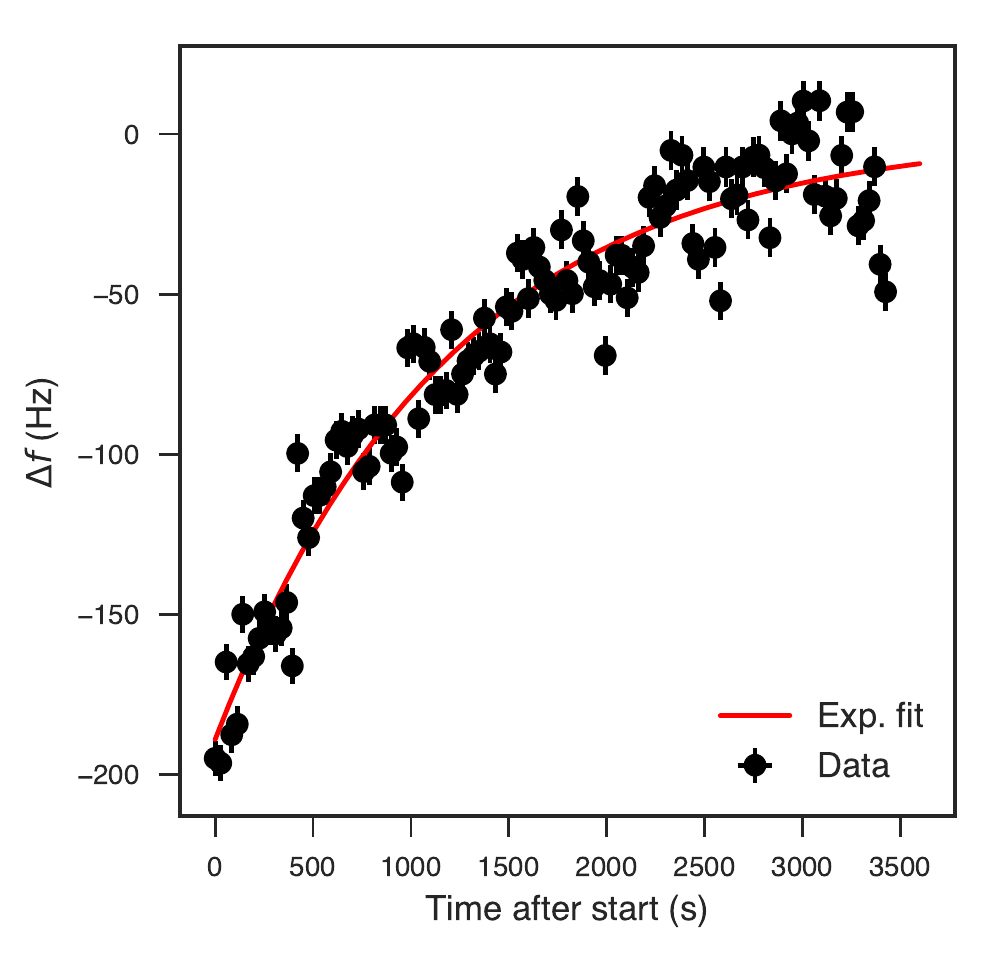}\label{fig:trigger_rate_vs_time}}
\subfigure[]{\includegraphics[width=0.32 \textwidth]{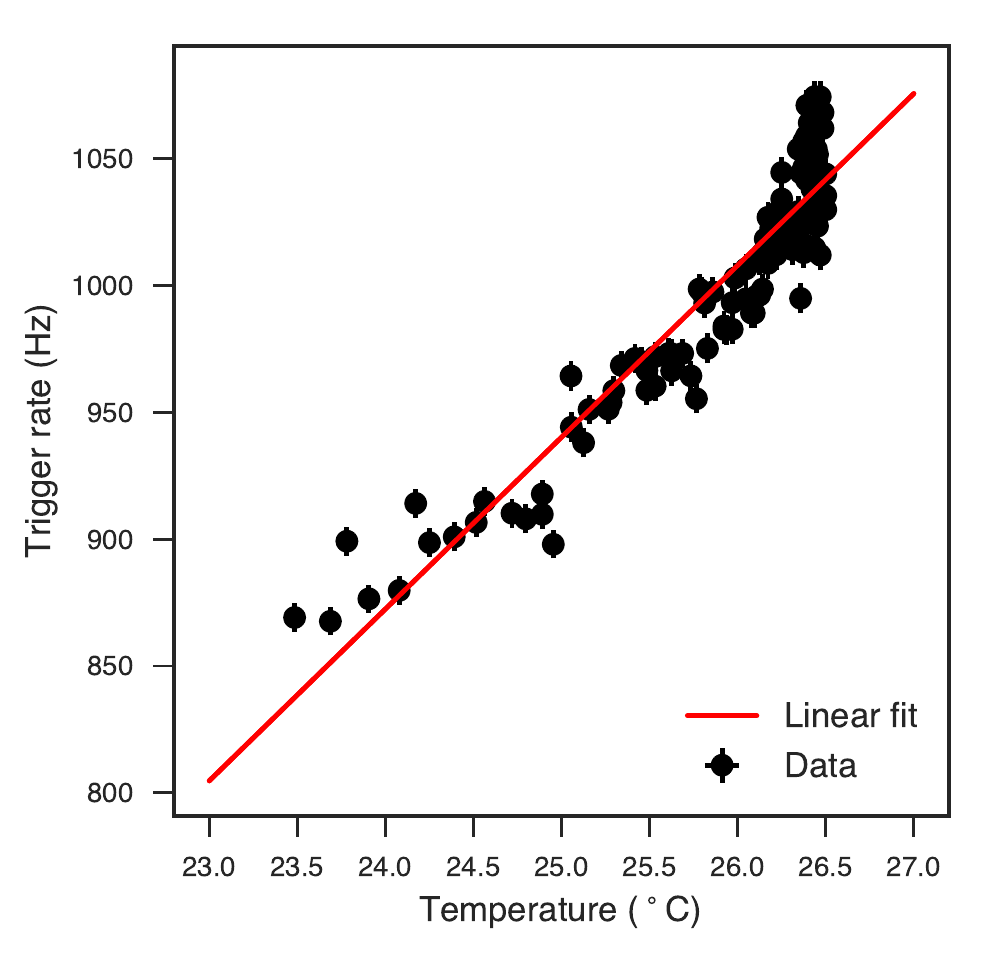}\label{fig:trigger_rate_vs_temp}}
\subfigure[]{\includegraphics[width=0.32 \textwidth]{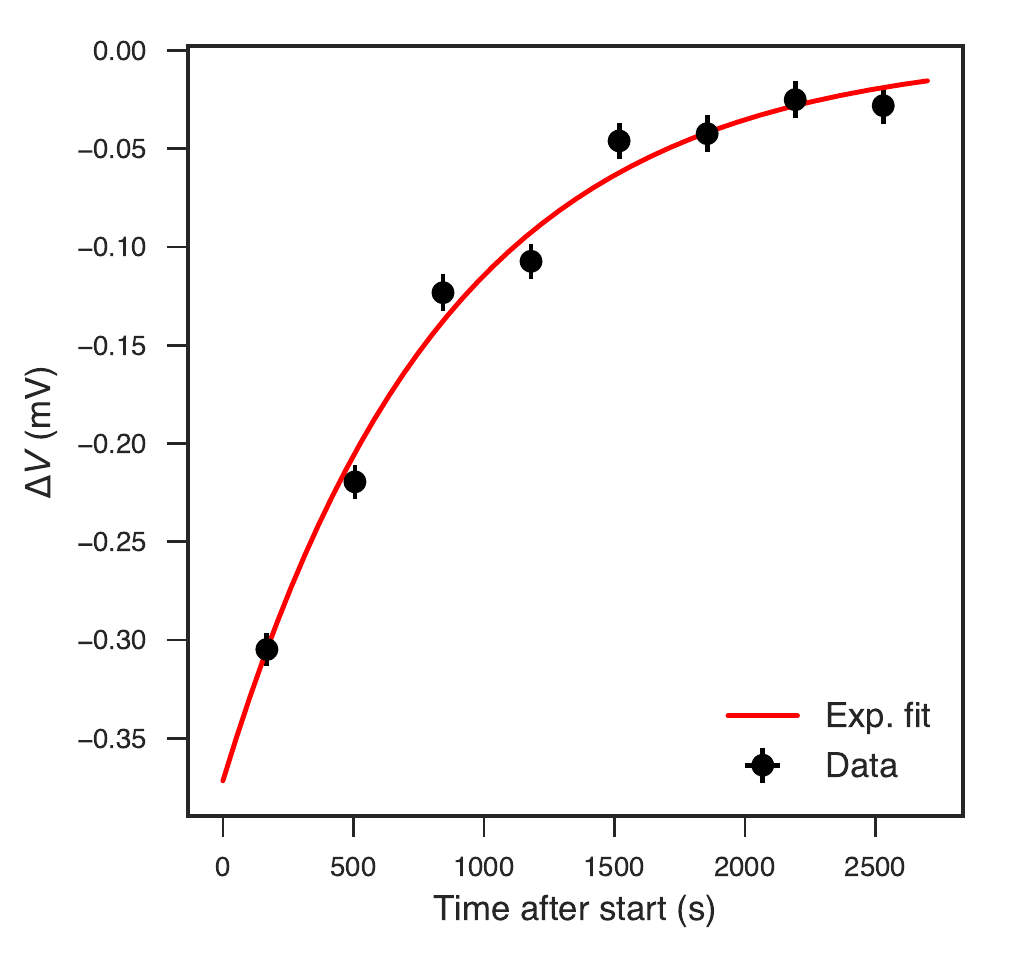}\label{fig:pedestal_vs_time}}
\caption[]{(a) Trigger rate change $\Delta f$ as function of time after camera booting, fitted with an exponential function $g(t)=a\,(1-\exp(-t/\tau_{\rm{t}}))+b$ with time $t$ and resulting fit parameters of $a\sim189$~Hz, $b\sim875$~Hz, and $\tau_{\rm{t}}\sim1193$~s. The trigger rate change at time $i$ is defined as $\Delta f_i = f_i-\lim\limits_{t \rightarrow \infty}{g(t)}$ with $f_i$ being the measured rate at time $i$. (b) Trigger rate as function of temperature for the same data as shown in (a), fitted with a linear function $g(T)=a\,T+b$ with temperature $T$ and resulting fit parameters of $a\sim67.74$~Hz/$^\circ$C and $b~\sim-753.19$~Hz. (c) Camera mean baseline drift $\Delta V$ as function of time after camera booting, fitted with an exponential function $g(t)=a\,(1-\exp(-t/\tau_{\rm{b}}))+b$ with time $t$ and resulting fit parameters of $a\sim0.37$~mV, $b\sim0.61$~mV, and $\tau_{\rm{b}}\sim849$~s. The camera mean baseline drift at time $i$ is defined as $\Delta V_i = V_i-\lim\limits_{t \rightarrow \infty}{g(t)}$ with $V_i$ being the measured camera mean baseline at time $i$.}
\end{figure*}

To investigate the camera warm-up, two relevant quantities were investigated: first, the change in trigger rate with closed lid and HV off, indicating whether the trigger threshold is stable and the camera is ready for triggering, and second, a possible shift in the baseline indicating whether the camera is ready for data taking. To measure the trigger rate change during warm-up, the trigger threshold was set in the electronic noise causing the trigger rate to be very sensitive to electronic noise changes expected to occur during warm-up\footnote{In observing mode, the trigger threshold will be set well above the electronic noise level so the trigger rate will be less sensitive to electronic noise changes. It will be dominated by NSB and Cherenkov events.}. Fig.~\ref{fig:trigger_rate_vs_time} shows the trigger rate change as function of the time after camera power up. It is fitted with an exponential function $g(t)=a\,(1-\exp(-t/\tau_{\rm{t}}))+b$ with time $t$ and fit parameters $a$, $b$, and $\tau_{\rm{t}}$, showing an increase over time with a time constant of $\tau_{\rm{t}}\approx1193$s.
This increase is connected with a temperature increase during warm-up (as shown by Fig.~\ref{fig:trigger_rate_vs_temp}) and very likely caused by the baseline drift with temperature explained above. An increasing baseline in fact reduces the threshold, causing a higher trigger rate. Thus, the mean camera baseline is expected to show an increase during warm-up which is shown by Fig.~\ref{fig:pedestal_vs_time}. 
The resulting time constant is $\tau_{\rm{b}}\sim849$s with a mean baseline drift of about 0.35~mV over $\sim$40~min being less than 1~p.e. In total, the camera is assumed to be stabilised/warmed up when the trigger rate change or the mean camera baseline drift is less than 5\% compared to the asymptotic value $\lim\limits_{t \rightarrow \infty}{g(t)}$. This is the case after $\sim$3\,$\tau$ ($\sim$1~h for the trigger rate change and $\sim$45~min for the mean camera baseline drift).

The camera parameters investigated in this section (temperature, trigger rate change, and baseline drift) are the most basic ones which allow one to characterise the general functionality of the camera and to answer the most basic questions like: Can the temperature be controlled and stabilised? By how much does the temperature change over a day? What is the temperature distribution and its spread inside of the camera? Can the camera be triggered reliably? Can data be taken reliably and is it affected by temperature or time? Other measurements for more detailed stability and temperature analyses (like stability and temperature dependence of timing and gain) are planned for the next camera prototype, especially since due to the use of SiPMs the performance of CHEC-S is expected to show more serious temperature dependencies than the performance of CHEC-M (see Sec.~\ref{outlook}).

\section{Cherenkov events}
\label{cherenkov}
%
%
First Cherenkov light with CHEC-M was observed in November 2015 during a first campaign with the camera deployed on the GCT telescope prototype located at the Paris Observatory in Meudon near Paris \cite{Watson:2016zyk}. A second campaign was carried out in Spring 2017.

\begin{figure*}[tb]
\centering
\subfigure[]{\includegraphics[width=0.33 \textwidth]{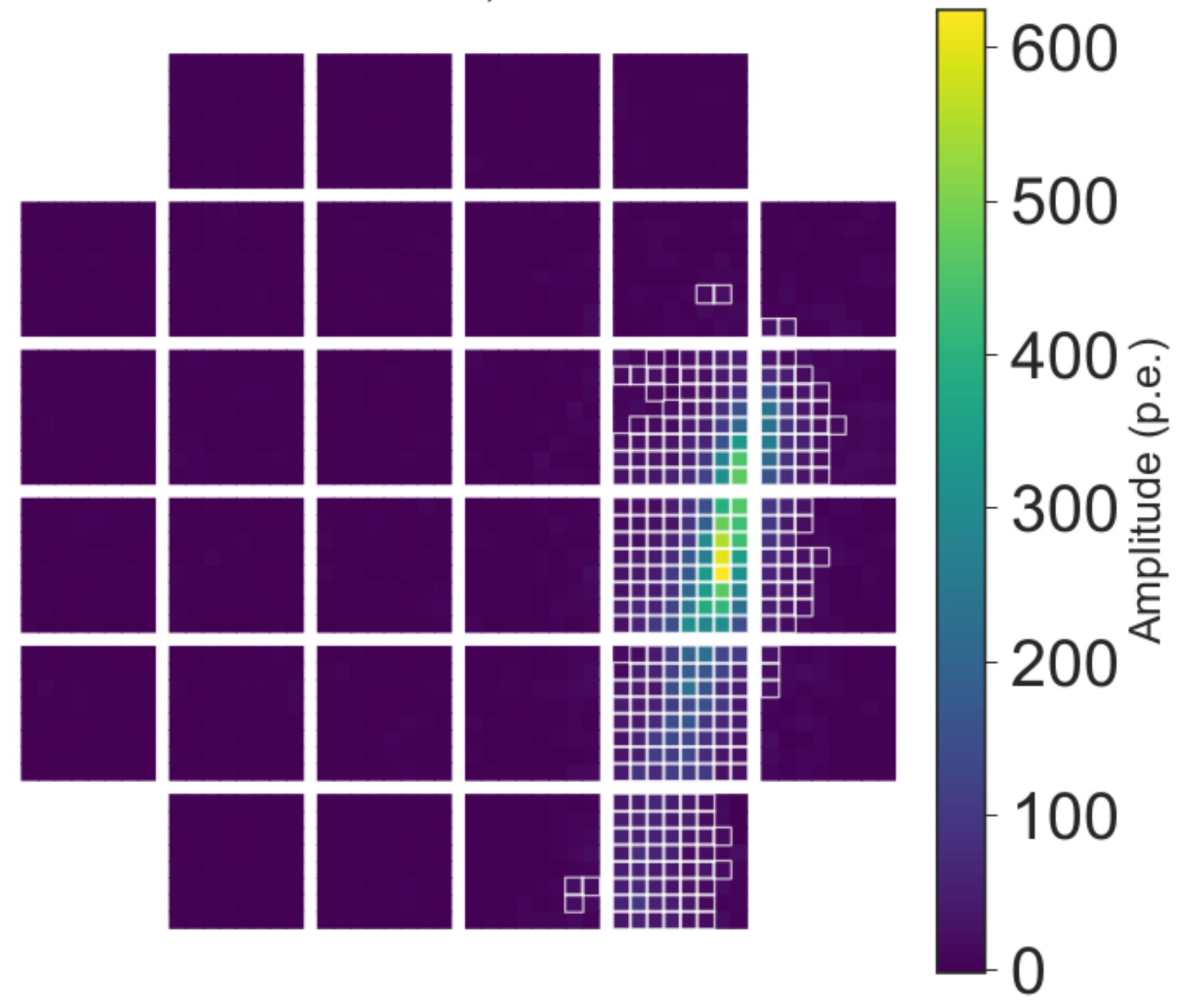}\label{fig:ch_ev11}}
\subfigure[]{\includegraphics[width=0.33 \textwidth]{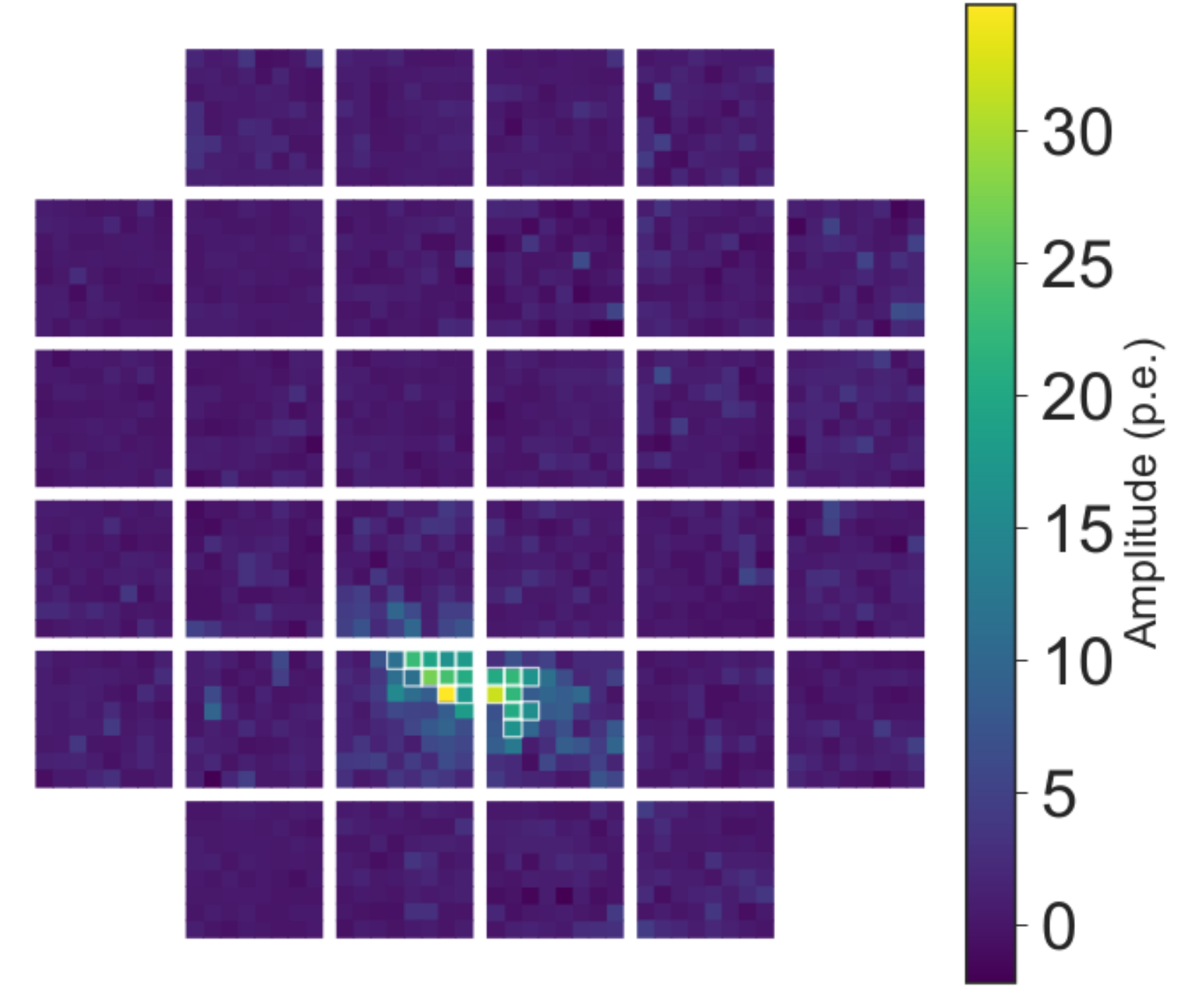}\label{fig:ch_ev236}}
\subfigure[]{\includegraphics[width=0.33 \textwidth]{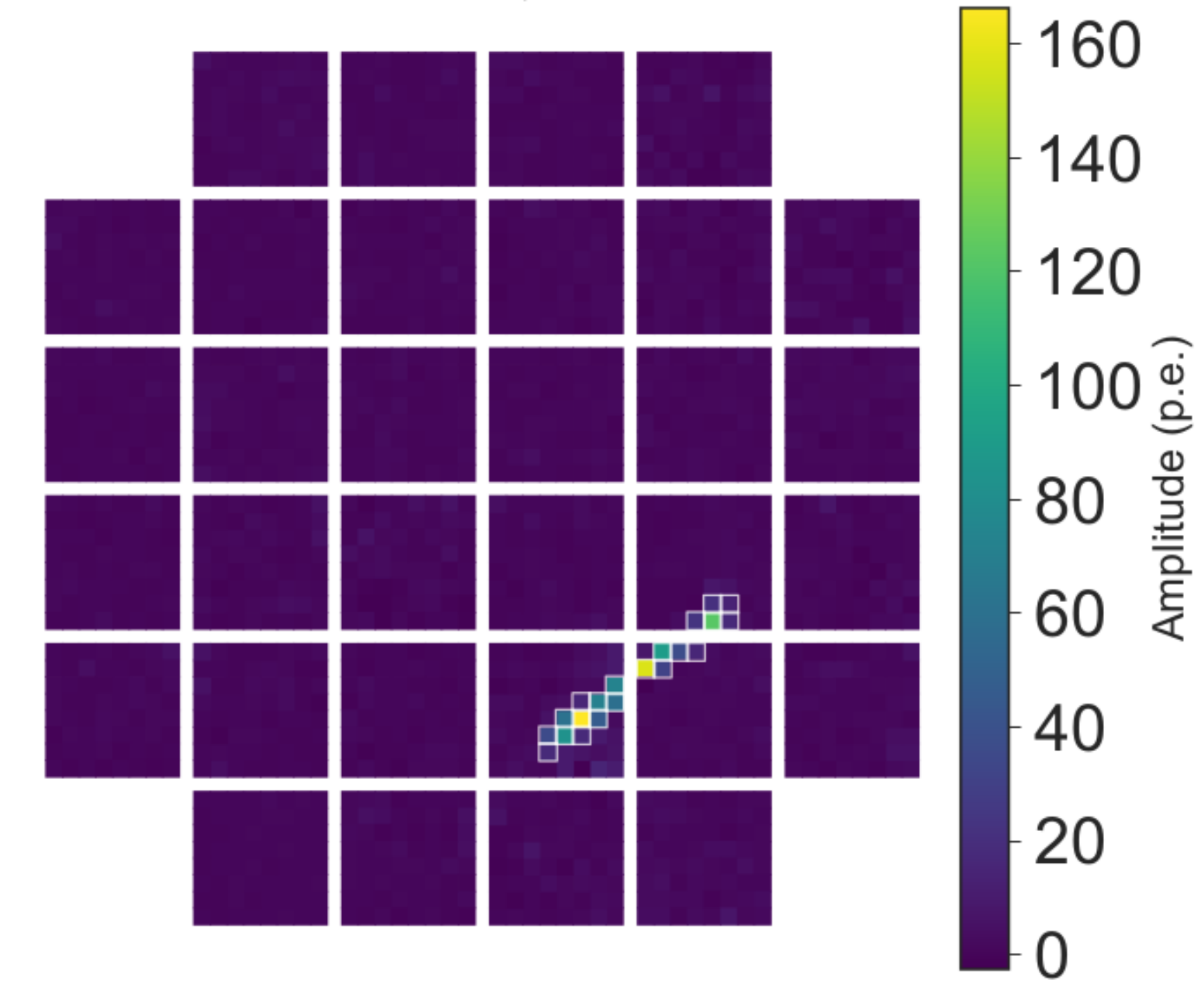}\label{fig:ch_ev126}}\\
\subfigure[]{\includegraphics[width=0.33 \textwidth]{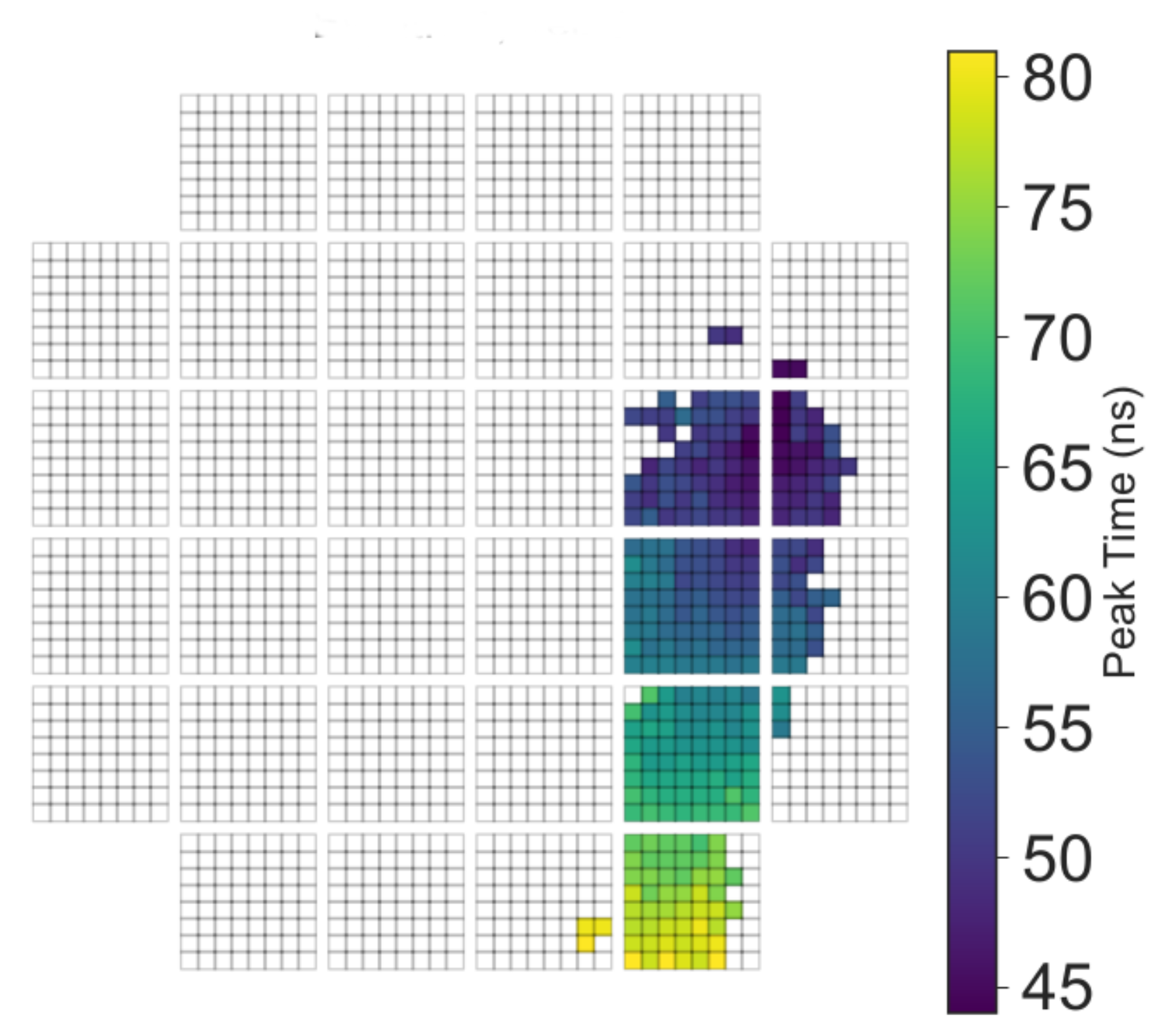}\label{fig:ch_ev11_timing}}
\subfigure[]{\includegraphics[width=0.33 \textwidth]{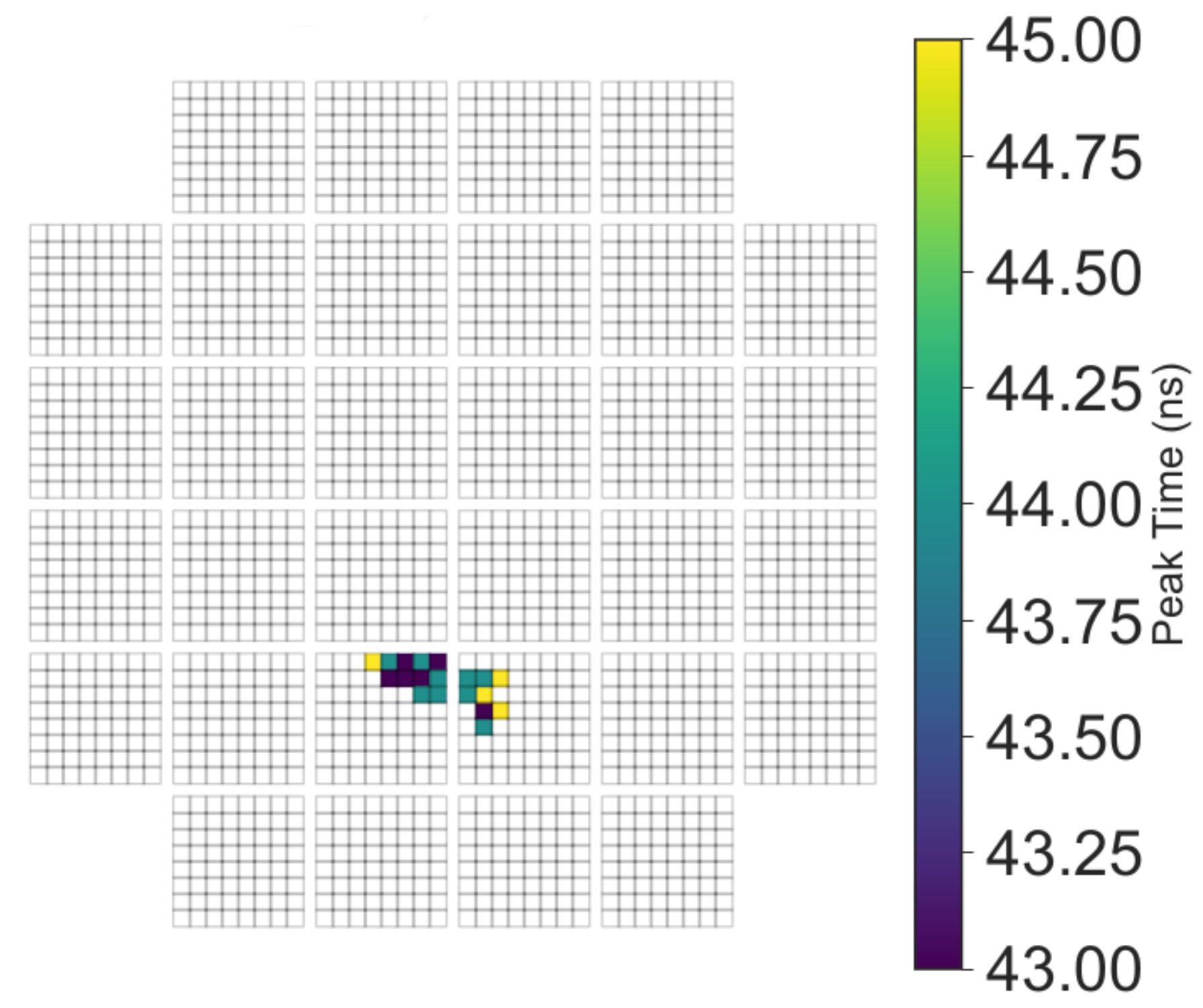}\label{fig:ch_ev236_timing}}
\subfigure[]{\includegraphics[width=0.33 \textwidth]{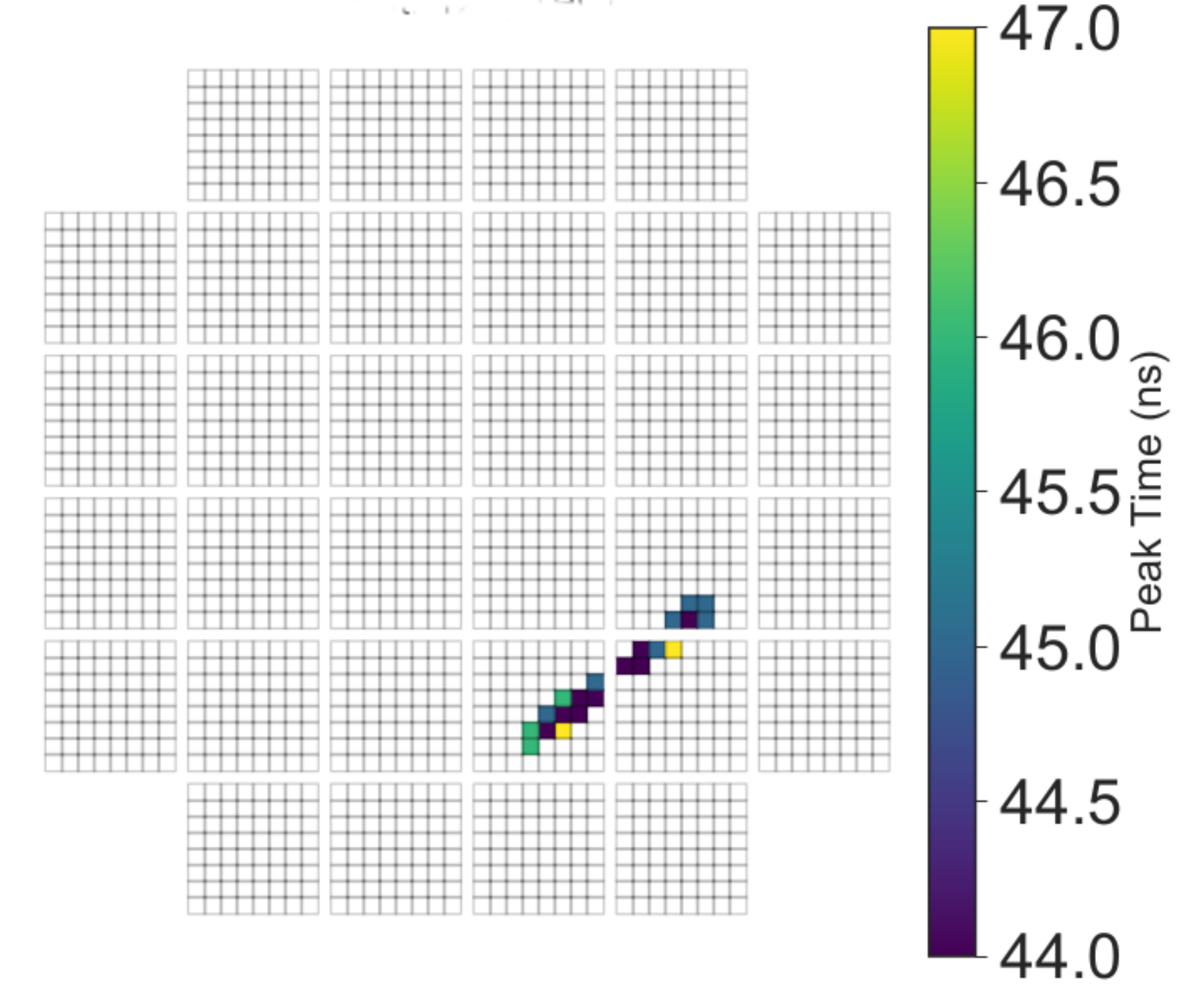}\label{fig:ch_ev126_timing}}
\caption[]{Camera images of three different events, showing the intensity (a, b, and c) and the peak arrival time (d, e, and f) for each pixel. The white squares in (a), (b), and (c) indicate the pixels that remain after the tail-cut cleaning. All modules were active in this observation run. For further explanation refer to the text.}
\label{fig:ch_events}
\end{figure*}

Due to the high NSB light level in Meudon, estimated to be 20 to 100 times brighter than at the CTA site, the camera was operated at a low gain (mean HV of 800\,V) and at trigger threshold setting 3 (ref.~Fig.~\ref{fig:camera_rate_laser_setting}), corresponding to roughly 11~p.e., pushing the trigger rate down to only $\sim$0.1~Hz. Two examples of on-sky events (telescope pointing to the sky, camera lid open, HV on) and one event with telescope in park position (0$^\circ$ elevation), camera lid closed, but HV switched on, all three recorded during the second campaign are shown in Fig.~\ref{fig:ch_events}.
The upper images show the intensity in p.e.~(integrated charge) for each pixel, while the lower ones indicate the peak arrival time (after trigger) for each pixel for the same events. Furthermore, the white boxes indicate pixels surviving image cleaning (cf.~Sec.~\ref{waveform_processing}). As expected for a Cherenkov flash from a shower, the timing plots of the Cherenkov events (Fig.~\ref{fig:ch_ev11_timing} and \ref{fig:ch_ev236_timing}) show the image propagating across the focal plane in time. Whereas the first event (Fig.~\ref{fig:ch_ev11} \& \ref{fig:ch_ev11_timing}) could have been a shower with a large impact distance thus showing a rather large time gradient of about 35~ns, the second event (Fig.~\ref{fig:ch_ev236} \& \ref{fig:ch_ev236_timing}) could have been an inclined shower with the telescope being at the edge of the Cherenkov light pool causing all pixels with Cherenkov signal being illuminated at a very similar time. The event shown by Fig.~\ref{fig:ch_ev126} and \ref{fig:ch_ev126_timing} was recorded with closed lid in park position. It must have been a cosmic ray induced particle travelling through the curved MAPM array of the camera. The unique geometry and fast time profile of such an event make it easy to be isolated from Cherenkov events in the analysis afterwards.

The additional timing information in both the Cherenkov and direct cosmic ray events is only possible due to the waveform sampling nature of the camera electronics and is useful for advanced image cleaning, background rejection, and event reconstruction algorithms. Additionally, images at the highest energies can take many tens of nanoseconds to cross the camera, as can be seen in Figure \ref{fig:ch_ev11_timing}. Without a $\sim$100~ns readout window, such images would appear truncated, negatively impacting the analysis.

With the on-sky data taken with the CHEC-M camera we proved both the technical functionality of the camera and the existence of a data calibration and analysis chain, both aspects being crucial for the camera to be used as an IACT camera. Both campaigns helped to verify interfaces and to improve operation procedures. Furthermore, the regular operation was used for understanding the system stability and reliability. In total, a few hundred meaningful Cherenkov and direct EAS particle events were recorded and analysed.

\section{Summary and outlook}
\label{outlook}
%
%
CHEC-M is an invaluable step towards a reliable and high-performance product for CTA. Regular operation of CHEC-M has shown that the camera control and data acquisition using the software CHECInterface and the calibration and waveform processing chain are robust and reliable. Intensive lab tests led to a detailed characterisation of the camera performance as well as a detailed understanding of the factors limiting the performance, thus being a critical input to the design of the next camera prototype CHEC-S, see below). Main results are:
\begin{itemize}
\item CHEC-M can be read out at an efficiency of 95\% at a mean random rate of 600~Hz. The efficiency is expected to be 100\% at this rate with the next TARGET-5 ASIC generation used in CHEC-S.
\item While commissioning and testing of CHEC-M, 128 TARGET ASICs have been tested and tuned simultaneously confirming the results obtained in single TARGET ASIC tests (cf.~\cite{2017APh....92...49A}).
\item Even after gain matching, the spread in gain between pixels is $\sim$30\%  -- the limiting factor in trigger threshold uniformity. This is due to the fact that the HV can only be set individually for each MAPM but not for each pixel -- a fundamental feature of the MAPM design. The gain spread is reduced significantly when using SiPMs as photosensors (which is the case in CHEC-S).
\item Camera trigger thresholds are characterised by 512 pairs of two ASIC parameters for each threshold which can be determined in lab measurements with a laser. In this way, five different threshold sets were defined for on-site tests between $\sim$2 and $\sim$170~p.e., intermediate thresholds can be identified by interpolation. A trigger rate scan over these settings can be used to identify the camera operating point on site.
\item Reading out TARGET-5 modules leads to additional, false, triggers resulting in increased dead-time and the need of an increased trigger threshold, both issues being addressed by decoupling sampling and triggering into two separate ASICs in the next TARGET module generation.
\item The pulse shape characteristics (FWHM and 10--90\% risetime) fulfil the requirements for trigger performance optimisation (5--10~ns and 2--6~ns, respectively).
\item The time resolution between different pixels hit simultaneously by the same laser flash is better than 1~ns for illumination levels $>$6~p.e. and thus also fulfils the requirements. 
\item The crosstalk reaches a maximum of 6\% between neighbouring pixels, affecting charge resolution, trigger efficiency, and camera image reconstruction.
\item The dynamic range of the signal recording chain (MAPM, preamplifier, and TARGET module) covers a range from the sub-p.e.~level to $\sim$1000 p.e.~at the highest possible HV of 1100~V and can be shifted towards higher signal amplitudes (factor $\sim$6) by reducing the gain.
\item The LED calibration flashers were shown to be appropriate devices for regular camera calibration and monitoring in terms of absolute gain and dynamic range determination. For the future, a slight change in their design is considered to avoid different temperature dependencies and to improve the predictability of absolute and relative brightnesses of different LED patterns.
\item The camera and temperature stability were assessed showing that
\begin{itemize}
\item the camera temperature can be controlled and (if required) kept at a constant level by adapting the chiller temperature,
\item the camera warm-up time is of the order of 1 hour, and
\item a mean baseline drift with temperature of 0.8~ADC/$^\circ$C is observed and is mitigated by maintaining a constant camera temperature within 1$^\circ$C or by taking regular 30~s pedestal calibration runs.
\end{itemize}
\end{itemize}

The on-telescope campaigns have not only provided a useful test-bed to assess operational and maintenance procedures, but have also -- for the first time -- demonstrated the use of Schwarzschild--Couder optics to collect atmospheric Cherenkov light. Data taken during the two campaigns have proven useful in the development of the data analysis chain and in understanding the levels of calibration that will be required for CTA.

Despite the level of success achieved with the prototype, CHEC-M does not meet all CTA performance requirements, with the non-uniformity in gain and the trigger noise incurred with sampling enabled being of greatest concern. CHEC-S, a second full-camera prototype based on SiPMs, is currently being commissioned and tested, and will address the limiting factors in the CHEC-M performance by design. 

The use of SiPMs allows gain measurements to be made easily for a range of bias voltages and input illumination levels (and even in the absence of light from the dark counts intrinsic to the SiPM) thereby improving calibration and charge reconstruction accuracy. The SiPM gain spread between pixels in a camera module is intrinsically smaller than with MAPMs, and the bias voltage is adjustable per superpixel, allowing gain matching to much higher precision than in CHEC-M. The gain of SiPMs is temperature sensitive, and for the devices used in CHEC-S will drop by approximately 10--20\% over a 10$^\circ$C increase. A liquid-cooled focal plane plate will stabilise the temperature to within $\pm$1$^\circ$C over time scales for which the gain may easily be re-measured in-situ. The average detection efficiency in the focal plane will also increase, due to improved photo-detection efficiency, better angular response\footnote{MAPMs need a protective glass worsening their angular response compared to SiPMs.}, and a reduced level of dead space between photodetectors. NSB rates are expected to be higher than in the MAPM case, due to the different wavelength dependence of the SiPM response, however, this background increase should be compensated for by the improvement in efficiency for signal photons. Dark count rates from the SiPMs at the nominal operating gain and temperature have been measured to be less than 20\% of the expected dark sky NSB rate, ensuring a negligible impact on performance. 

The FEE of CHEC-S will also see a substantial upgrade from CHEC-M. Due to the undesirable coupling between sampling and triggering in the TARGET 5 ASICs, these functionalities have been split into two separate ASICs. The first ASIC, T5TEA, provides triggering based on the same concept as TARGET 5, with a sensitivity reaching the single p.e.~level and a trigger noise of 0.25~p.e.~for the CHEC-S gain. The second ASIC, TARGET C, performs sampling and digitisation, with a $\sim$70\% larger dynamic range and with an improvement in charge resolution by a factor $>$2 with respect to TARGET 5~\cite{2017AIPC.1792h0012F}. The operational requirement for an 80~$\mu$s hold-off time has therefore been removed, allowing operation dead time free at the required event trigger rate of 600~Hz.

In the final CHEC design, the absolute timing will be provided by a unified clock and trigger timestamping board. The board acts as an interface between the camera and the CTA timing system and is based on White Rabbit technology \citep{WR2009}. It provides clock signals to the camera with the required precision that are phase-locked to the central master clock. In return, it adds absolute timestamps to the camera events. Furthermore, this board will have the capability to trigger the backplane and LED calibration flashers (synchronised with the internal clock), thus replacing the need of an external trigger device. A prototype of this board will be integrated into CHEC-S.

Once prototyping is complete, we plan to construct and deploy three CHEC cameras on the southern-hemisphere CTA site during a pre-production phase. During the production phase of CTA we aim to provide cameras for a significant fraction of the 70 baseline SSTs. It is expected that the majority of components used in CHEC-S will also be used in the final production design of CHEC with the exception of the photosensors. SiPM technology is rapidly evolving and the latest devices offer significant performance improvements compared to the SiPMs used in CHEC-S, including increased photo-detection efficiency, lower optical crosstalk and a reduced dependency of the gain on temperature~\cite{tajima_sipm,Otte:2016aaw,ASANO2017}. Additionally, there may be performance advantages associated with an enlarged FoV that may be obtained by using 7~mm rather than 6~mm pixels. Laboratory tests of the latest SiPMs and simulations with different pixel sizes are ongoing.

\section*{Acknowledgements}
\label{ackn}
%
This work was conducted in the context of the CTA GCT project. We gratefully acknowledge financial support from the agencies and organisations listed here: http://www.cta-observatory.org/consortium\_acknowledgments

Furthermore, we thank the Paris Observatory as well as the DT-INSU for their support during the on-sky campaigns in Meudon. This study was also supported by JSPS KAKENHI Grant Numbers JP17H04838, JP25610040, JP15H02086, and JP23244051. A.~Okumura was supported by a Grant-in-Aid for JSPS Fellows.

\vspace{0.5cm}
This paper has gone through internal review by the CTA Consortium.


\section*{References}
\bibliographystyle{elsarticle-num-names}
\bibliography{mybibfile}

\end{document}